\documentclass[12pt]{article}
\usepackage{amsmath}
\usepackage{graphicx}
\usepackage{enumerate}
\usepackage{natbib}
\usepackage{url} % not crucial - just used below for the URL

\usepackage{mathrsfs}
\usepackage{amsfonts}
\usepackage{graphicx}
\usepackage{mathrsfs}
\usepackage{amssymb}
\usepackage{rotating,color}
\usepackage{xcolor}
\usepackage{xr}

\allowdisplaybreaks[4]

\newcommand{\be}{\begin{equation}}
\newcommand{\ee}{\end{equation}}
\newcommand{\beaa}{\begin{eqnarray*}}
\newcommand{\eeaa}{\end{eqnarray*}}
\newcommand{\bea}{\begin{eqnarray}}
\newcommand{\eea}{\end{eqnarray}}
\newcommand{\lbl}{\label}

\newcommand{\ml}{\mathcal}

\newcommand{\bd}{\bold}

\newcommand{\red}[1]{\color{red}}

\newtheorem{theorem}{ \noindent T{\footnotesize HEOREM}}
\newtheorem{prop}{ \noindent P{\footnotesize ROPOSITION}}
\newtheorem{lemma}{ \noindent L{\footnotesize EMMA}}
\newtheorem{coro}{ \noindent C{\footnotesize OROLLARY}}
\newtheorem{remark}{ \noindent R{\footnotesize EMARK}}

\def\var{\mathrm {var}}

\def\diag{\mathrm {diag}}

\newcommand{\bm}{\boldsymbol}

\def\tr{\mathrm {tr}}

\def\M{{\bf M}}
\def\B{{\bf B}}
\def\z{{\bm z}}

\def\S{{\bf S}}
\def\R{{\bf R}}
\def\I{{\bf I}}

\def\bmu{{\bm \mu}}

\def\H{{\bf H}}

\def\X{{\bm X}}

\def\bme{{\bm \varepsilon}}
\def\Y{{\bm Y}}

\def\W{{\bm W}}
\def\b{{\bm b}}

\def\D{{\bf D}}

\def\tr{\mathrm {tr}}

\def\bms{{\bm\Sigma}}

\def\mR{\mathbb{R}}

\def\bb{\bm \beta}
\def\mY{\bm Y}
\def\wX{\tilde{\bf X}}
\def\mX{{\bf X}}

\newcommand{\blind}{1}

\begin{document}

\def\spacingset#1{\renewcommand{\baselinestretch}%
{#1}\small\normalsize} \spacingset{1}

%%%%%%%%%%%%%%%%%%%%%%%%%%%%%%%%%%%%%%%%%%%%%%%%%%%%%%%%%%%%%%%%%%%%%%%%%%%%%%

\if1\blind
{
  \title{\bf Asymptotic Independence of the Sum and Maximum of Dependent Random Variables with Applications to High-Dimensional Tests}
  \author{Long Feng\\
    School of Statistics and Data Science, LPMC \& KLMDASR, Nankai University\\
    and \\
    Tiefeng Jiang\thanks{
    Corresponding author.}\hspace{.2cm} \\
    School of Statistics, University of Minnesota\\
        and \\
    Xiaoyun Li \\
    Cognitive Computing Lab, Baidu Research\\
        and \\
    Binghui Liu \\
    School of Mathematics and Statistics \& KLAS,  Northeast Normal University}
  \maketitle
} \fi

\if0\blind
{
  \bigskip
  \bigskip
  \bigskip
  \begin{center}
    {\LARGE\bf Asymptotic Independence of the Sum and Maximum of Dependent Random Variables with Applications to High-Dimensional Tests}
\end{center}
  \medskip
} \fi

\bigskip
\begin{abstract}
For a set of dependent random variables, without stationary or the strong mixing assumptions, we derive the asymptotic independence between their sums and maxima. Then we apply this result to high-dimensional testing problems, where we combine the sum-type  and max-type tests and propose a novel test procedure for the one-sample mean test, the two-sample mean test and the regression coefficient test in high-dimensional setting. Based on the asymptotic independence between sums and maxima, the asymptotic distributions of test statistics are established. Simulation studies show that our proposed tests have good performance regardless of data being sparse or not. Examples on real data are also presented to  demonstrate the advantages of our proposed methods.
\end{abstract}

\vspace{0.2in}
\noindent%
\vfill

\newpage
\spacingset{1.9} % DON'T change the spacing!
\section{Introduction}\label{sec1}

Statistical independence is a very simple structure and is convenient in statistical inference and applications. In this paper, we study the asymptotic independence between two  common statistics: the extreme-value statistic $M_p=\max_{1\le i\le p}X_i$ and the sum  $S_p=\sum_{i=1}^p X_i$, where $\{X_i\}_{i=1}^{p}$ is a sequence of dependent random variables. This theoretical results will be applied to three high-dimensional testing problems with numerical examples.

\subsection{Independence Between Sum and Maximum}

In the past few decades, great efforts have been devoted in understanding the asymptotic joint distribution of $M_p$ and $S_p$. In an early research, \cite{chow1978sum} established the asymptotic independence between $M_p$ and $S_p$ for independent and identically distributed random variables. To overcome the limitation of the required assumptions, \cite{anderson1991joint}, \cite{anderson1993limiting}, \cite{anderson1995sums} and \cite{hsing1995note} generalized the asymptotic result  to the case that $\{X_i\}_{i=1}^{p}$ is strong mixing; for the concept  ``strong mixing" and its properties, see, for example, the survey paper \cite{bradley2005basic} and the literature therein.   In particular, \cite{hsing1995note} showed that for a stationary sequence, strong mixing property and asymptotic normality of $S_p$ are basically enough to guarantee the asymptotic independence of the sum and maximum. However, it is shown in \cite{davis1995point} that in the case of infinite variance, $M_p$ and $S_p$ are not asymptotically independent because the asymptotic behavior of $S_p$ is dominated by that of the extreme order statistic.
In addition, \cite{ho1996asymptotic}, \cite{ho1999asymptotic}, \cite{mccormick2000asymptotic} and \cite{peng2003joint} considered the joint limit distribution of the maximum and sum of stationary Gaussian sequence  $\{X_i\}_{i=1}^{p}$ in which $E(X_i)=0$, $\mbox{Var}(X_i)=1$ and $r(p)=E(X_iX_{i+p})$. Under different conditions on $r(p)$, the joint limiting distributions of maxima and sums are different. Specifically, \cite{ho1996asymptotic} showed that $M_p$ and $S_p$ are asymptotically independent as long as $\lim_{p\to \infty}r(p)\log p=0$; the two statistics are not independent  provided $\lim_{p\to \infty}r(p)\log p=\rho \in (0,\infty)$. For the rest situations, by assuming $\lim_{p\to \infty} \frac{\log p}{p}\sum_{i=1}^p|r(i)-r(p)|=0$,   \cite{ho1999asymptotic} and \cite{mccormick2000asymptotic} obtained the asymptotic independence of $M_p-(S_p/n)$ and $S_p$.

All these results are based on the stationary assumption that the covariance structure among  $\{X_i\}_{i=1}^{p}$ has the property that $E(X_iX_{i+h})=E(X_1X_{1+h})$ for each integer $h$ and $i=1,\cdots,p-h$. This is a common assumption in research, however, it is not easy to be checked. Even though it can be verified by hypothesis testing, the stationary property still may not hold up to certain statistical errors. In fact, in many scenarios this assumption is not true.
For example, for stock data of US S\&P 500 index in which stock returns are considered as variables, if stocks are ordered alphabetically by names, the two stocks, such as AAPL and MSFT, may have both far distance and strong correlation, which does not satisfy the stationary assumption.
%On this ground, it is very necessary to establish the asymptotic independence between sum and maximum under a more general setting of covariance structure.

In this work, we study the asymptotic independence between $\tilde{S}_p=\sum_{i=1}^p Z_i^2$ and $\tilde{M}_p=\max_{1\le i\le p}Z_i^2$ without stationary assumption. In our case each $Z_i$ is marginally $N(0,1)$ and the covariance matrix of $Z_i$'s, denoted by  $\bd{\Sigma}_p=(\sigma_{ij})_{1\le i,j\le p}$, satisfies certain conditions.
Specifically, we first establish the asymptotically normality of $\tilde{S}_p$ if $[\mbox{tr}(\bd{\Sigma}_p^{2+\delta})]^2\cdot
[\mbox{tr}(\bd{\Sigma}_p^2)]^{-2-\delta}\to 0$ for some $\delta>0$. Then, we show that the limit distribution of the maximum $\tilde{M}_p-2\log p+\log \log p$ is a Gumbel distribution under conditions on the covariance matrix $\bd{\Sigma}_p$.
%a weaker condition than Lemma 6 in \citep{cai2014two}.
Finally, we prove the asymptotic independence between $\tilde{S}_p$ and $\tilde{M}_p$ under the conditions $\max_{1\leq i<j \leq p}|\sigma_{ij}|\leq \varrho$ and $\max_{1\leq i \leq p}\sum_{j=1}^p\sigma_{ij}^2\leq (\log p)^C$, together with two additional conditions on the maximum and minimum eigenvalues of $\bms_p$.
These theoretical results are novel and essentially different from the existing ones in which the stationary property is required.
Since these results are universal, they may provide many useful implications. In this paper, we will apply the above asymptotic independence results to three high-dimensional hypothesis testing problems: one-sample mean test, two-sample mean test and the  regression coefficient test.

\subsection{High-Dimensional Hypothesis Testing}

High-dimensional hypothesis testing is an important research area in modern statistics. It has been frequently used in many application fields, such as genomics, medical imaging, risk management and web search. The motivation of studying high-dimensional test is that traditional tests, such as the Hotelling $T$-squared test, do not work in general when the data dimension is larger than the sample size due to the singularity of sample covariance matrix. A nature way to amend this problem is replacing the sample covariance matrix appearing in the Hotelling $T$-squared test statistic with a nonsingular matrix, such as the identity matrix and the diagonal matrix of sample covariance matrix. In this way, for example, \cite{srivastava2009test}, \cite{park2013test}, \cite{wang2015high}, %\citep{feng2016spatial},
\cite{feng2016multivariate}, \cite{feng2015two} and \cite{feng2017composite} developed tests for one-sample mean problem, while \cite{bai1996effect}, \cite{srivastava2008test}, \cite{chen2010two},
%\citep{feng2015note}
and \cite{gregory2015two}  developed  tests for two-sample mean problem. In addition, \cite{goeman2006testing} %\citep{feng2013rank}
and \cite{lan2014testing}, for instance, considered testing regression coefficients in high-dimensional linear models. All these tests are sum-type tests, based on the summation of parameter estimators. It is well known that the sum-type tests generally perform well as data are dense, i.e. most of the parameters are nonzero under the local alternative.
%, where the parameters are dense under local alternative.
However, it may be inefficient when data are sparse, where only a few parameters are nonzero under local alternative. To establish high-dimensional tests for sparse data, \cite{cai2014two}, \cite{zhong2013tests} and \cite{chen2019two} proposed  some max-type tests, which typically perform well on sparse data, but worse when the data become dense.

In practice, it is often difficult to determine whether data are sparse or not.  Thus, many efforts have been devoted to develop tests with good and robust performance under both data conditions. For example,  \cite{fan2015power} proposed a power enhancement procedure by a screening technique for high-dimensional tests. They combined the power enhancement component with an asymptotically pivotal statistic to strengthen powers under sparse alternatives. \cite{xu2016adaptive} initiated  an adaptive test for high-dimensional two-sample mean test. It combines information across a class of sum-of-powers tests, including tests based on the sum-of-squares of the mean differences and the supremum mean difference. \cite{wu2019adaptive} extended the adaptive test to generalized linear models. In \cite{he2021asymptotically}, the authors constructed $U$-statistics of different orders that are asymptotically independent of the max-type test statistics in high-dimensional tests, upon which an adaptive testing procedure is proposed. However,  these results are based on \cite{hsing1995note}, which require data to be sampled from stationary and $\alpha$-mixing random variables. In fact, the $\alpha$-mixing property is hardly checked in practice, which greatly limits the application of these methods. In this paper, by using the novel asymptotic independence analysis between the sum and maximum aforementioned, we solve the problem without the stationary assumption or the $\alpha$-mixing property and propose a series of high-dimensional tests including one-sample mean test, two-sample mean test and the regression coefficient test. Numerical results demonstrate strong robustness of the proposed tests regardless data being sparse or not.

The main contributions of this paper are listed as follows.
(1) We establish the asymptotic distribution of the maximum of dependent Gaussian random variables under a general assumption.
(2) We prove the asymptotic independence between the sum and  maximum of dependent Gaussian random variables without the stationary or the $\alpha$-mixing property.
(3) We propose three high-dimensional combo-type tests based on the above asymptotic properties. They are one-sample mean test, two-sample mean test and the regression coefficient test. Numerical examples on simulated and real-world data demonstrate strong robustness of our tests, on both sparse and dense datasets.

The rest of the paper is organized as follows. In Section \ref{theory_results}, we state our theoretical results, including the asymptotic distributions of the sum and maximum statistics, and the asymptotic independence between them.
In Section \ref{applications_HDT}, we propose a series of tests for high-dimensional data  based on these theoretical results. Then, we demonstrate the simulation
results of the proposed tests in comparison with some existing ones in Section \ref{simulation_results}, followed by two applications in Section \ref{real_data_application}. Finally, we present some concluding remarks in Section \ref{conclusion}, while providing some extended results and technical proofs in the supplementary material.

\section{Asymptotic Independence of Sum and Maximum of Dependent Random Variables}\lbl{theory_results}

First, in this section, we study the asymptotic normality of the sum of dependent random variables. For each $p\geq 2$, let $Z_{p1}, \cdots, Z_{pp}$ be $N(0,1)$-distributed random variables with $p\times p$ covariance matrix $\bd{\Sigma}_p$. If there is no danger of  confusion, we simply write
``$Z_1, \cdots, Z_p$" for ``$Z_{p1}, \cdots, Z_{pp}$" and ``$\bd{\Sigma}$" for ``$\bd{\Sigma}_p$".
 The following assumption is needed:
\bea\lbl{condition_1}
\lim_{p\to\infty}\frac{[\mbox{tr}(\bd{\Sigma}^{2+\delta})]^2}
{[\mbox{tr}(\bd{\Sigma}^2)]^{2+\delta}}=0\ \ \mbox{for some}\ \delta>0.
\eea
Assumption  \eqref{condition_1} with $\delta=2$ is the same as condition (3.7) in \cite{chen2010two}, and here we make it more general. Although in applications the true covariance matrix $\bm\Sigma$ is usually unknown, this condition assures the practitioners that our results would be applicable to a wide range of problems. For instance, if all  eigenvalues of $\bms$ are bounded above and are bounded below from zero, it is trivial to see that \eqref{condition_1} holds.

\begin{theorem}\lbl{theorem_1}  Under Assumption \eqref{condition_1},
$\frac{Z_1^2+\cdots + Z_p^2-p}{\sqrt{2\mbox{tr}(\bd{\Sigma}^2)}} \to N(0, 1)$
in distribution as $p\to \infty.$
\end{theorem}

Theorem \ref{theorem_1} shows that the sum of squares of the dependent Gaussian random variables has the asymptotic normality if the covariance matrix satisfies Assumption (\ref{condition_1}). %$\mbox{tr}(\bd{\Sigma}^4)/[\mbox{tr}(\bd{\Sigma}^2)]^2\to 0$ as $p\to \infty$.

Next, for the same Gaussian random variables, we consider the asymptotic distribution of  $\max_{1\leq i \leq p}Z_i^2$. The following assumption will be imposed:
\bea\lbl{condition_2}
&&\mbox{Let } \bms=(\sigma_{ij})_{1\le i,j\le p}. \mbox{ For some } \varrho\in (0,1), \mbox{ assume } |\sigma_{ij}|\leq \varrho \mbox{ for all } 1\leq i<j \leq p
\mbox{ and } \nonumber\\
&& p\geq 2. \mbox{ Suppose }\{\delta_p;\, p\geq 1\} \mbox{ and } \{\kappa_p;\, p\geq 1\} \mbox{ are  positive constants with }
 \delta_p=o(1/\log p)\nonumber\\
 && \mbox{and } \kappa=\kappa_p\to 0
\mbox{ as } p\to\infty.\mbox{ For }1\leq i \leq p,\mbox{ define }
 B_{p,i}=\big\{1\leq j \leq p;\, |\sigma_{ij}|\geq \delta_p\big\}\nonumber\\
&& \mbox{and}\  C_p=\big\{1\leq i \leq p;\, |B_{p,i}|\geq p^{\kappa}\big\}.\mbox{ We assume that } |C_p|/p\to 0\ \mbox{as}\ p \to\infty.\ \ \ \ \ \ \
\eea

\begin{theorem}\lbl{theorem_2} Suppose Assumption \eqref{condition_2} holds. Then $\max_{1\leq i \leq p}Z_i^2-2\log p +\log\log p$
converges to a Gumbel distribution with cdf $F(x)=\exp\{-\frac{1}{\sqrt{\pi}}e^{-x/2}\}$ as $p\to \infty$.
\end{theorem}

% \noindent{\bf Remark 1}.
\begin{remark}\lbl{remark_1}
\cite{cai2014two} obtained the above limiting distribution of $\max_{1\leq i \leq p}Z_i^2$ under the assumption that   $\max_{1\leq i \leq p}\sum_{j=1}^p\sigma_{ij}^2\leq C_0$ for each $p\geq 1$, where $C_0$ is a constant free of~$p$. In the following we will see that their result is a special case of Theorem \ref{theorem_2}. In fact, let $\delta_p=(\log p)^{-2}$ for $p\geq e^e$,  then for each $1\leq i \leq p$,
$\delta_p^2\cdot |B_{p,i}|\leq \sum_{j=1}^p\sigma_{ij}^2\leq C_0$.
Hence, $|B_{p,i}|\leq C_0\cdot (\log p)^2< p^{\kappa}$ where  $\kappa=\kappa_p:=5(\log\log p)/\log p$ for large $p$. As a result, $|C_p|=0$, which implies the results of Theorem \ref{theorem_2}.
\end{remark}

A closely related but not exactly the same result by \cite{fan2019largest} shows that $\delta_p=o(1/\log p)$ in Assumption \eqref{condition_2} can not be relaxed. Their statistic is $\max_{1\leq i \leq p}Z_i$ in contrast to $\max_{1\leq i \leq p}|Z_i|$ here. We expect that $\delta_p=o(1/\log p)$ is also the critical threshold for $\max_{1\leq i \leq p}|Z_i|$.

Theorem \ref{theorem_2} is proved by using the spirit of the proof of Lemma 6 from \cite{cai2014two}. There are two purposes to derive the result. First, the conditions imposed in our theorem is weaker than those required in Lemma 6 from \cite{cai2014two}, which has been discussed in Remark \ref{remark_1}. This allows us to apply this type of results to a   more general covariance matrix $\bd{\Sigma}$. Secondly, part of the steps in the proof of Theorem \ref{theorem_2} will also be used in the proof of Theorem \ref{theorem_3} stated next.

To proceed, we need more notations and an additional assumption. For two sequences of numbers $\{a_p\geq 0;\, p\geq 1\}$ and $\{b_p>0;\, p\geq 1\}$, we write $a_p\ll b_p$ if $\lim_{p\to\infty}\frac{a_p}{b_p}=0.$ The following assumption will be used:
\bea\lbl{assumption_A3}
&& \mbox{There exist } C>0 \mbox{ and } \varrho\in (0, 1) \mbox{ so that }
\max_{1\leq i<j \leq p}|\sigma_{ij}|\leq \varrho \mbox{ and }  \max_{1\leq i \leq p}\sum_{j=1}^p\sigma_{ij}^2\leq (\log p)^C\nonumber\\
&& \mbox{for all}\ p\geq 3;\ p^{-1/2}(\log p)^C \ll \lambda_{min}(\bd{\Sigma})\leq \lambda_{max}(\bd{\Sigma})\ll \sqrt{p}(\log p)^{-1}\ \mbox{and}\ \nonumber\\
&& \lambda_{max}(\bd{\Sigma})/\lambda_{min}(\bd{\Sigma})=O(p^{\tau})\ \mbox{for some}\ \tau \in (0, 1/4).
%\end{itemize}
\eea

Assumption \eqref{assumption_A3} is actually stronger than both  \eqref{condition_1} and \eqref{condition_2}.  To see this, assume \eqref{assumption_A3} holds now. To derive \eqref{condition_1}, observe that $\mbox{tr}(\bd{\Sigma}^{2+\delta})\leq p\cdot\lambda_{max}(\bd{\Sigma})^{2+\delta}$ and $\mbox{tr}(\bd{\Sigma}^2)\geq p\cdot \lambda_{min}(\bd{\Sigma})^2$. Then
$
\frac{[\mbox{tr}(\bd{\Sigma}^{2+\delta})]^2}
{[\mbox{tr}(\bd{\Sigma}^2)]^{2+\delta}}\leq \frac{1}{p^{\delta}}\cdot \Big(\frac{\lambda_{max}(\bd{\Sigma})}{\lambda_{min}(\bd{\Sigma})}\Big)^{4+2\delta}
=O\Big(\frac{1}{p^{\delta-(4+2\delta)\tau}}\Big)\to 0
$
by choosing $\delta=2$ and using the assumption $\tau\in (0, 1/4)$ stated in \eqref{assumption_A3}. We then get  \eqref{condition_1} with $\delta=2$. To deduce \eqref{condition_2}, we replace ``$C_0$" in Remark \ref{remark_1} with ``$(\log p)^C$". By the same argument as that in Remark \ref{remark_1} and choosing $\delta_p=(\log p)^{-2}$, we see $|B_{p,i}|\leq C_0\cdot (\log p)^{C+2}< p^{\kappa},$ where  $\kappa=\kappa_p:=(C+3)(\log\log p)/\log p$ for $p\geq e^e$. Hence, $|C_p|=0$ and Assumption \eqref{condition_2} holds.

\begin{theorem}\lbl{theorem_3} Under Assumption  \eqref{assumption_A3}, the following holds:
$\frac{Z_1^2+\cdots + Z_p^2-p}{\sqrt{2\mbox{tr}(\bd{\Sigma}^2)}}$ and $\max_{1\leq i \leq p}Z_i^2-2\log p +\log\log p$ are asymptotically independent as $p\to \infty.$
\end{theorem}

Importantly, notice that the above asymptotic independence result holds without the stationary assumption or the $\alpha$-mixing condition. Regarding the assumption on the spectrum, in high-dimensional statistics literature, it is common to assume
$
[\lambda_{min}(\bd{\Sigma}),  \lambda_{max}(\bd{\Sigma})]\subset [a, b],
$
with $0<a<b<\infty$. Note that this is stronger than our assumption on the eigenvalues of $\bd{\Sigma}$ in \eqref{assumption_A3}. In fact, Assumption \eqref{assumption_A3} allows that the largest eigenvalue goes to infinity and the smallest eigenvalue goes  to zero. Thus, Theorem \ref{theorem_3} provides more general result and more freedom and practicality in application.

\section{Application: High-Dimensional Testing Problems}\lbl{applications_HDT}

In this section, we will apply the theoretical results derived in Section \ref{theory_results} to three high-dimensional testing problems: one-sample mean test, two-sample mean test and the regression coefficient test. The first and third tests will be presented in the following two subsections, while  two-sample mean test will be presented in the supplementary material.

\subsection{One-Sample Mean Test}\lbl{one_sampl_me}

 Assume $\X_1,\cdots,\X_n$ are
independent and identically distributed $p$-dimensional random vectors from $N(\bmu, \bms)$. The
classical one-sample mean testing problem considers
\begin{align}\label{one}
H_0: \bmu=\bm 0 ~~\textrm{versus}~~H_1: \bmu\not=\bm 0.
\end{align}
In the traditional setting where $p$ is fixed, this topic is covered in classic textbooks on multivariate analysis such as in \cite{anderson2003introduction}, \cite{eaton1983multivariate} and \cite{muirhead1982aspects}. Starting from this century,
a tremendous effort has been made for the test towards  the high-dimensional setting, where both $n$ and $p$ go to infinity.
In the following we will highlight part of these work en route to a problem we are interested in: the test \eqref{one} under the situation $n\leq p$. This  is a typical problem of interest in high-dimensional statistics with small $n$ and large $p$.

Let $\bar{\X}=\frac{1}{n}\sum_{i=1}^n\bd{X}_i$ and $\hat{\S}=\frac{1}{n}\sum_{i=1}^n (\X_i-\bar{\X})(\X_i-\bar{\X})^T$ be the sample mean and the sample covariance matrix of  $\X_1,\cdots,\X_n$, respectively. The Hotelling $T^2$-statistic is defined by $n\bar{\X}^T\hat{\S}^{-1}\bar{\X}$; see \cite{hotelling1931generalization}. For the case with $n>p$, \cite{bai1996effect} studied the Hotelling statistic. When $n\leq p$, however, the matrix $\hat{\S}$ is no longer invertible, which motivates the design of new statistics. By replacing $\hat{\S}$ with its diagonal matrix in the Hotelling $T^2$-statistic,
\cite{srivastava2008test} and \cite{srivastava2009test} proposed a scale-invariant test for (\ref{one}), defined by
\begin{align}\lbl{jin_wuzu}
T^{(1)}_{sum}=\frac{n\bar{\X}^{T} \hat{\D}^{-1}\bar{\X}-(n-1)p/(n-3)}{\sqrt{2[\tr(\hat{\R}^2)-p^2/(n-1)]}},
\end{align}
where $\hat{\D}$ is the diagonal matrix of the sample covariance matrix $\hat{\S}$, and $\hat{\R}=\hat{\D}^{-1/2}\hat{\S}\hat{\D}^{-1/2}$ is the sample correlation matrix.
The major ingredient of $T^{(1)}_{sum}$ can be written as a sum of random variables, so we sometimes call it a ``sum-type'' statistic.  In general, the performance of sum-type statistics are not ideal in sparse cases when only a few entries in $\bm\mu$ in the sum are non-zero; see \cite{cai2014two} for more detailed discussion. \cite{zhong2013tests} proposed  two  alternative tests  by first thresholding two statistics based on the sample means and then maximizing over a range of thresholding levels. Denote $\bar{\X}=(\bar{\X}_1,\cdots,\bar{\X}_p)^T$. The $L_2$-version of the thresholding statistic is
\begin{align}\lbl{qiche}
T_{HC2}=\max_{s\in \mathcal{S}}\frac{T_{2n}(s)-\hat{\mu}(s)}{\hat{\sigma}(s)},
\end{align}
where $\mathcal{S}$ is a subset of the interval $(0,1)$,
\begin{align*}
T_{2n}(s)&=\sum_{j=1}^pn\left(\bar{\X}_j/\sigma_j\right)^2I \left(|\bar{\X}_j|\ge \sigma_j\sqrt{\lambda_s/n}\right),\\
\hat{\mu}(s)&=p\left\{2\lambda_p^{1/2}(s)\phi(\lambda_p^{1/2}(s))+2\bar{\Phi}(\lambda_p^{1/2}(s))\right\},\\
\hat{\sigma}^2(s)&=p\left\{2\left[\lambda_p^{3/2}(s)+3\lambda_p^{1/2}(s)\right]\phi(\lambda_p^{1/2}(s))+6\bar{\Phi}(\lambda_p^{1/2}(s))\right\}.
\end{align*}
Here $\lambda_s(p)=2s\log p$, and $\phi(\cdot)$, $\bar{\Phi}(\cdot)$ are the density and survival functions of the standard normal distribution, respectively.
\cite{fan2015power} proposed a novel procedure by adding a power enhancement component which is asymptotically zero under the null and diverges under some specific regions of alternatives. Their test statistic is
\bea\lbl{keyia}
J=J_0+J_1,
\eea
 where the power enhancement component $J_0$ is
$
J_0=\sqrt{p}\sum_{j=1}^p\bar{\X}_j^2\hat{\sigma}_j^{-2}I (|\bar{\X}_j|>\hat{\sigma}_j\delta_{p,n}),
$
and $J_1$ is the standard Wald statistic
$
J_1=\frac{\bar{\X}^T \widehat{\var}^{-1}(\hat{\X})\bar{\X}-p}{2\sqrt{p}}.
$
Here $\hat{\sigma}_j^2$ is the sample variance of the $j$th coordinate of the population vector, $\delta_{p,n}$ is a thresholding parameter and $\widehat{\var}^{-1}(\hat{\X})$ is a consistent estimator of the asymptotic inverse covariance matrix of $\bar{\X}$. However, the power enhancement component would be negligible if the signal is not very strong. As we mentioned before, \cite{cai2014two} showed that extreme-value statistics are particularly powerful against sparse alternatives and possess certain optimal properties. Hence, we propose a statistic by compromising the sum-type statistic from \eqref{jin_wuzu} and an extreme-value statistic, based on our results in Section~\ref{theory_results}, which will be compared with aforementioned baselines numerically in Section \ref{one_sample_simulation}. As will be confirmed later, our method performs very well regardless of the sparsity of the alternative hypothesis.

We now formally introduce our approach. Define
\begin{align}\lbl{hero_grass}
T^{(1)}_{max}=n\cdot\max_{1\le i\le p}\frac{\bar{\X}_i^2}{\hat{\sigma}_{ii}^2},
\end{align}
where $\bar{\X}_i$ is the $i$th coordinate of $\bar{\X}=\frac{1}{n}(\bd{X}_1+\cdots \bd{X}_n) \in \mathbb{R}^p$ and $\hat{\sigma}_{ii}^2$ is the sample variance of the $i$th coordinate of the population vector, that is, if we write $\bd{X}_j=(x_{1j}, \cdots, x_{pj})^T$ for each $1\leq j\leq n$, then $\hat{\sigma}_{ii}^2$ is the sample variance of the i.i.d. random variables $x_{i1}, x_{i2}, \cdots, x_{in}.$ Firstly, the asymptotic distribution of $T^{(1)}_{max}$ will be presented which needs more notations. Let $\R=\D^{-1/2}\bms\D^{-1/2}=(\rho_{ij})_{1\le i,j\le p}$
denote the population correlation matrix, where  $\D$ is the diagonal matrix of $\bms$.
The following assumption will be imposed:
\bea\lbl{(C1)}
&& \mbox{There exists}\ \epsilon \in \Big(\frac{1}{2}, 1\Big]\ \mbox{and}\ K>1\ \mbox{such that}\ K^{-1}p^{\epsilon}\leq n\leq Kp^{\epsilon}\ \mbox{and}\nonumber\\
&&\sup_{p\geq 2}\frac{1}{p}\tr(\R^i)<\infty\, \mbox{ for } i=2,3,4.
\eea
Note that \eqref{(C1)} is the same as assumptions (3.1) and (3.2) from  \cite{srivastava2009test}. If the eigenvalues of the correlation matrix $\R$ are  bounded, the second condition of \eqref{(C1)} will hold automatically. For rigor of mathematics, we assume $n$ depends on $p$ and sometimes write $n_p$ when there is a  possible confusion.

\begin{theorem}\label{thone}
Under the null hypothesis in \eqref{one}, the following holds as $p\to\infty$:
\begin{itemize}
\item[(i)] If \eqref{(C1)} holds, then $T^{(1)}_{sum}\to N(0,1)$ in distribution;
\item[(ii)] If  \eqref{condition_2} holds with ``$\bd{\Sigma}$'' replaced by ``$\bd{R}$'' and $\log p=o(n^{1/3})$, then
$T^{(1)}_{max}-2\log p+\log\log p$ converges weakly to a Gumbel distribution with cdf $F(x)=
 \exp\{-\frac{1}{\sqrt{\pi}}\exp(-x/2)\}$;
\item[(iii)] Assume \eqref{(C1)} is true. If  \eqref{assumption_A3} holds with ``$\bd{\Sigma}$'' replaced by ``$\bd{R}$'',  then  $T^{(1)}_{sum}$ and $T^{(1)}_{max}-2\log p+\log\log p$ are asymptotically independent.
\end{itemize}
\end{theorem}

Part (i) of the above theorem is from \cite{srivastava2009test}, which is also a corollary of the recent work by \cite{jiang2021mean}.  For the sum-type test, a level-$\alpha$ test will
be performed through rejecting $H_0$ when $T^{(1)}_{sum}$ is larger than the $(1-\alpha)$-quantile $z_{\alpha}= \Phi^{-1}(1-\alpha)$ where $\Phi(y)$ is the cdf of $N(0,1)$.
For the max-type test, a level-$\alpha$ test will
then be performed through rejecting $H_0$ when $T^{(1)}_{max}-2\log p+\log\log p$ is larger than the $(1-\alpha)$-quantile
$q_{\alpha}= -\log \pi-2\log\log(1-\alpha)^{-1}$ of the Gumbel distribution $F(x)$.

Based on Theorem \ref{thone}, we propose a combo-type test statistic by combining the max-type and the sum-type tests. It is defined by
\begin{align}\lbl{Victor_1}
T_{com}^{(1)}=\min\{P^{(1)}_S, P^{(1)}_M\},
\end{align}
where
$
P^{(1)}_{S}=1-\Phi\left\{T^{(1)}_{sum}\right\}$ and $P^{(1)}_{M}=1-F(T^{(1)}_{max}-2\log p+\log\log p).
$
Note that $P^{(1)}_{S}$ and $P^{(1)}_{M}$ are the $p$-values for the tests by using statistics $T^{(1)}_{sum}$ and $T^{(1)}_{max}$, separately, and $T_{com}^{(1)}$ is defined by the smaller one, whose asymptotic distribution can be characterized by the minimum of two standard uniform random variables.

\begin{coro}\lbl{coro2} Assume the conditions in Theorem \ref{thone}(iii) hold.
Then $T_{com}^{(1)}$ from \eqref{Victor_1} converges weakly to a distribution with density  $G(w)=2(1-w)I(0\leq w \leq 1)$ as $p\to\infty.$
\end{coro}

According to Corollary \ref{coro2}, the proposed combo-type test allows us to perform a level-$\alpha$ test by rejecting the null hypothesis
when $T_{com}^{(1)}<1-\sqrt{1-\alpha}\approx \frac{\alpha}{2}$ as $\alpha$ is small. We now discuss the power functions. First, the power function of our combo-type test is
\bea
\beta^{(1)}_C(\bmu,\alpha)&=&P\left(T_{com}^{(1)}<1-\sqrt{1-\alpha}\right) =P\left(P^{(1)}_{M}<1-\sqrt{1-\alpha}~or~ P^{(1)}_{S}<1-\sqrt{1-\alpha}\right) \nonumber\\
&\ge & \max\left\{P\left(P^{(1)}_{S}<1-\sqrt{1-\alpha}\right),P\left(P^{(1)}_{M}<1-\sqrt{1-\alpha}\right)\right\}\nonumber\\
&\approx & \max\left\{\beta^{(1)}_S(\bmu,\alpha/2),\beta^{(1)}_M(\bmu,\alpha/2)\right\} \label{pms}
\eea
when $\alpha$ is small, where $\beta^{(1)}_M(\bmu,\alpha)$ and $\beta^{(1)}_S(\bmu,\alpha)$ are the power functions of $T_{max}^{(1)}$ and $T_{sum}^{(1)}$  with significant level $\alpha$, respectively. From \cite{srivastava2009test}, the power function of $T_{sum}^{(1)}$ is
\begin{align}\lbl{powerS}
\beta^{(1)}_S(\bmu,\alpha)=\lim_{p\to\infty}\Phi\left(-z_{\alpha}+\frac{n\bmu^T \D^{-1}\bmu}{\sqrt{2\tr(\R^2)}}\right),
\end{align}
where $z_{\alpha}=\Phi^{-1}(1-\alpha)$ is the  $(1-\alpha)$-quantile of $N(0, 1)$.  Due to (\ref{pms}), we have
$\beta^{(1)}_C(\bmu,\alpha)\ge \lim_{p\to\infty}\Phi\left(-z_{\alpha/2}+\frac{n\bmu^T \D^{-1}\bmu}{\sqrt{2\tr(\R^2)}}\right)$.
Denote $\D=\diag(\sigma_{11}^2,\cdots,\sigma_{pp}^2)$. By the same argument at that from Theorem 2 in \cite{cai2014two}, the asymptotic power of $T_{max}^{(1)}$ converges to one if $\max_{1\le i\le p} |\mu_i/{\sigma_{ii}}|\ge c\sqrt{\log p/n}$ for a certain constant $c$, and also the nonzero $\mu_i$ are randomly uniformly sampled with sparsity level $\gamma<1/4$, i.e., the number of nonzero $\mu_i$ is less than $p^\gamma, \gamma<1/4$.  Thus, according to (\ref{pms}), the power function of our proposed test $T_{com}^{(1)}$ also converges to one in this case. Similarly, according to Theorem 3 in \cite{cai2014two}, the condition $\max_{1\le i\le p} |\mu_i/{\sigma_{ii}}|\ge c\sqrt{\log p/n}$ is minimax rate optimal for testing against sparse alternatives. If $c$ is sufficiently small, then any $\alpha$-level test is unable to reject the null hypothesis with probability tending to one. It is shown in \cite{cai2014two}  that $T_{max}^{(1)}$ enjoys a certain optimality against sparse alternatives. By (\ref{pms}), our test $T_{com}^{(1)}$ also has this optimality.

In order to get a rough picture of the asymptotic power comparison between $T_{sum}^{(1)}, T_{max}^{(1)}$ and $T_{com}^{(1)}$, now we simply assume that $\bms=\I_p$. There are $m$ nonzeros $\mu_i$ and they are all equal to $\delta \ne 0$. Equation \eqref{powerS} gives
$\beta^{(1)}_S(\bmu,\alpha)=\lim_{p\to\infty}\Phi\left(-z_{\alpha}+\frac{nm\delta^2}{\sqrt{2p}}\right)$.

We consider two special cases:
\begin{itemize}
\item[(1)] {\it Dense case}: $\delta=O(n^{-\xi})$ and $m=O(p^{1/2}n^{2\xi-1})$ with  $\xi\in (1/2, 5/6]$. We also assume $\log p=o(n^{\xi-\frac{1}{2}})$, hence $\log p=o(n^{1/3})$. As a consequence, the requirement on $p$ vs $n$ imposed in Theorem \ref{thone}(ii) is fulfilled. Obviously, the number of nonzero $\mu_i$ goes to infinity. The power function for $T_{max}^{(1)}$ is given by
$\beta^{(1)}_{M}(\bmu,\alpha)=P\left(T_{max}^{(1)}-2\log p+2\log\log p>q_{\alpha}\right)$.
In this case, we will show in Section \ref{jingshen} of the supplementary material that
$\beta^{(1)}_{M}(\bmu,\alpha)\approx \alpha$,
    which means that $T_{max}^{(1)}$ is not effective or useful.
    Consequently, we have $\beta^{(1)}_C(\bmu,\alpha)\approx \beta^{(1)}_S(\bmu,\alpha/2)$. When the significant level $\alpha$ is small, the difference between $\beta^{(1)}_S(\bmu,\alpha)$ and $\beta^{(1)}_S(\bmu,\alpha/2)$ is negligible. So our proposed test $T_{com}^{(1)}$ has similar performance as $T_{sum}^{(1)}$ in this dense case.

\item[(2)] {\it Sparse case}: $\delta=c\sqrt{\log p /n}$ for sufficient large constant $c$ and $m=o((\log p)^{-1}p^{1/2})$. Here the value of $m$ is much smaller than that in (1) and hence confirms the notion of ``sparse". In this case, $\frac{nm\delta^2}{\sqrt{2p}}\to 0$, so $\beta^{(1)}_{S}(\bmu,\alpha)\approx \alpha$ and $T_{sum}^{(1)}$ is not effective or useful. Yet, $\beta^{(1)}_M(\bmu, \alpha)\to 1$ by an argument similar to Theorem 2 from  \cite{cai2014two} as discussed above, which also leads to $\beta^{(1)}_C(\bmu,\alpha)\to 1$ in this sparse case. Additionally,  in \cite{fan2015power} the quantity $\delta_{p,n}$ is chosen to be $\log\log n\sqrt{\log p/n}$, which implies that the screening set $\{i:\sqrt{n}|\bar{\X}_i|>\log\log n\sqrt{\log p}\}$ would be empty as probability tending to one. Thus, the power enhancement component of \cite{fan2015power} would be negligible in this case, which makes the standardized Wald test statistic the same as $T_{sum}^{(1)}$ since $\bms=\I_p$. That is, their test is also ineffective in this sparse case.

\end{itemize}
The above theoretical results and analysis, together with the simulation in the next section, indicate that our proposed test $T_{com}^{(1)}$ performs very well regardless of the sparsity of the alternative hypothesis, which is more convenient to use in various practical scenarios.

Due to space limitation, we will present the combo-type two-sample mean test as well as its simulation results in the supplementary material.

\subsection{Test for Regression Coefficients}\lbl{regress_3}
%There are two parts in this section: the theory and the simulation.
%%\subsubsection{Tests for High-Dimensional Regression Coefficients: Theory}
%\subsubsection{The Theory}
Then, we apply our theory to the testing problem in high-dimensional regression. Let $\X=(\X_1,\cdots,\X_n)^T$ be independent and identically distributed $p$-dimensional covariates, and  $\Y=(Y_1,\cdots,Y_n)^T$ be the corresponding  independent responses. For simplicity, we assume $E\X_i=0$, $\forall i$. We introduce a decomposition of sample point $\X_i$ as $\X_i=(\X_{ia}^{T},\X_{ib}^{T})^{T}$ with $\X_{ia}=(X_{i1},\cdots,X_{iq})^{T} \in\mR^{q}$ and
$\X_{ib}=(X_{i(q+1)},\cdots,X_{ip})^{T}\in\mR^{p-q}$,
 where $q$ is
smaller than the sample size $n$, and $n$ is much smaller than $p$.
We consider the following standard linear regression model:
\begin{align}\label{eq:modelv}
Y_i=\X_i^{T} \bb+\varepsilon_i=\X_{ia}^{T} \bb_{a}+\X_{ib}^{T} \bb_b+\varepsilon_i,
\end{align}
where $\bb=(\bb_a^{T},\bb_b^{T})^{T} \in \mathbb{R}^p$, $\bb_a\in \mathbb{R}^q$ and $\bb_b \in  \mathbb{R}^{p-q}$ are the regression coefficient vectors. The random noises
$\{\varepsilon_i;\, 1\leq i \leq n\}$ are independent with $E\varepsilon_i=0$ and $\mbox{Var}(\varepsilon_i)=\sigma^2$ for each $i$, and are also independent of the data $\X$.
In this section, we consider the following testing problem:
\begin{align} \label{h1}
H_0:\bb_b=0\mbox{~~vs.~~} H_1:\bb_b\not=0
\end{align}
under the situation that $p$ is much larger than $q$ and the sample size $n$ is small.
When $(\X_{1a}^T, \cdots, \X_{na}^T)^T$ is a null vector, (\ref{h1}) is equivalent to test
$H_0:\bb=0$ vs. $H_1:\bb\not=0$.

For this problem,  \cite{goeman2006testing} and \cite{goeman2011testing} proposed an empirical Bayes test. It is formulated via a score test on the hyper parameter of a prior distribution on the regression coefficients. By excluding the inverse term in the classical $F$-statistic, \cite{zhong2011tests} proposed a $U$-statistic to extend their results in factorial designs.

Denote $\bb=(\beta_1, \cdots, \beta_p)^T$. To motivate our test procedure, let us consider a set of related but not exactly the same test as follows:
\begin{align} \label{h1j}
H_{0j}:\beta_j=0\mbox{~~vs.~~} H_{1j}:\beta_j\not=0
\end{align}
for each $j=q+1, \cdots, p$. Some notations are needed before we proceed. For the set of features included in the index set of ``$a$'', let
$\X_{a}=(\X_{1a}, \cdots, \X_{na})^T,\ \ \H_a=\mX_a(\mX_a^{T} \mX_a)^{-1}\mX_a^{T}$ and
$\wX_j=(\bd{I}_n-\H_a)(X_{1j}, \cdots, X_{nj})^T$,
for each $j$. Notice $\X_{a}$ is $n\times q$, $\H_a$ is $n\times n$ and $\wX_j\in \mathbb{R}^n$. For the ``$b$" part, we define
\begin{align}
\X_{b}=(\X_{1b}, \cdots, \X_{nb})^T\ \ \ \mbox{and}\ \ \ \wX_b=(\wX_{q+1}, \cdots, \wX_p)=(\bd{I}_n-\H_a)\X_{b};~\hat{\bms}_{b|a}=n^{-1}\wX_b^{T} \wX_b, \lbl{hondulasi}
\end{align}
where both $\X_{b}$ and $\wX_b$ are $n\times (p-q)$ and $\hat{\bms}_{b|a}$ is $(p-q)\times (p-q)$. Regarding the response vector, the residual vector and the sample variance, we denote
$\hat{\bme}=(\I_n-\H_a)\Y$ and $\hat{\sigma}^2= (n-q)^{-1}\hat{\bme}^{T} \hat{\bme}$.
For each test in (\ref{h1j}), the classical partial F-test is given by
$ {F}_j=\frac{\mY^{T} {\wX_j}(\wX_j^{T} \wX_j)^{-1}\wX_j^{T} \mY
}{\mY^{T}[\I_n-\mX(\mX^{T} \mX)^{-1}\mX^{T}]\mY/(n-p)}$.
However, as $p> n$, the statistic $F_j$ becomes problematic since $\mX^{T} \mX $ is not invertible. To overcome this issue, we replace the denominator of $F_j$ by $\mY^{T}(\I_n-\H_a )\mY/(n-q)$. Under the null hypothesis, $\hat{\sigma}^2$ is an unbiased estimator of $\sigma^2$; see \eqref{xiaopingguo} in the supplementary material.  Thus, the test statistic for  (\ref{h1j}) becomes
$\tilde{F}_j=\frac{\mY^{T} \wX_j(\wX_j^{T} \wX_j)^{-1}\wX_j^{T} \mY
}{\mY^{T}(\I_n-\H_a)\mY/(n-q)}$.
Back to the testing problem (\ref{h1}) of interest, we will handle it by combining $\tilde{F}_j$ together. There are two classical ways to synthesize them. The first one is the sum-type test statistic:
$S_F=\sum_{j=q+1}^p \tilde{F}_j$.
Obviously, we have $S_F=T_1/\hat{\sigma}^2$ with $T_1=\sum_{j=q+1}^p\mY^{T} \wX_j(\wX_j^{T} \wX_j)^{-1}\wX_j^{T} \mY$. By normalization, without loss of generality, assume $\wX_j^{T} \wX_j=n$ for each $j=q+1, \cdots, p.$ Then
$T_1=\frac{1}{n}\sum_{j=q+1}^p\mY^{T} \wX_j\wX_j^{T} \mY=\frac{1}{n}\hat{\bm\varepsilon}^{T} \X_b\X_b^{T} \hat{\bme}$.
By standardizing $T_1$ with estimators of the mean and standard deviation of $T_1$, we propose the sum-type statistic
\bea\lbl{Stat_3}
T^{(3)}_{sum}=\frac{n^{-1}\hat{\bm\varepsilon}^{T} \X_b\X_b^{T} \hat{\bme} -n^{-1}(n-q)(p-q)\hat{\sigma}^{2}}{\sqrt{2\hat{\sigma}^{4}\,\widehat{\mbox{tr}\big(\bms_{b|a}^2\big)}}},
\eea
where the denominator is
$\widehat{\tr(\bms^2_{b|a})}=\frac{n^2}{(n+1-q)(n-q)}\Big\{\tr\big(\hat{\bms}_{b|a}^2\big)-\frac{1}{n-q}
\tr^2(\hat{\bms}_{b|a})\Big\}$.
The second statistic we are interested in is of the max-type, defined by
\bea\lbl{3rdmax}
T^{(3)}_{max}=\max_{q+1\le j\le p} \tilde{F}_j.
\eea
Then, we will study the asymptotic distributions of $T^{(3)}_{sum}$ and $T^{(3)}_{max}$, as well as their asymptotic independence, based on which a combo-type test will be proposed. To begin with, we introduce some additional assumptions that will be used.

Define $\bms_{b|a}=E[\mbox{Cov}(\X_{ib}|\X_{ia})]=(\sigma_{jk}^*)$ as a $(p-q)\times
(p-q)$ matrix. Without loss of generality, we assume that $\bms_{b|a}$ is normalized such that its diagonal entries are equal to one, i.e., $\sigma_{jj}^*=1$ for each $j$.
%$j\in S=\{q+1,\cdots,p\}$.
Next, we introduce moment conditions on
a conditional predictor. For each $i=1,\cdots, n$, by regressing
$\X_{ib}$ on $\X_{ia}$, the residual vector is given by
$\X_{ib}^*=\X_{ib}-\B \X_{ia}\in \mR^{p-q}$,
 where
 %$\X_{ib}^*\in \mR^{p-q}$ is the residual vector obtained by regressing
%$\X_{ib}$ on $\X_{ia}$ and
$\B:=\mbox{Cov}(\X_{ib},\X_{ia})\cdot [\mbox{Cov}(\X_{ia})]^{-1}$ is a $(p-q)\times q$ matrix.
By the previous assumption $E\X_i=0$, we immediately have that
$E(\X_{ib}^*)=0$ and $\mbox{Cov}(\X_{ib}^*)=\bms_{b|a}=(\sigma_{ij}^*)$.
Define
$\mX_b^{*}=(\X_{1b}^{*},\cdots,\X_{nb}^{*})^{T}$, which is a $n\times
(p-q)$ matrix. The conditions we will use later on are stated below.
\bea
% &&\mbox{\it Independent Component Model}.\
&& \X_{ib}^*\sim N(\bd{0}, \bms_{b|a})\ \mbox{and the diagonal entries of } \bms_{b|a}\ \mbox{are all}\  \mbox{equal to}\ 1;\lbl{con_(C2)}\\
&&\mbox{There exists a constant}\  \tau>1\ \mbox{such that}\   \tau^{-1}<\lambda_{min}(\bms_{b|a})
 \leq\lambda_{max}(\bms_{b|a})<\tau.\lbl{con_(C3)}
\eea

Assumption \eqref{con_(C3)} is the same as condition (C1) from \cite{lan2014testing}, which is also a common assumption in literature for research on high-dimensional data; see, for example, \cite{fan2008high}, \cite{rothman2008sparse}, \cite{zhang2008sparsity} and \cite{wang2009forward}. We assume that both $n$ and $q$ depend on $p$ and limits will be taken as $p\to \infty.$ The random errors $\varepsilon_1, \cdots, \varepsilon_n$ are assumed to be i.i.d. with $E\varepsilon_i=0$ and $\mbox{Var}(\varepsilon_i)=\sigma^2$ for each $i$, and no Gaussian assumption is needed for $\varepsilon_i$. Our main result for high-dimensional regression coefficient test is as follows.

\begin{theorem} \label{thlm}
Assume  \eqref{con_(C2)} and \eqref{con_(C3)} hold and \eqref{assumption_A3} also holds with ``$\bd{\Sigma}$'' replaced by ``$\bms_{b|a}$''. Suppose $p=o(n^3)$, $q=o(p)$, $q\leq n^{\delta}$ for some $\delta \in (0,1)$ and $E(|\epsilon_1|^{\ell}) < \infty$ with  $\ell=14(1-\delta)^{-1}$.
% Suppose  $n\to\infty$, $q/n^{1/4}\to 0$, $n/p\to 0$ and $\log p=o(n^{1/3})$ as $p\to\infty$.
Under $H_0$ from \eqref{h1}, as $p\to\infty$ we have:
\begin{itemize}
\item[(i)] $T^{(3)}_{sum}\to N(0,1)$ in distribution;
\item[(ii)] $T^{(3)}_{max}-2\log(p-q)+\log\log(p-q)$ converges weakly to a distribution with cdf $F(x)=\exp\{-\frac{1}{\sqrt{\pi}}
\exp(-\frac{x}{2})\}$, $x \in \mathbb{R}$;
\item[(iii)] $T^{(3)}_{sum}$ and $T^{(3)}_{max}-2\log(p-q)+\log\log(p-q)$ are asymptotically independent.
\end{itemize}
\end{theorem}
It is known from \cite{lan2014testing} that
$
\frac{n^{-1}\hat{\bm\varepsilon}^{T} \X_b\X_b^{T} \hat{\bme}-\hat{\sigma}^2n^{-1}(p-q)\tr(\M(\I_n-\H_a))}
{\sqrt{2\sigma^{4}\,\mbox{tr}\big(\bms_{b|a}^2\big)}}
$
converges to $N(0, 1)$ in distribution, with $\M:=(p-q)^{-1}\sum_{j=q+1}^p\wX_j\wX_j^{T}$. Although this is not a statistic (since unknown parameters appear in the denominator), it describes the asymptotic behavior of $\hat{\bm\varepsilon}^{T} \X_b\X_b^{T} \hat{\bme}$. The numerator of our statistic $T^{(3)}_{sum}$ in \eqref{Stat_3} is simpler, because no computation of $\tr(\M(\I_n-\H_a))$ is needed.

We now study the implications of Theorem \ref{thlm} and discuss the rejection rules. For the sum-type test, a level-$\alpha$ test will
be performed through rejecting $H_0$ when $T^{(3)}_{sum}$ is larger than the $(1-\alpha)$-quantile $z_{\alpha}= \Phi^{-1}(1-\alpha)$ of
$N(0,1)$.
For the max-type test, a level-$\alpha$ test will
then be performed through rejecting $H_0$ when $T^{(3)}_{max}-2\log(p-q)+\log\log(p-q)$ is larger than the $(1-\alpha)$-quantile
$q_{\alpha}=-\log \pi-2\log\log(1-\alpha)^{-1}$ of $F(x)$.

Analogously, a combined test is defined through
\bea\lbl{3comtest}
T_{com}^{(3)}=\min\big\{P^{(3)}_S, P^{(3)}_M\big\},
\eea
where $P^{(3)}_{S}=1-\Phi(T_{sum}^{(3)})$ and
$P^{(3)}_{M}=1-F(T_{max}^{(3)}-2\log(p-q)+\log\log(p-q))$ with $F(x)=\exp\{-\frac{1}{\sqrt{\pi}}\exp\left(-\frac{x}{2}\right)\}.$
Similar to Corollary \ref{coro2}, the proposed combo-type test allows us to perform a level-$\alpha$ test by rejecting the null hypothesis
when $T_{com}^{(3)}<1-\sqrt{1-\alpha}\approx \frac{\alpha}{2}$.

Again, using the same argument as (\ref{pms}), the power function of our combo-type test $\beta_C^{(3)}(\bmu,\alpha)$ is larger than $\max\{\beta_M^{(3)}(\bmu,\alpha),\beta_S^{(3)}(\bmu,\alpha)\}$, where $\beta^{(3)}_M(\bmu,\alpha)$ and $\beta^{(3)}_S(\bmu,\alpha)$ are the power functions of $T_{max}^{(3)}$ and $T_{sum}^{(3)}$ at significant level $\alpha$, respectively.
To demonstrate the power of the tests, assume the simple case where $\X_a$ is null vector and $\mbox{Cov}(\X_b)=\I_p$. From \cite{zhong2011tests}, the power function of $T_{sum}^{(3)}$ is
$\beta_S^{(3)}(\bb,\alpha)=\lim_{p\to\infty}\Phi\left(-z_{\alpha}+\frac{n\bb^T \bb}{\sqrt{2p}\sigma^2}\right)$.
In addition, assuming $\bb$ only contains $m$ non-zeros all equal to $\delta\not=0$, leading to
$\beta_S^{(3)}(\bb,\alpha)=\lim_{p\to\infty}\Phi\left(-z_{\alpha}+\frac{nm\delta^2}{\sqrt{2p}\sigma^2}\right)$.
For the non-sparse case with $\delta=O(n^{-\xi})$ and $m=O(p^{1/2}n^{2\xi-1})$, we have $\beta_M^{(3)}(\bmu,\alpha)\approx \alpha$ and $\beta_C^{(3)}(\bmu,\alpha)\approx \beta_S^{(3)}(\bmu,\alpha/2)$. For the sparse case where $\delta=c\sqrt{\log p/n}$ with sufficient large $c$ and $m=o((\log p)^{-1}p^{1/2})$, we have $\beta_S^{(3)}(\bmu,\alpha)\approx \alpha$ and $\beta_C^{(3)}(\bmu,\alpha)\approx \beta_M^{(3)}(\bmu,\alpha/2)\to 1$. Again, in this testing problem for high-dimensional regression, the combined test statistics also exhibits good performance under both sparse and dense alternative hypotheses.

\vspace{-0.1in}
\section{Simulation Results}\label{simulation_results}

In this section, we carry out a series of simulation study on the testing problems studied in the previous section, to compare different test statistics and validate the advantage of the proposed combo-type tests.

\subsection{One-Sample Test Problem}\lbl{one_sample_simulation}
Firstly, we conduct numerical examples on the one-sample test problem. We compare our  combo-type test $T_{com}$ in \eqref{Victor_1} (abbreviated as COM) with
the sum-type test $T^{(1)}_{sum}$ in \eqref{jin_wuzu}  by \cite{srivastava2009test} (abbreviated as SUM), the max-type test $T^{(1)}_{max}$ in \eqref{hero_grass} (abbreviated as MAX),
the Higher Criticism test $T_{HC2}$ from \eqref{qiche} by \cite{zhong2013tests} (abbreviated as HC2) and
the power enhancement test $J$  from \eqref{keyia} by \cite{fan2015power} (abbreviated as FLY).
The dataset is simulated as follows.

\noindent{\bf EXAMPLE 1.}
We consider $\X_i=\bmu+\bms^{1/2}\z_i$ for
$i=1\cdots,n$, and each component of $\z_i$ is independently
generated from three distributions:
(1) the normal distribution $N(0,1)$;
(2) the t distribution $t(3)/\sqrt{3}$;
(3) the mixture normal random variable $V/\sqrt{1.8}$, where
 $V\ \mbox{has  density function}\ 0.1f_1(x)+0.9f_2(x)$ with $f_1(x)$ and $f_2(x)$ being  the densities  of $N(0,9)$ and $N(0,1)$, respectively.
We will work on two different sample sizes with $n=100,200$ and three different dimensions with $p=200, 400, 600$. Under the null hypothesis, we set $\bmu=\bm 0$ and the significance level $\alpha=0.05$. The following  three scenarios of covariance matrices will be considered.
\begin{itemize}
\item[(I)] AR(1) model: $\bms=(0.5^{|i-j|})_{1\le i,j\le p}$.
\item[(II)] $\bms=\D^{1/2}\R\D^{1/2}$ with
$\D=\diag(\sigma_1^2,\cdots,\sigma_p^2)$ and $\R=\I_p+\bm b\bm b
^{T}-\check{\B}$,
where $\sigma_i^2$ are generated independently from $Uniform(1,2)$,
$\bm b=(b_1,\cdots,b_p)^{T}$ and
$\check{\B}=\diag(b_1^2,\cdots,b_p^2)$.
The first $[p^{0.3}]$ entries of $\b$ are
independently sampled from $Uniform(0.7,0.9)$, and the remaining entries  are set to be zero, where $[\cdot]$ denotes taking integer part.

\item[(III)]
$\bms=\bm \gamma \bm
\gamma^{T}+(\I_p-\rho_{\epsilon}\W)^{-1}(\I_p-\rho_{\epsilon}\W^{T})^{-1},$ where
$\bm \gamma=(\gamma_1,\cdots,\gamma_{[p^{\delta_{\gamma}}]},0,0,\cdots,0)^{T}$.
Here  $\gamma_i$ with $i=1,\cdots, [p^{\delta_{\gamma}}]$ are generated independently from $Uniform(0.7,0.9)$. Let
$\rho_{\epsilon}=0.5$ and $\delta_{\gamma}=0.3$. Let $\W=(w_{i_1i_2})_{1\le i_1,i_2\le p}$ have a so-called rook form, i.e.,
all elements of $\W$ are zero except that
$w_{i_1+1,i_1}=w_{i_2-1,i_2}=0.5$ for $i_1=1,\cdots,p-2$ and
$i_2=3,\cdots,p$, and $w_{1,2}=w_{p,p-1}=1$.
\end{itemize}

\begin{table}[!tb]
\begin{center}
\small
\caption{\label{tt1} Sizes of tests for Example 1 with Scenario (I), $\alpha=0.05$.}
                     \vspace{0.2cm}
                     \renewcommand{\arraystretch}{0.8}
                     \setlength{\tabcolsep}{2pt}{
\begin{tabular}{cc|ccc|ccc|ccc}
\hline
\hline
  &Distribution& \multicolumn{3}{c}{(1)} & \multicolumn{3}{c}{(2)} & \multicolumn{3}{c}{(3)} \\ \hline
&$p$&200&400&600&200&400&600&200&400&600\\
\hline
$n=100$&MAX&0.053&0.062&0.082&0.026&0.052&0.045&0.044&0.039&0.061\\
&SUM&0.064&0.064&0.060&0.052&0.050&0.059&0.063&0.058&0.064\\
&COM&0.063&0.069&0.059&0.040&0.059&0.055&0.056&0.047&0.061\\
&HC2&0.028&0.044&0.034&0.033&0.029&0.032&0.038&0.025&0.044\\
&FLY& 0.014& 0.009& 0.004& 0.003& 0.003& 0.002& 0.025& 0.018& 0.014\\ \hline
$n=200$&MAX&0.046&0.060&0.049&0.045&0.041&0.045&0.042&0.045&0.032\\
&SUM&0.065&0.068&0.058&0.053&0.057&0.062&0.056&0.054&0.056\\
&COM&0.056&0.068&0.048&0.042&0.047&0.052&0.043&0.050&0.039\\
&HC2&0.019&0.027&0.030&0.031&0.024&0.023&0.029&0.020&0.029\\
&FLY& 0.005& 0.000& 0.000& 0.003& 0.000& 0.000& 0.017& 0.012& 0.005\\
\hline \hline
\end{tabular}}
\end{center}
\end{table}

\begin{table}[!htb]
\begin{center}
\small
\caption{\label{tt12} Sizes of tests for Example 1 with Scenario (II), $\alpha=0.05$.}
                     \vspace{0.2cm}
                     \renewcommand{\arraystretch}{0.8}
                     \setlength{\tabcolsep}{2pt}{
\begin{tabular}{cc|ccc|ccc|ccc}
\hline
\hline
  &Distribution& \multicolumn{3}{c}{(1)} & \multicolumn{3}{c}{(2)} & \multicolumn{3}{c}{(3)} \\ \hline
&$p$&200&400&600&200&400&600&200&400&600\\
\hline
$n=100$&MAX&0.058&0.070&0.065&0.044&0.037&0.039&0.048&0.042&0.047\\
&SUM&0.053&0.067&0.056&0.054&0.052&0.048&0.054&0.055&0.045\\
&COM&0.055&0.057&0.061&0.054&0.044&0.040&0.043&0.047&0.047\\
&HC2&0.022&0.011&0.013&0.005&0.015&0.005&0.011&0.011&0.006\\
&FLY& 0.022& 0.011& 0.011& 0.013& 0.010& 0.006& 0.024& 0.015& 0.007\\ \hline
$n=200$&MAX&0.053&0.054&0.076&0.025&0.042&0.025&0.044&0.040&0.041\\
&SUM&0.053&0.057&0.060&0.053&0.051&0.052&0.055&0.065&0.060\\
&COM&0.058&0.061&0.066&0.037&0.045&0.044&0.043&0.053&0.055\\
&HC2&0.003&0.011&0.006&0.010&0.006&0.003&0.004&0.005&0.008\\
&FLY& 0.037& 0.033& 0.025& 0.030& 0.022& 0.011& 0.032& 0.026& 0.015\\
\hline \hline
\end{tabular}}
\end{center}
\end{table}

\begin{table}[!htb]
\begin{center}
\small
\caption{\label{tt13} Sizes of tests for Example 1 with Scenario (III), $\alpha=0.05$.}
                     \vspace{0.2cm}
                     \renewcommand{\arraystretch}{0.8}
                     \setlength{\tabcolsep}{2pt}{
\begin{tabular}{cc|ccc|ccc|ccc}
\hline
\hline
  &Distribution& \multicolumn{3}{c}{(1)} & \multicolumn{3}{c}{(2)} & \multicolumn{3}{c}{(3)} \\ \hline
&$p$&200&400&600&200&400&600&200&400&600\\
\hline
$n=100$&MAX&0.054&0.066&0.059&0.053&0.040&0.033&0.049&0.039&0.043\\
&SUM&0.052&0.055&0.059&0.053&0.048&0.060&0.059&0.064&0.061\\
&COM&0.053&0.066&0.059&0.053&0.050&0.040&0.062&0.046&0.051\\
&HC2&0.034&0.038&0.035&0.032&0.030&0.025&0.036&0.030&0.030\\
&FLY& 0.013& 0.003& 0.005& 0.013& 0.001& 0.000& 0.020& 0.013& 0.010\\ \hline
$n=200$&MAX&0.053&0.058&0.063&0.034&0.027&0.038&0.049&0.039&0.050\\
&SUM&0.061&0.065&0.062&0.044&0.058&0.068&0.063&0.058&0.057\\
&COM&0.065&0.075&0.069&0.033&0.048&0.047&0.059&0.051&0.053\\
&HC2&0.035&0.032&0.032&0.019&0.029&0.019&0.029&0.023&0.024\\
&FLY& 0.001& 0.001& 0.000& 0.004& 0.001& 0.000& 0.016& 0.011& 0.002\\
\hline \hline
\end{tabular}}
\end{center}
\end{table}

Tables \ref{tt1}, \ref{tt12}, \ref{tt13} report the empirical sizes of the five tests. SUM, MAX and COM can control the empirical sizes very well in most cases. However, the empirical sizes of HC2 and FLY can be much smaller than the nominal level in some cases.

\begin{figure}[h]
    \caption{Power vs. number of variables with nonzero means for Example 4.1. The $x$-lab $m$ denotes the number of variables with non-zero means; the $y$-lab is the empirical power. }
    \label{f1}
	\begin{center}
		\mbox{			    \includegraphics[width=2.15in]{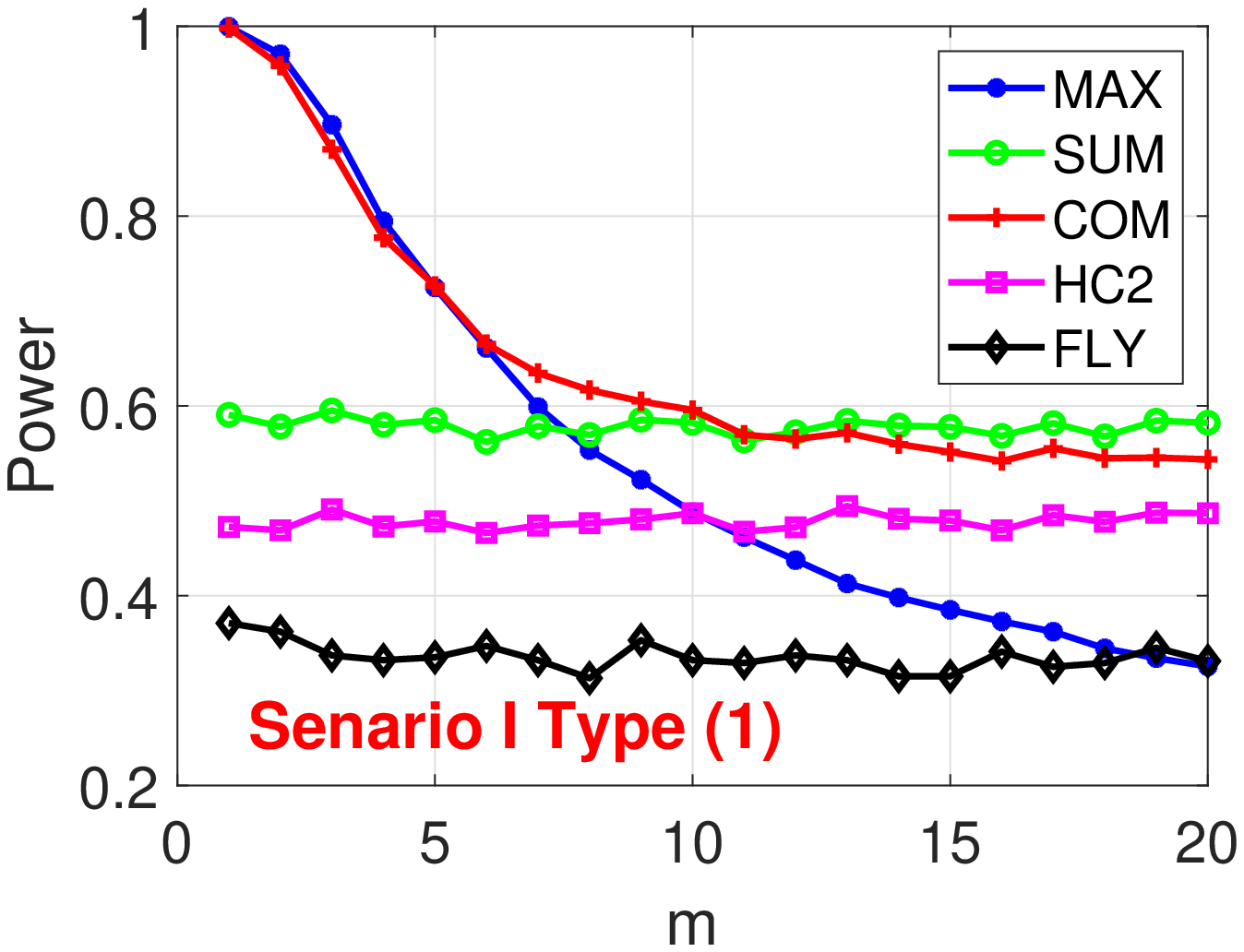}\hspace{-0.1in}
		\includegraphics[width=2.15in]{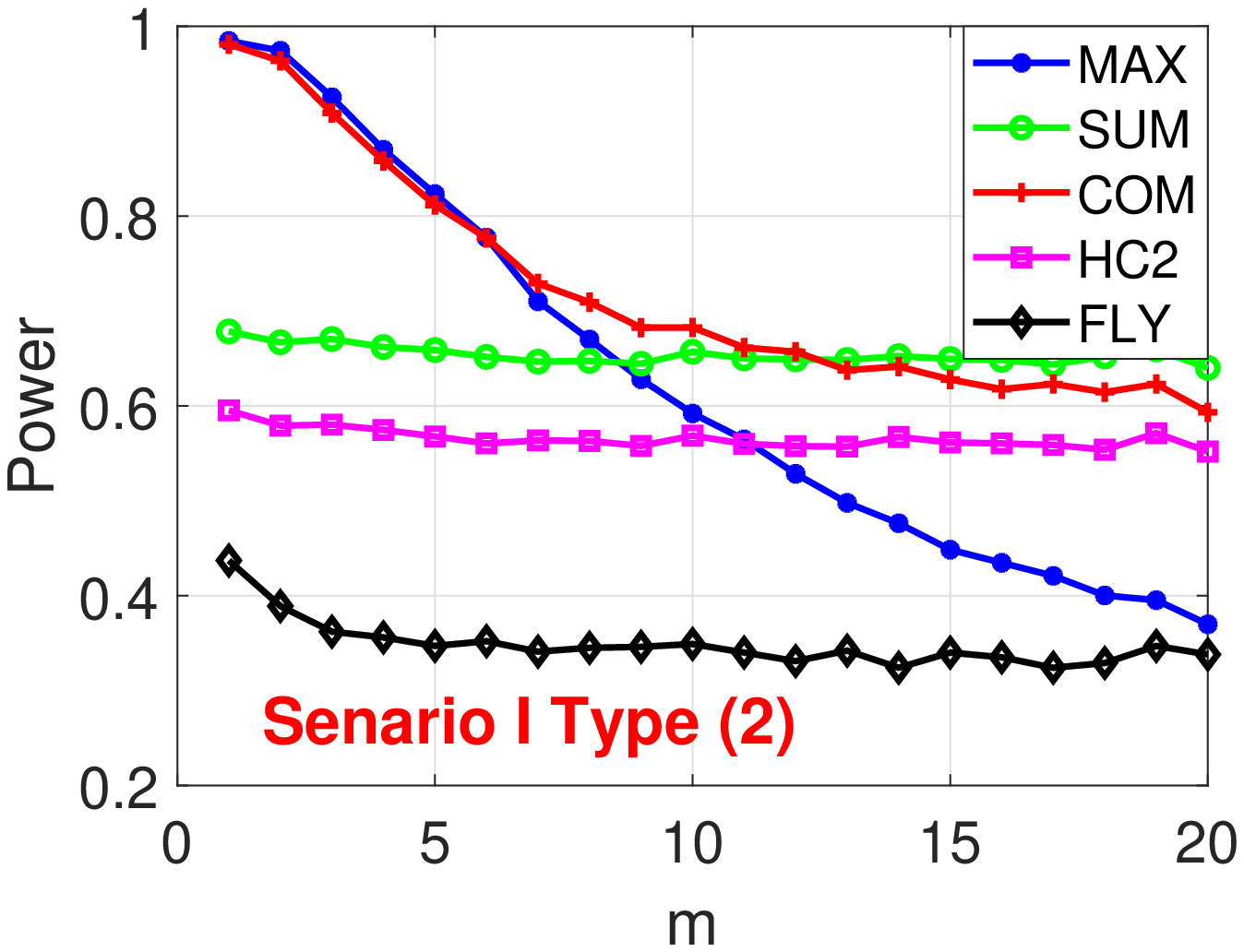}\hspace{-0.1in}
		\includegraphics[width=2.15in]{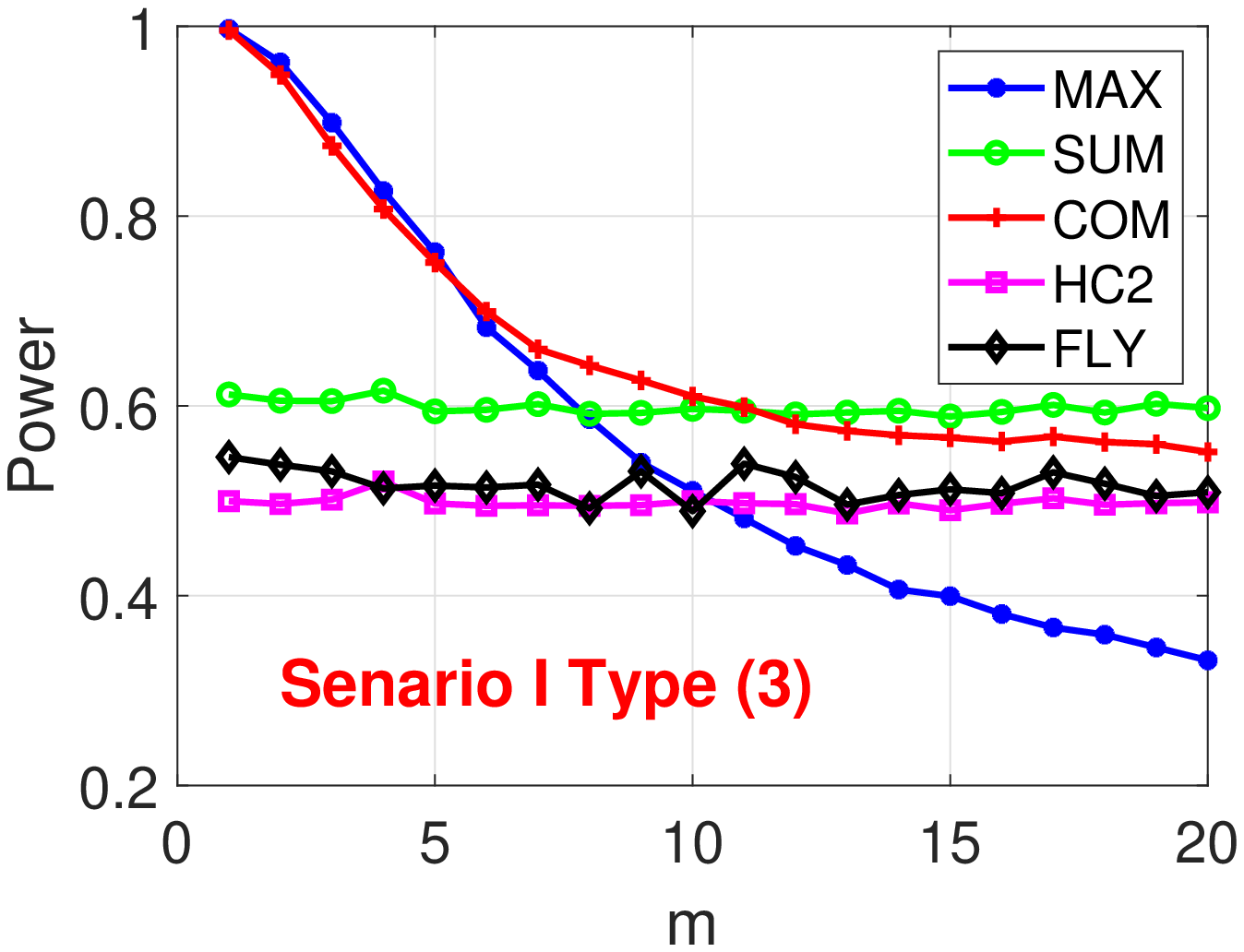}
		}
		\mbox{			    \includegraphics[width=2.15in]{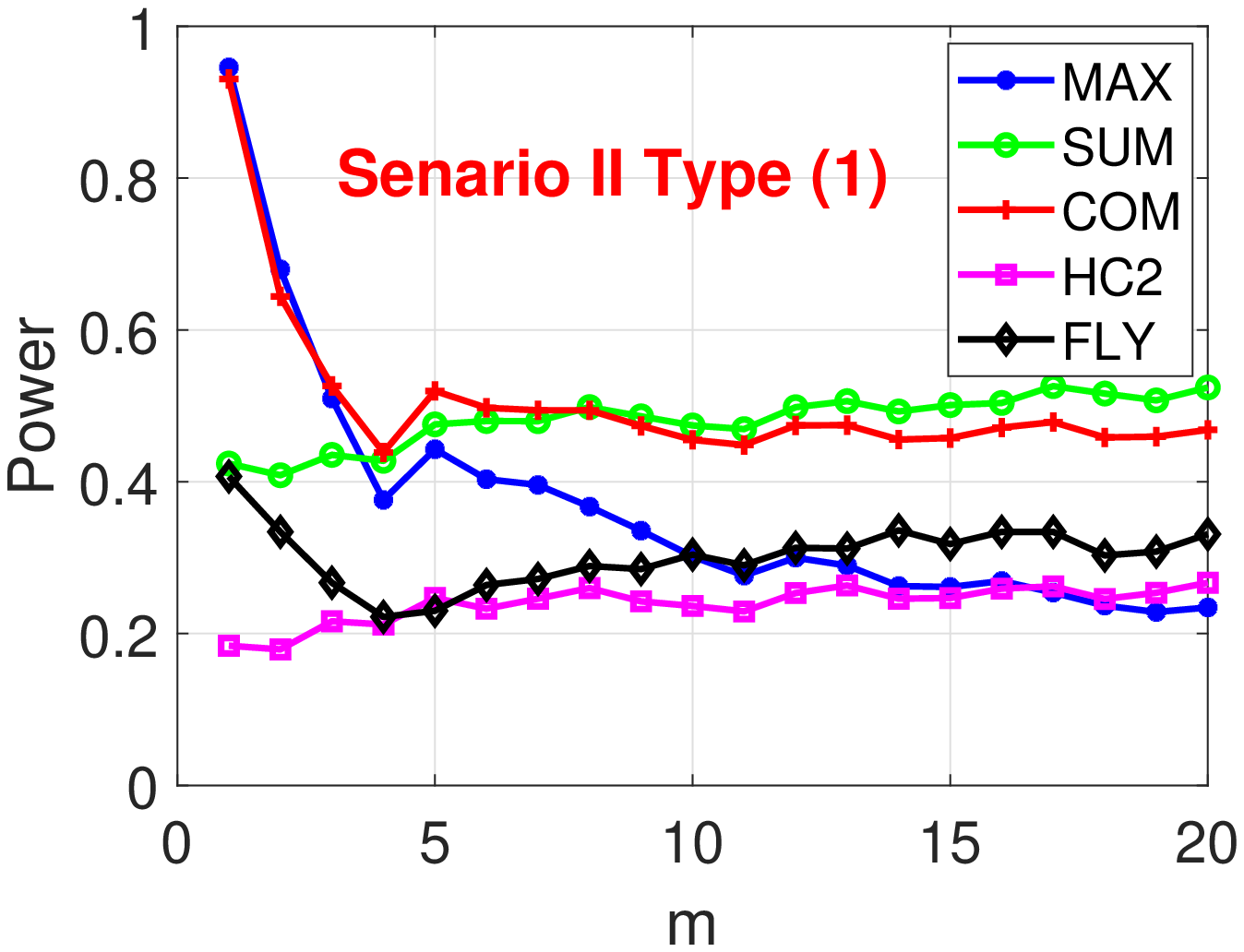}\hspace{-0.1in}
		\includegraphics[width=2.15in]{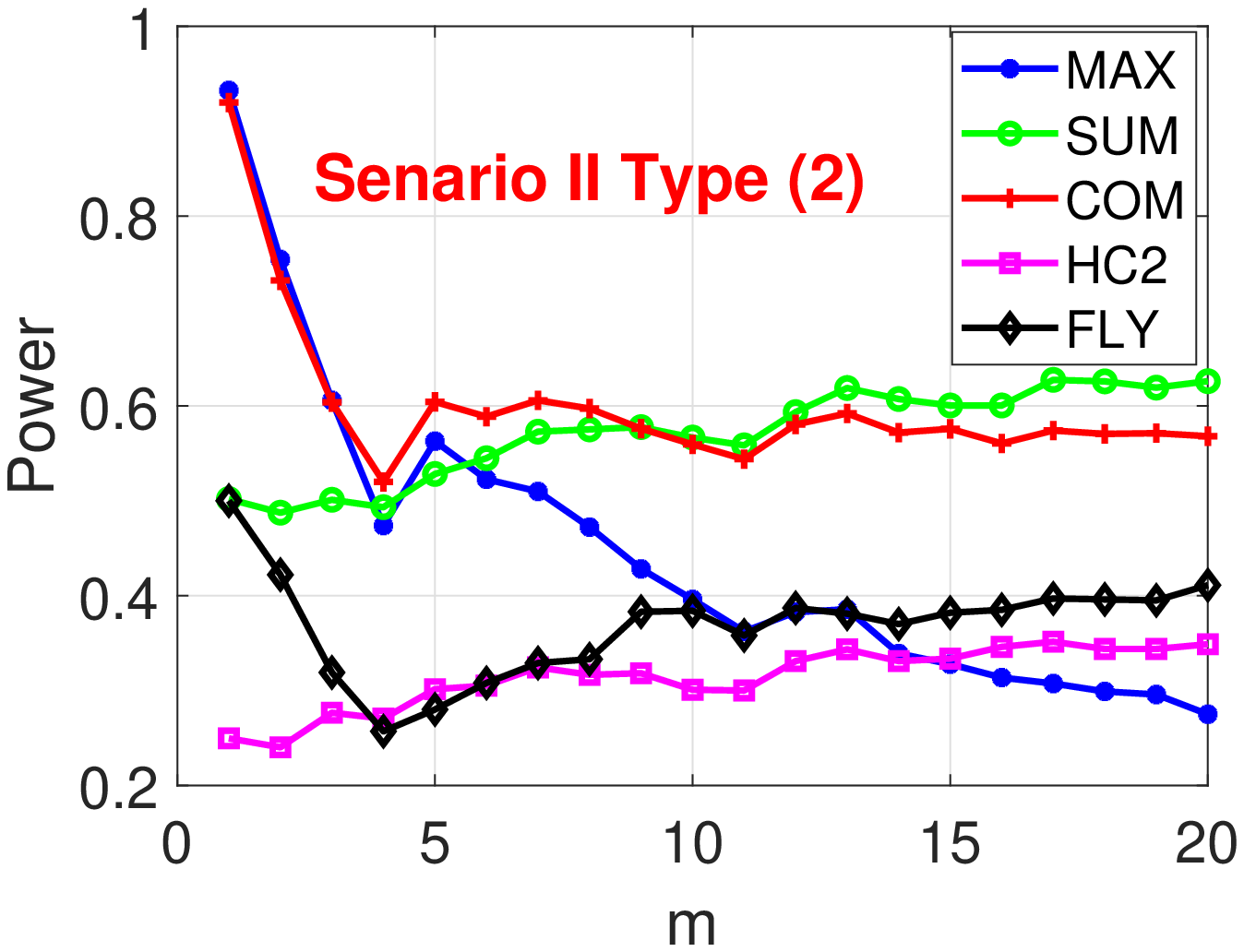}\hspace{-0.1in}
		\includegraphics[width=2.15in]{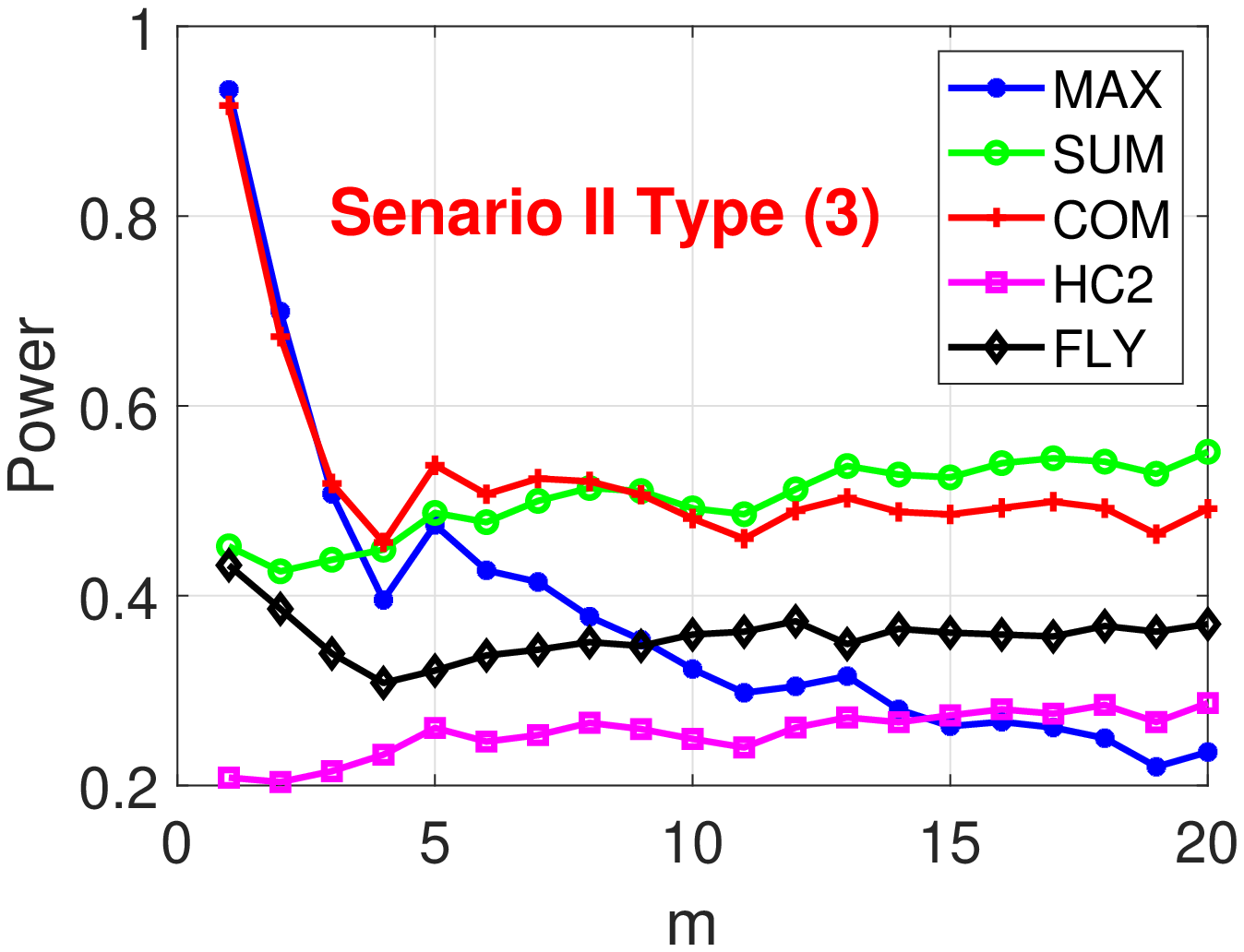}
		}
		\mbox{			    \includegraphics[width=2.15in]{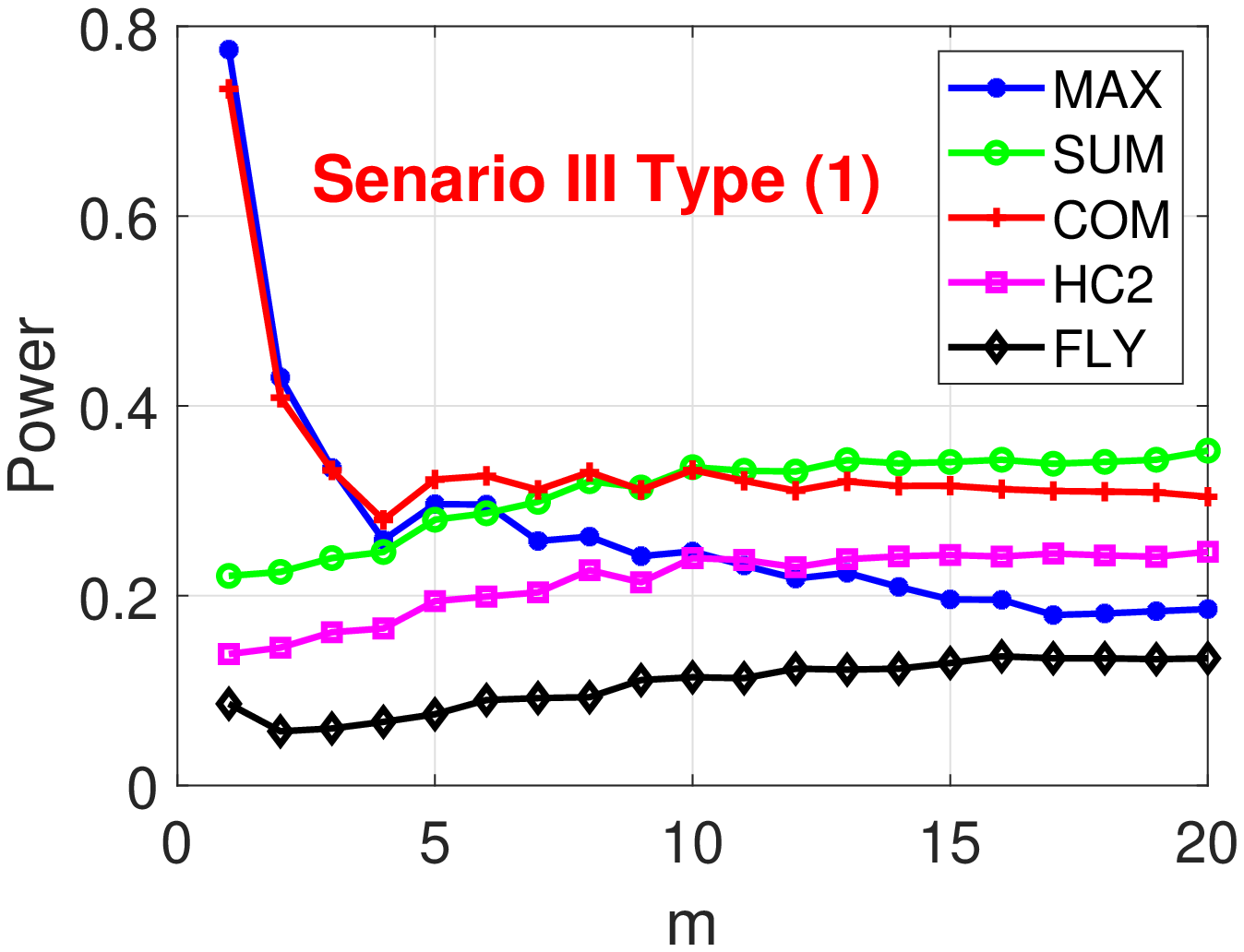}\hspace{-0.1in}
		\includegraphics[width=2.15in]{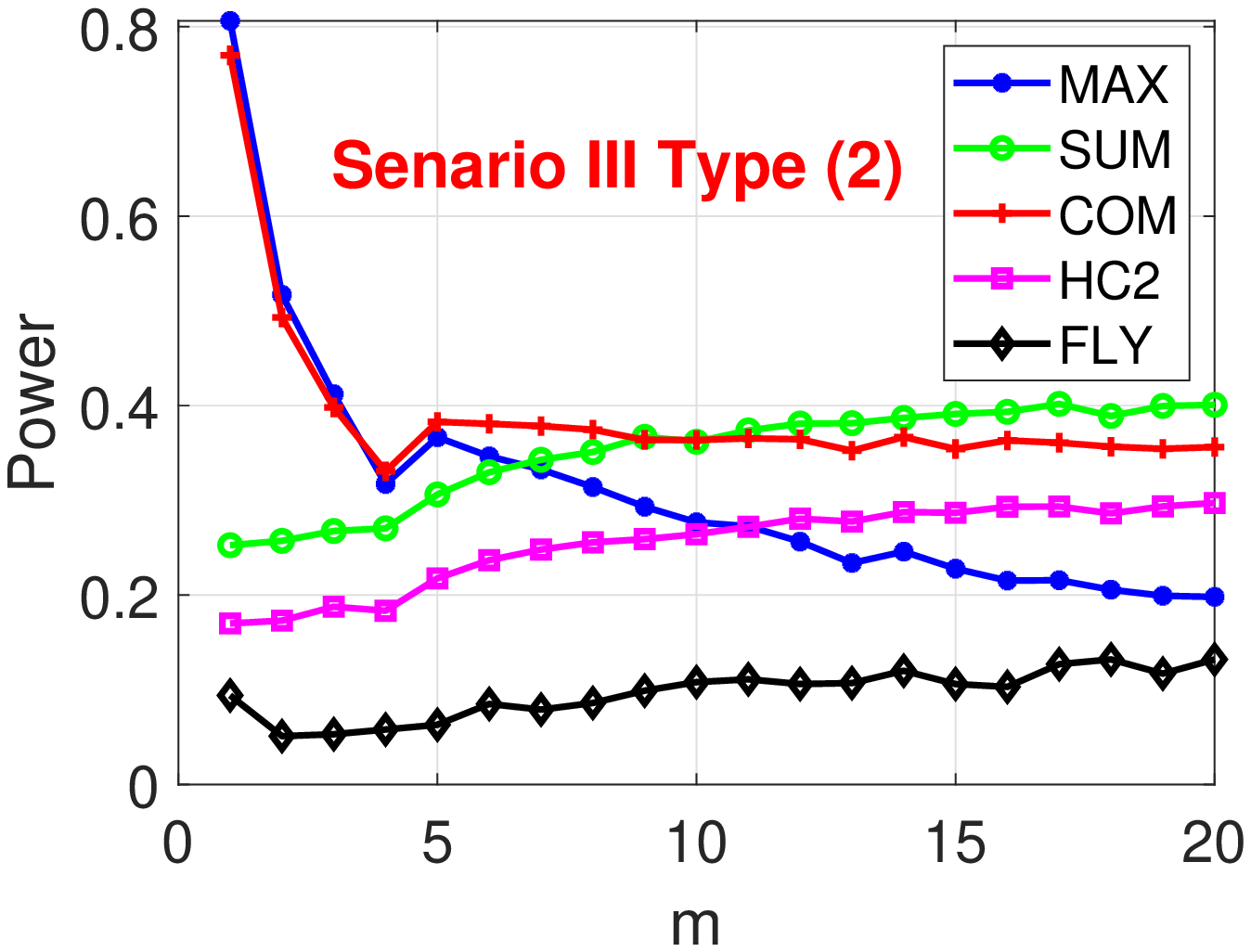}\hspace{-0.1in}
		\includegraphics[width=2.15in]{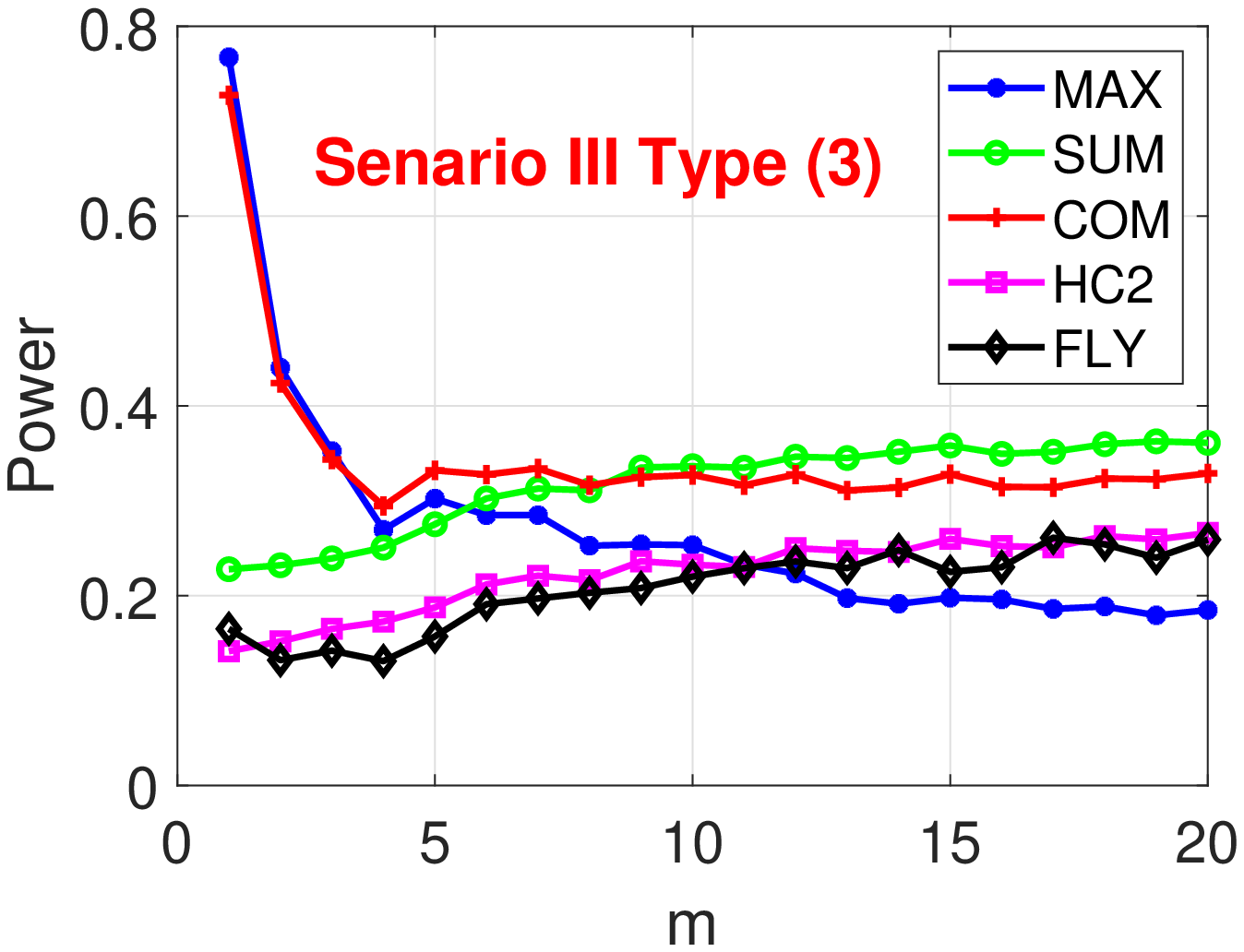}
		}
	\end{center}
\end{figure}

Next, we examine the power of each test. Our simulation shows that the power comparisons are similar for any combination of $(n, p)$ with $n=100, 200$ and $p=200, 400, 600$. Hence, we present the case $n=100$ and $p=200$ for conciseness. Define $\bmu=(\mu_{1},\cdots,\mu_p)^T$. For different number of nonzero-mean variables $m=1,\cdots,20$, we consider $\mu_j=\delta$ for $0<j\le m$ and $\mu_j=0$ for $j>m$. The parameter $\delta$ is chosen as $||\bmu||^2=m\delta^2=0.5$. Figure~\ref{f1} reports the power of the five tests. The power of MAX decreases as the number of nonzero-mean variables increases, which is as expected because, generally speaking, the max-type test is more powerful in  sparse case and less powerful in non-sparse case. The power of SUM slightly increases with $m$ and is higher than the power of HC2 and FLY in all cases. The proposed COM is as powerful as MAX when the number of variables with nonzero means is small (sparse case), and almost has the same power as SUM when the number of variables with nonzero means grows. In general, COM possesses the advantages of both MAX (in sparse case) and SUM (in non-sparse case), and outperforms HC2 and FLY in all scenarios. Observe that all the tests, except for COM, favor either the sparse or non-sparse case. Since in practice it is hard to justify whether the true underlying model is sparse or not, our proposed COM test, with its strong robustness, should be a more favorable choice over the competing approaches.

% \begin{figure}[tbph]
% \begin{center}
% \caption{\label{f1} Power vs. the number of variables of nonzero means for Example 4.1. The $x$-lab $m$ denotes the number of variables with non-zero means; the $y$-lab is the empirical power. }
% \includegraphics[width=6 in]{fig/onetotal.pdf}
% \end{center}
% \end{figure}

\subsection{Regression Coefficient Test Problem}  \label{sec:regression_simulation}

Then, we present our simulation results on the regression coefficient testing problem. We will compare our combo-type test $T_{com}^{(3)}$ (abbreviated as COM) from \eqref{3comtest} against
the test $T^{(3)}_{sum}$  (abbreviated as SUM) from \eqref{Stat_3}, the  test $T^{(3)}_{max}$ (abbreviated as MAX) from \eqref{3rdmax} and the empirical Bayes test proposed by \cite{goeman2006testing} (abbreviated as EB).

\noindent{\bf EXAMPLE 2.} We generate data from
(\ref{eq:modelv}), where the regression coefficients $\beta_{j}$ for
$j\in\{1,2,\cdots,q\}$ are simulated from a standard normal
distribution, and then we set  $\beta_j=0$ for $j>q$. In addition, the
predictor vector is given by $\X_i=\bms^{1/2}\z_i$ for
$i=1\cdots,n$, and each component of $\z_i$ is independently
generated from three distributions:
(1) the normal distribution $N(0,1)$;
(2) the exponential distribution $\exp(1)-1$;
(3) the mixture normal distribution $V/\sqrt{1.8}$, where $V$ is as in Example 1.

Moreover, the random error $\varepsilon_i$ is
independently generated from a standard normal distribution. We report the Scenarios (I) result where
 $\mbox{Cov}(\X_i)=\bms=(\sigma_{jk})\in
\mR^{p\times p}$ with $\sigma_{jk}=0.5^{|j-k|}$, while the results in the other two cases are similar. We consider two different sample sizes $n=100,200$, three different dimensions $p=200,400,600$ and two dimension of predictors in the reduced model $q=0$ or $5$.

\begin{table}[!htb]
\begin{center}
\small
\caption{\label{tt3} Sizes of tests for Example 2, $\alpha=0.05$.}
                     \vspace{0.2cm}
                     \renewcommand{\arraystretch}{0.8}
                     \setlength{\tabcolsep}{2pt}{
\begin{tabular}{cc|ccc|ccc|ccc}
\hline
\hline
  &Distribution& \multicolumn{3}{c}{(1)} & \multicolumn{3}{c}{(2)} & \multicolumn{3}{c}{(3)} \\ \hline
&$p$&200&400&600&200&400&600&200&400&600\\
\hline
\multicolumn{11}{c}{$q=0$}\\ \hline
$n=100$&MAX&0.032&0.024&0.027&0.026&0.032&0.026&0.027&0.036&0.036\\
&EB&0.047&0.047&0.047&0.049&0.052&0.050&0.059&0.044&0.044\\
&SUM&0.061&0.064&0.065&0.060&0.072&0.062&0.079&0.058&0.060\\
&COM&0.044&0.043&0.047&0.046&0.05&0.043&0.055&0.047&0.044\\ \hline
$n=200$&MAX&0.032&0.042&0.038&0.030&0.032&0.033&0.045&0.035&0.041\\
&EB&0.063&0.052&0.049&0.038&0.042&0.036&0.053&0.041&0.055\\
&SUM&0.069&0.06&0.057&0.054&0.052&0.045&0.063&0.049&0.064\\
&COM&0.053&0.051&0.048&0.036&0.039&0.044&0.058&0.042&0.050\\ \hline
\multicolumn{11}{c}{$q=5$}\\ \hline
$n=100$&MAX&0.032&0.029&0.024&0.024&0.030&0.020&0.031&0.037&0.029\\
&EB&0.048&0.050&0.055&0.056&0.058&0.061&0.061&0.054&0.045\\
&SUM&0.046&0.049&0.048&0.063&0.059&0.064&0.063&0.052&0.045\\
&COM&0.038&0.037&0.031&0.043&0.047&0.041&0.048&0.048&0.026\\ \hline
$n=200$&MAX&0.030&0.031&0.030&0.037&0.033&0.033&0.034&0.032&0.030\\
&EB&0.068&0.051&0.047&0.045&0.052&0.046&0.058&0.066&0.057\\
&SUM&0.067&0.051&0.049&0.049&0.054&0.045&0.070&0.068&0.061\\
&COM&0.048&0.045&0.036&0.040&0.040&0.037&0.051&0.049&0.049\\ \hline \hline
\end{tabular}}
\end{center}
\end{table}

We report the empirical sizes of the tests in Table \ref{tt3}. We observe that the empirical size of MAX tends to be smaller than the nominal level. EB and SUM, as well as the proposed COM, control the empirical sizes well for most of the times.
% EB and SUM tests can control the empirical sizes very well. Our COM test also has good performance of the empirical sizes in most cases.

We compare the power of the tests with $n=100$, $p=200$. Each entry of $\z_i\in \mathbb{R}^p$ is generated from the standard normal distribution (i.e. case (1) in Example 2).  Define $\bb_b=\kappa\cdot (\beta_{q+1},\cdots,\beta_p)^T$. Let $m$ denote the number of nonzero coefficients. For $m=1,\cdots,50$, we consider $\beta_j \sim N(0,1), q<j\le q+m$ and $\beta_j=0$, $j>q+m$. The parameter $\kappa$ is chosen  so that $||\bb_b||^2=0.5$. As we see from the plots, EB performs similarly to SUM. When the number of nonzero coefficients is small, MAX is more powerful than EB and SUM. In contrast, when the number of nonzero coefficients is large, EB and SUM outperform MAX. Once again, the proposed COM has same power as MAX in the sparse case, and has similar performance to EB and SUM in the non-sparse case. As we mentioned earlier, the results show the benefits of COM, as the true model is usually unknown in practical applications. The proposed COM provides good testing power in all cases.

\begin{figure}[h]
    \caption{Power vs. the number of nonzero coefficients for Example 2. The $x$-lab $m$ denotes  the number of non-zero coefficients; the $y$-lab is the empirical power.}
    \label{f3}
	\begin{center}
		\mbox{			    \includegraphics[width=2.8in]{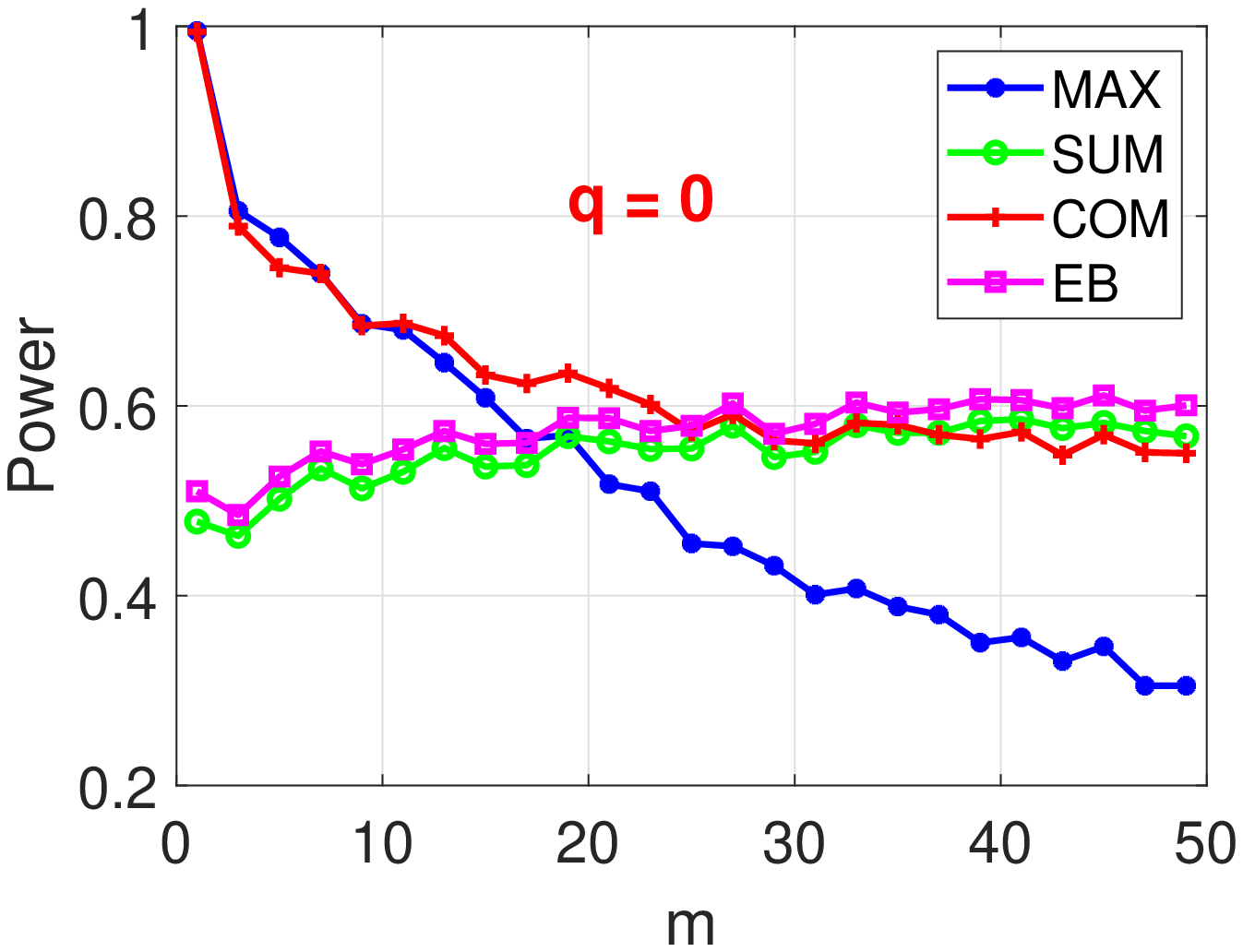}
		\includegraphics[width=2.8in]{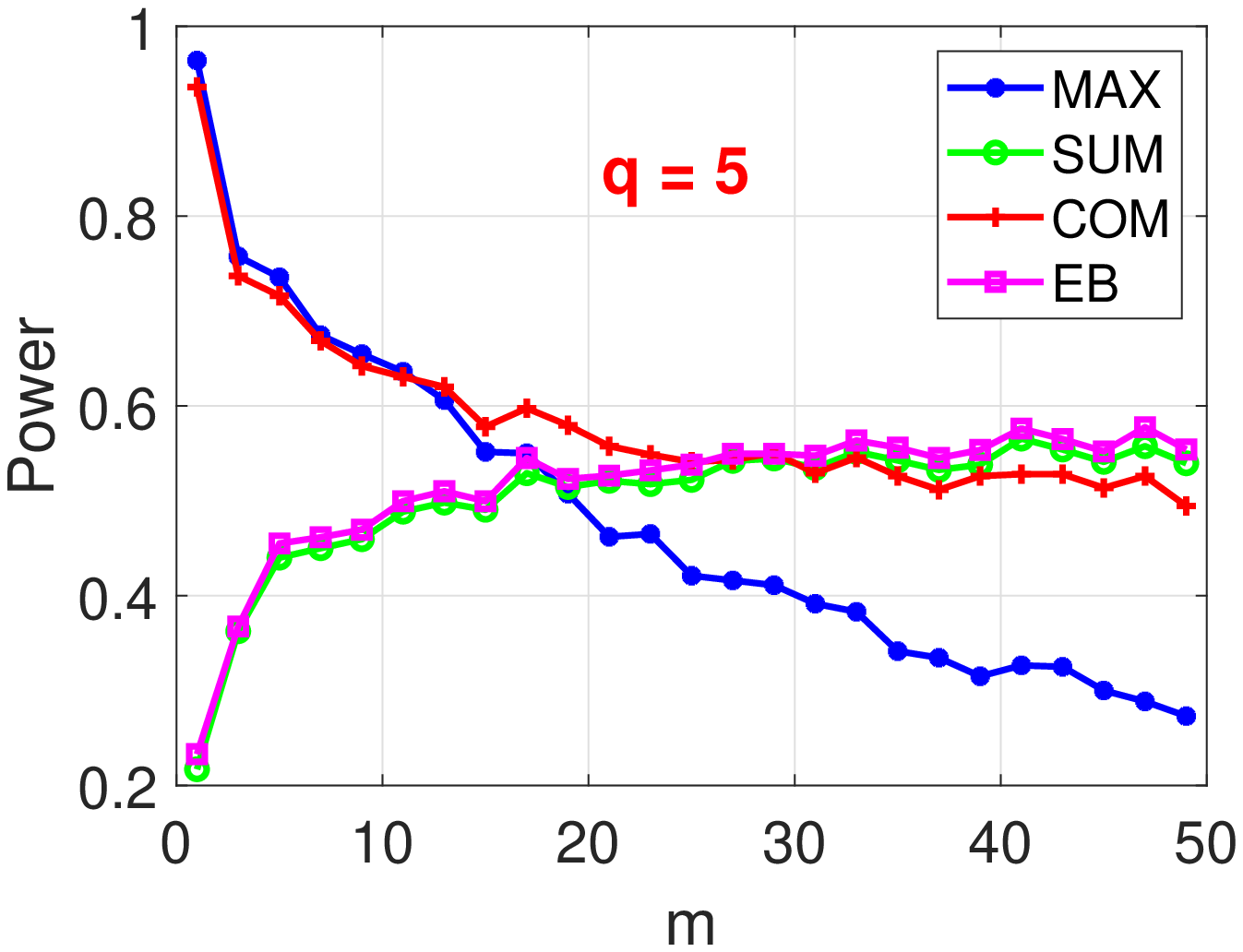}
		}
	\end{center}
\end{figure}

% \begin{figure}[ht]
% \begin{center}
% \caption{Histogram of sample means of stocks in S\&P 500.}
% \label{f4}
% \includegraphics[width=5 in]{matlab_fig/mean_histogram.eps}
% \end{center}
% \end{figure}

\begin{figure}[ht]
\begin{center}
\caption{Histogram of sample means of stock monthly return rates in S\&P 500.}
\label{f4}
\includegraphics[width=3.5 in]{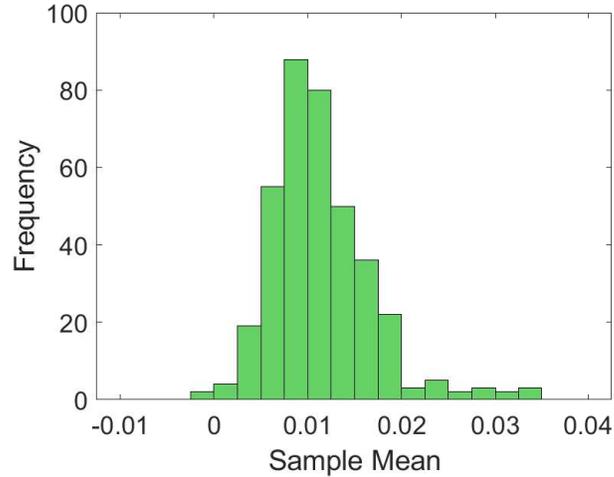}
\end{center}
\end{figure}

\section{Real Data Application}\lbl{real_data_application}

In this section, we further apply the results and test statistics obtained in Section \ref{applications_HDT} to two real data: a US stock data (dense model) and a search engine data (sparse model). As will be seen, the proposed combo-type test, COM, performs well on both datasets. Thus, it could serve as a ``universal'' test in practice no matter the true model is sparse or not.

\subsection{US Stock Data}

We apply the methods developed for one-sample mean test in Section \ref{one_sampl_me} to a pricing problem in finance. Specifically, we  investigate how financial
returns of assets are related to their risk-free returns.
Let $X_{ij}= R_{ij}-\textrm{rf}_i$ denote the excess return of the $j$th asset at
time $i$ for $i=1,\cdots,n$ and $j=1,\cdots,p$, where $R_{ij}$ is the
return on asset $j$ during period
$i$ and $\textrm{rf}_i$ is the risk-free return rate of all assets during period $i$. We
study the following pricing model
\begin{align}\label{rf}
X_{ij}=\mu_j+\xi_{ij},
\end{align}
for $i=1,\cdots,n$ and $j=1,\cdots,p$, or, in vector form,
$\X_i=\bmu+\bm\xi_i$,  where $\X_i=(X_{i1}, \dots, X_{ip})^{T}$,
$\bmu=(\mu_1, \dots, \mu_p)^{T}$, and $\bm\xi_i=(\xi_{i1}, \dots,
\xi_{ip})^{T}$ is the zero-mean error vector. The pricing model
in (\ref{rf}) is  the zero-factor model within the well known
Arbitrage Pricing Theory \citep{ross1976arbitrage}, where ``zero-factor" means
that no additional factor is used to model the price.
A common null hypothesis to be considered under the pricing
model (\ref{rf}) is $H_0: \bmu=\bm 0$, which means that the excess return of any asset is
zero on average, i.e. the return rate of any asset
$R_{ij}$ is equal to the risk-free return rate $\textrm{rf}_i$ on average.

We consider the monthly return rates of the stocks that constitute
the S\&P 500 index over the period from January 2005 to
November 2018.  Since the stocks that made up
the index changed over time and some stocks were created during this period,
we only consider a total of 374 stocks that were
included in the index during the entire time range. Figure \ref{f4} shows the sample mean of each stock in this period. We observe that most average returns are positive. In fact, as we enlarge the time range (increasing sample size $n$), the p-values of MAX, SUM and COM are eventually smaller than 0.05. These results suggest rejection of the null that the asset return does not only comes from the risk-free rates (on average), which is consistent with the views of many economists \citep{fama1993common,fama2015five}.

\begin{table}[!htb]
\begin{center}
\small
\caption{\label{tt6} Rejecting rates of each test in US stock Data.}
                     \vspace{0.2cm}
                     \renewcommand{\arraystretch}{0.8}
                     \setlength{\tabcolsep}{2pt}{
\begin{tabular}{cccc} \hline \hline
& MAX & SUM  & COM\\ \hline
$n=30$ &0.35 &0.39 &0.40 \\
$n=50$ &0.40 &0.51 &0.51 \\
$n=70$ &0.44 &0.67 &0.62 \\
$n=100$&0.52 &0.86 &0.83 \\
\hline \hline
\end{tabular}}
\end{center}
\end{table}

We further evaluate the tests by a random sampling procedure. Specifically, we randomly choose $n$ samples from the whole dataset and apply MAX, SUM and COM on this new sample. For each $n$, we repeat this experiment for 1000 times. Table \ref{tt6} reports the rejecting rates for each method with different $n$. From Table \ref{tt6}, we observe that SUM outperforms MAX in all cases by providing higher rejection rates. This is not surprising because for this data, the number of variables with nonzero means (assets with non-zero expected excess return) might be large, which is the case where sum-type tests could typically perform better than max-type tests. On the other hand, the combo-type test COM performs similarly as SUM overall. Therefore, COM does not lose efficiency in this problem.

% \begin{figure}[ht]
% \begin{center}
% \caption{\label{f4} Histogram of sample mean of each stock in S\&P 500.}% {\color{red}change ``sample means" to ``sample mean"}}
% \includegraphics[width=5 in]{fig/stockmean.pdf}
% \end{center}
% \end{figure}

\subsection{Search Engine Data}
We now use the data from \cite{lan2014testing} to make a case study on the regression coefficient test.
The dataset is obtained from an online mobile phone retailer. It contains a total of $n=98$ daily records. The response $Y$ is the revenue from the retailer's online sales. The explanatory variable  $V$ stands for the advertising spending on each of $p=164$ different keywords that were bid for Baidu, the leading search engine in China. We sort these explanatory variables by the correlation with the response, from high to low, and denote $V_1, V_2, \cdots$, etc. Since the sales vary with each day of the week, we introduce a 6-dimension indictor variables $W$ to represent Sunday to Friday. So there are 170 explanatory variables $X=(W,V)$ in our model. We will analyze it via the theory established in Section \ref{regress_3}.

For different values of $k$, set $\X_{a}^{(k)}=(W,V_1,\cdots,V_k)^T$ and the rest variables $\X_{b}^{(k)}=(V_{k+1},\cdots, V_p)^T$. We consider the linear model
$\Y=(\X_{a}^{(k)})^T\bm\beta_a^{(k)}+(\X_{b}^{(k)})^T\bm\beta_b^{(k)}+\bm\varepsilon$,
to test whether advertising spending on the rest of keywords $X_{b}^{(k)}$ could provide a significant contribution to online sales, conditional on the effect of $X_{ak}$, i.e. we test $H_0: \bm\beta_{b}^{(k)}=\bm 0$. We adopt the tests introduced in Example 2
in Section~\ref{sec:regression_simulation}, i.e. MAX, SUM, EB and COM.

\begin{table}[!htb]
\begin{center}
\small
\caption{\label{tt5} $p$-values of each test in Search Engine Data.}
                     \vspace{0.2cm}
                     \renewcommand{\arraystretch}{0.8}
                     \setlength{\tabcolsep}{2pt}{
\begin{tabular}{ccccc} \hline \hline
& MAX & SUM & EB & COM\\ \hline
$k=0$& 0.0056 &0.1432 &0.1100 &0.0112 \\
$k=1$& 0.0025 &0.2295 &0.1776 &0.0050 \\
$k=2$& 0.2083 &0.1319 &0.2464 &0.2465\\
$k=3$& 0.1508 &0.1387 &0.2877 &0.2583\\ \hline \hline
\end{tabular}}
\end{center}
\end{table}

Table \ref{tt5} reports the $p$-values of each tests with different $k$, which controls the sparsity of the true model. The significant level is set to be $\alpha=0.05$. From the results, we see that there are two keywords that are significant to the response (revenue) because both MAX and COM reject the null hypothesis as long as $k<2$ (note that these are two powerful methods for sparse model). When $k \ge 2$, all these four tests do not reject the null hypothesis, suggesting that the rest of keywords are not significant to the response. Notice that when $k<2$, SUM and EB fail to reject the null hypothesis at the significant level $0.05$, which shows their poor performance with sparse model, consistent with our theoretical claims and simulation results. On the contrary, COM succeeds in identifying the significant keywords in this problem, illustrating its edge over SUM and EB.

\vspace{-0.2in}

\section{Concluding Remarks}\label{conclusion}

In this paper, we prove the asymptotic independence between the sum and maximum of dependent random variables without stationary assumptions or strong mixing conditions. Then we apply our results to high-dimensional testing problems. Our proposed combo-type tests perform well regardless data being sparse or not. Now we make some comments.

1. The normal assumption is essential in the proof of asymptotic independence. Hence, we also assume the Gaussian assumption in the high-dimensional test problems. In literature, we may not need the Gaussian assumption to analyze the asymptotic distribution of the sum-type and max-type test statistics, e.g., \cite{cai2014two,chen2010two}. To prove the asymptotic independence between the sum and maximum of non-normal dependent random variables deserves further investigation.

2. To obtain the asymptotic distribution of the sum and maximum of dependent random variables, we assume the correlations between the random variables are not very strong. Recently, there has been much literature that consider high-dimensional testing problems without the weak correlation assumption, such as \cite{wang2021approximate,zhang2020simple_JASA}. The analogue of our asymptotic independence result  between the sum and maximum of dependent random variables with arbitrary covariance structures is also a very interesting and challenging problem.

3. The asymptotic independence results in Theorem \ref{theorem_3} is universal. We believe it can be generalized and applied to many other applications, such as change point detection and  statistical process controls.\\

\noindent\textbf{Acknowledgement}. We thank Professor Wei Xiong for very helpful discussions. The authors in this paper are listed in alphabetical order.

\newpage

\setcounter{section}{0}
\setcounter{equation}{0}
\setcounter{lemma}{0}
\renewcommand{\theequation}{S.\arabic{equation}}
\renewcommand{\thelemma}{S.\arabic{lemma}}
\def\thetable{S\arabic{table}}
\def\thefigure{S\arabic{figure}}

\renewcommand\thesection{S\arabic{section}}
\renewcommand\thefigure{S\arabic{figure}}
\renewcommand\thetable{S\arabic{table}}

\begin{center}
    {\LARGE\bf SUPPLEMENTARY MATERIAL of ``Asymptotic Independence of the Sum and Maximum of Dependent Random Variables with Applications to High-Dimensional Tests"}
\end{center}

\begin{description}
\item[Two-sample mean test] We propose the combo-type two-sample mean test and present its simulation
results in comparison with some of its competitors.
\item[Technical proofs] We provide the technical proofs of the theoretical results in Sections \ref{theory_results}, \ref{applications_HDT} of the text and Section \ref{twosample} of the supplementary material.
\end{description}

\newpage
\spacingset{1.9} % DON'T change the spacing!

\section{Two-sample Mean Test}\label{twosample}

\subsection{Testing Procedure}

Here, we consider the two-sample mean testing problem
in the high-dimensional setting. Assume that
$\{\X_{i1},\cdots,\X_{in_i}\}$ for $i=1,2$ are two independent
random samples with sizes $n_1$ and $n_2$, and from $p$-variate
normal distributions $N(\bmu_1,\bms)$ and $N(\bmu_2,\bms)$, respectively. Consider
\begin{align}\label{ht}
H_0:\bmu_1=\bmu_2\ \ \mbox{versus}\ \  H_1:\bmu_1\neq\bmu_2.
\end{align}
For the case where dimension $p$ is fixed, the classic Hotelling's $T^2$ test statistic is
\begin{align} \lbl{hong_red}
\frac{n_1n_2}{n_1+n_2}(\bar{\X}_1-\bar{\X}_2)^{T} \hat{\S}^{-1}(\bar{\X}_1-\bar{\X}_2),
\end{align}
 where $\bar{\X}_i$ is the sample mean vector of the $i$th sample and $\hat{\S}$ is the pooled sample covariance matrix defined by
\begin{align}\lbl{duck_yazi}
\hat{\S}=\frac{1}{n_1+n_2}\Big[\sum_{j=1}^{n_1}
(\bd{X}_{1j}-\bar{\bd{X}}_1)(\bd{X}_{1j}-\bar{\bd{X}}_1)^T+\sum_{j=1}^{n_2}
(\bd{X}_{2j}-\bar{\bd{X}}_2)(\bd{X}_{2j}-\bar{\bd{X}}_2)^T\Big].
\end{align}
Let $n=n_1+n_2$. In the high-dimensional case with $p>n$, $\hat{\S}$ is not guaranteed to be invertible. Under the assumption that $p/n\to c\in (0,\infty)$, \cite{bai1996effect} proposed a test statistic $(\bar{\X}_1-\bar{\X}_2)^{T}(\bar{\X}_1-\bar{\X}_2)$ by replacing $\hat{\S}$ in \eqref{hong_red} with  the identity matrix. Without the restriction on $n$ and $p$,  \cite{chen2010two} constructed a different test statistic by excluding the term $\sum_{j=1}^{n_i}\X_{ij}^{T} \X_{ij}$ for  $i=1$ and $2$ from  $(\bar{\X}_1-\bar{\X}_2)^{T}(\bar{\X}_1-\bar{\X}_2)$.
%{\red in \citep{bai1996effect}'s test statistic, which does not require explicit conditions in relationship between $p$ and $n$.}
However, the above two tests are not scale-invariant. A statistic $T(\X_{11},\cdots,\X_{1n_1}, \X_{21},\cdots,\X_{2n_2})$ is said to be location-scale invariant if the corresponding value of $T$ is not changed provided ``$\X_{ij}$" is replaced by ``$a\X_{ij}+b$" for all $i,j$, where $a$ and $b$ are arbitrary constants free of $i$ and $j$.  We say $T$ is scale-invariant if the above holds with $b=0$. For this reason, many efforts have been devoted to construct location-scale invariant test procedures, including  \cite{srivastava2008test,gregory2015two,feng2015two}, to name a few. In particular, \cite{srivastava2008test} considered the following sum-type test statistic
\begin{align}\lbl{sum2test}
T^{(2)}_{sum}=\frac{\frac{n_1n_2}{n_1+n_2}(\bar{\X}_1-\bar{\X}_2)^{T} \hat{\D}^{-1}(\bar{\X}_1-\bar{\X}_2)-\frac{(n_1+n_2-2)p}{(n_1+n_2-4)}}
{\sqrt{2\big[\tr(\hat{\R}^2)-\frac{p^2}{(n_1+n_2-2)}\big]c_{p,n}}},
\end{align}
where $\hat{\D}$ is the diagonal matrix of $\hat{\S}$ in \eqref{duck_yazi} and $\hat{\R}=\hat{\D}^{-1/2}\hat{\S}\hat{\D}^{-1/2}$ is the pooled sample correlation matrix, and  $c_{p,n}=1+\frac{\tr(\hat{\R}^2)}{p^{3/2}}$. The statistic $T^{(2)}_{sum}$ is location-scale invariant. Similar to the discussion in the paragraph above \eqref{hero_grass},
 the above sum-type tests usually do not perform well for sparse data. For the sparse alternative, \cite{chen2019two} extended the work of \cite{zhong2013tests} by studying the statistic
\begin{align}\lbl{chen_sta}
 M_{L_n}=\max_{s\in \ml{S}_n}\frac{L_n(s)-\hat{\mu}_{L_n(s), 0}}{\hat{\sigma}_{L_n(s), 0}}.
\end{align}
Note that this formula follows (4.3) from \cite{chen2019two}, where the notations ``$L_n(s)$, $\ml{S}_n$, $\hat{\mu}_{L_n(s), 0}$, $\hat{\sigma}_{L_n(s), 0}$" are quite involved; interested readers are referred to their paper for more details. Later on, we will compare our proposed test with $M_{L_n}$ in \eqref{chen_sta}.  Again, for sparse data, \cite{cai2014two} proposed the following max-type test statistic
\begin{align*}
T^{(2)}_{max}=\frac{n_1n_2}{n_1+n_2}\max_{1\le i\le p}\frac{(\bar{\X}_{1i}-\bar{\X}_{2i})^2}{\hat{\sigma}_{ii}^2},
\end{align*}
where $\bar{\X}_{ji}$ is the $i$th coordinate of $\bar{\X}_j\in \mathbb{R}^p$ for $j=1,2$ and $1\leq i \leq p$  and $\hat{\sigma}_{ii}^2$ is the $i$th diagonal element of $\hat{\S}$ in (\ref{duck_yazi}).  Similar to the one-sample test case, this test statistic is particularly powerful against sparse alternatives
with certain optimality. We will study the asymptotic behavior of the sum-type and max-type tests, and design a new test that takes advantage of both worlds.

Recall that $\bd{\Sigma}$ is the covariance matrix shared by two populations, and let $\bd{D}$ be the diagonal matrix of $\bd{\Sigma}$ such that $\bd{R}:=\bd{D}^{-1/2}\bd{\Sigma}\bd{D}^{-1/2}$ is the population correlation matrix.  For soundness, assume that the sample sizes $n_1$ and $n_2$ both depend on $p$. Now we apply the theoretical results in Section \ref{theory_results} to the two-sample mean test as follows.

\begin{theorem}\label{thtwo}
Assume the null hypothesis in \eqref{ht} holds and $\lim_{p\to \infty}n_1/n_2\to \kappa\in (0, \infty)$. The following are true as $p\to\infty$:
\begin{itemize}
\item[(i)] If \eqref{(C1)} holds, then $T^{(2)}_{sum}\to N(0,1)$ in distribution;
\item[(ii)]
If  \eqref{condition_2} holds with ``$\bd{\Sigma}$''  being replaced by ``$\bd{R}$''
 and $\log p=o(n^{1/3})$,  then
$T^{(2)}_{max}-2\log p+\log\log p$ converges weakly to a Gumbel distribution with cdf $F(x)=
 \exp\{-\frac{1}{\sqrt{\pi}}\exp(-x/2)\}$;
\item[(iii)]
Assume \eqref{(C1)} holds. If  \eqref{assumption_A3} is true with ``$\bd{\Sigma}$'' replaced by ``$\bd{R}$'',  then  $T^{(2)}_{sum}$ and $T^{(2)}_{max}-2\log p+\log\log p$ are asymptotically independent.
\end{itemize}
\end{theorem}

Part (i) of Theorem S\ref{thtwo} is from \cite{srivastava2008test}. Recently \cite{jiang2021mean} obtained a general theory, which also leads to the same conclusion.
Same as in the one-sample test, for test $T^{(2)}_{sum}$, a level-$\alpha$ test rejects $H_0$ when $T^{(2)}_{sum}> z_{\alpha}= \Phi^{-1}(1-\alpha)$ of
$N(0,1)$.
For the max-type test, a level-$\alpha$ test will
then be carried out through rejecting the null hypothesis when $T^{(2)}_{max}-2\log p+\log\log p> q_{\alpha}=-\log \pi-2\log\log(1-\alpha)^{-1}$ of the distribution function in Theorem S\ref{thtwo}(ii).

Relying on Theorem S\ref{thtwo}, we propose the following test statistic which utilizes the max-type and sum-type tests. Define
\begin{align}\lbl{shiba}
T_{com}^{(2)}=\min\{P^{(2)}_M, P^{(2)}_S\},
\end{align}
where
$P^{(2)}_{M}=1-F(T_{max}^{(2)}-2\log p+\log\log p)$ with $F(y)= e^{-\pi^{-1/2}e^{-y/2}}$ and $P^{(2)}_{S}=1-\Phi(T_{sum}^{(2)}).$ are the $p$-values of the two tests, respectively. Similar to Corollary \ref{coro2}, we immediately obtain the following result by the asymptotic independence.

\begin{coro}\lbl{coro2sample} Assume the condition in Theorem S\ref{thtwo}(iii) holds. Then
$T_{com}^{(2)}$ in \eqref{shiba} converges weakly to a distribution with density  $G(w)=2(1-w)I(0\leq w \leq 1)$ as $p\to\infty.$
\end{coro}

According to Corollary S\ref{coro2sample}, the proposed combo-type test leads us to perform a level-$\alpha$ test by rejecting the null hypothesis
when $T_{com}^{(2)}<1-\sqrt{1-\alpha}\approx \frac{\alpha}{2}$ when $\alpha$ is small. Now we analyze the power of the test $T_{com}^{(2)}$.

Write $\bmu=(\bmu_1, \bmu_2)^T$. Similar to (\ref{pms}), the power function of our combo-type test $\beta_C^{(2)}(\bmu,\alpha)$ is larger than $\max\{\beta_M^{(2)}(\bmu,\alpha),\beta_S^{(2)}(\bmu,\alpha)\}$, where $\beta^{(2)}_M(\bmu,\alpha)$ and $\beta^{(2)}_S(\bmu,\alpha)$ are the power functions of $T_{max}^{(2)}$ and $T_{sum}^{(2)}$ with significant level $\alpha$, respectively. Following \cite{srivastava2008test}, the power function of $T_{sum}^{(2)}$ is given by
\begin{align*}
\beta^{(2)}_S(\bmu,\alpha)=\lim_{p\to\infty}\Phi\left(-z_{\alpha}+\frac{\frac{n_1n_2}{n_1+n_2}(\bmu_1-\bmu_2)^{T} \D^{-1}(\bmu_1-\bmu_2)}{\sqrt{2\tr(\R^2)}}\right).
\end{align*}
Thus, we have
\begin{align*}
\beta^{(2)}_C(\bmu,\alpha)\ge\lim_{p\to\infty} \Phi\left(-z_{\alpha/2}+\frac{\frac{n_1n_2}{n_1+n_2}(\bmu_1-\bmu_2)^{T} \D^{-1}(\bmu_1-\bmu_2)}{\sqrt{2\tr(\R^2)}}\right).
\end{align*}
Further write $\bmu_1=(\mu_{11}, \cdots, \mu_{1p})^T$ and $\bmu_2=(\mu_{21}, \cdots, \mu_{2p})^T$, and define $\delta_i=\mu_{1i}-\mu_{2i}$ for $i=1,\cdots, p.$ We have an analogous set of analysis and claims as for the one-sample test. Firstly, by Theorem 2 from \cite{cai2014two}, the asymptotic power of $T_{max}^{(2)}$ converges to one if $\max_{1\le i\le p} |\delta_i/{\sigma_{ii}}|\ge c\sqrt{\log p/n}$ for a certain constant $c$ and if the sparsity level $\gamma<1/4$ and the locations of the non-zero variables are randomly and uniformly selected from $\{1, \cdots, p\}$, meaning that the power function of our proposed test $T_{com}^{(2)}$ also converges to one under this situation. Secondly, according to Theorem 3 from \cite{cai2014two}, the condition $\max_{1\le i\le p} |\delta_i/{\sigma_{ii}}|\ge c\sqrt{\log p/n}$ is minimax rate-optimal for testing against sparse alternatives, and such optimality also holds for our test $T_{max}^{(2)}$.

Similar to the one-sample test problem, we consider a special case with $\bms=\I_p$. There are $m$ nonzeros $\delta_i$ and they are all equal to $\delta\not=0$. Thus,
\begin{align*}
\beta_S^{(2)}=\lim_{p\to\infty}\Phi\left(-z_{\alpha}+\frac{n_1n_2m\delta^2}{n\sqrt{2p}}\right).
\end{align*}
Take $\xi>0$ such that $p^{1/2}n^{2\xi-1}\to\infty.$
For the non-sparse case: $\delta=O(n^{-\xi})$ and $m=o(p^{1/2}n^{2\xi-1})$, we have $\beta_M^{(2)}(\bmu,\alpha)\approx \alpha$ and $\beta_C^{(2)}(\bmu,\alpha)\approx \beta_S^{(2)}(\bmu,\alpha/2)$. For the sparse case: $\delta=c\sqrt{\log p/n}$ with sufficient large $c$ and $m=o((\log p)^{-1}p^{1/2})$, we have $\beta_S^{(2)}(\bmu,\alpha)\approx \alpha$ and $\beta_C^{(2)}(\bmu,\alpha)\approx \beta_M^{(2)}(\bmu,\alpha/2)\to 1$.

\subsection{Simulation Results}
Then, we present the simulation results for the two-sample test problem, where our test $T_{com}^{(2)}$ (abbreviated as COM) from \eqref{shiba} will be compared with
the sum-type test $T^{(2)}_{sum}$ from \eqref{sum2test}  proposed by \cite{srivastava2008test}  (abbreviated as SUM), the max-type test $T^{(2)}_{max}$ proposed by \cite{cai2014two} (abbreviated as MAX) and the Higher Criticism test proposed by \cite{chen2019two} (abbreviated as HC2).

Recall the three scenarios of covariance matrices appeared in (I), (II) and (III) after Example 1. Since the conclusions from all three scenarios are similar, here we only present the results when the covariance matrix follows Scenario (I), i.e.,  $\bms=(0.5^{|i-j|})_{1\le i,j\le p}$.

\noindent{\bf Example 3.}
We consider $\X_{ki}=\bmu_k+\bms^{1/2}\z_{ki}$ for $k=1,2$ and
$i=1\cdots,n$, and each component of $\z_{ki}$ is independently
generated from three distributions:
(1) $N(0,1)$; (2) $t$-distribution, $t(5)/\sqrt{5/3}$;
(3) the mixture normal random variable $V/\sqrt{1.8}$, where $V$ is as in Example 1.

%has  density function $0.1f_1(x)+0.9f_2(x)$ with $f_1(x)$ and $f_2(x)$ being  the density functions of  $N(0,9)$ and $N(0,1)$, respectively.

We consider two different sample sizes $n=100,200$ and three different dimensions $p=200,400,600$. Under the null hypothesis, we set $\bmu_1=\bmu_2=\bm 0$. The significance level is chosen such that $\alpha=0.05$. Again, for the alternative hypothesis, we only present on $n=100, p=200$ and $\bmu_2=\bm 0$ since the observations from different combinations of $n$ and $p$ are similar.  Define $\bmu_1=(\mu_{11},\cdots,\mu_{1p})^T$. For different number of nonzero-mean variables $m=1,\cdots,20$, we consider $\mu_{j1}=\delta$ for  $0<j\le m$ and $\mu_{1j}=0$ for $j>m$. The parameter $\delta$ is chosen such that $||\bmu_1||^2=m\delta^2=1$.

\begin{table}[!htb]
\begin{center}
\caption{\label{tt2} Sizes of tests for Example 3, $\alpha=0.05$.}
                     \vspace{0.2cm}
                     \renewcommand{\arraystretch}{1}
                     \setlength{\tabcolsep}{4pt}{
\begin{tabular}{cc|ccc|ccc|ccc}
\hline
\hline
  &Distribution& \multicolumn{3}{c}{(1)} & \multicolumn{3}{c}{(2)} & \multicolumn{3}{c}{(3)} \\ \hline
&$p$&200&400&600&200&400&600&200&400&600\\
\hline
$n=100$&MAX&0.057 &0.063 &0.058 &0.062 &0.061 &0.060 &0.060 &0.066 &0.064\\
&SUM&0.055 &0.058 &0.057 &0.060 &0.064 &0.065 &0.055 &0.061 &0.061\\
&COM &0.057 &0.074 &0.060 &0.064 &0.065 &0.068 &0.061 &0.061 &0.062\\
&HC2&0.043&0.035&0.044&0.047&0.037&0.052&0.042&0.043&0.053\\\hline
$n=200$&MAX&0.053 &0.053 &0.051 &0.049 &0.062 &0.061 &0.044 &0.065 &0.055\\
&SUM &0.052 &0.065 &0.059 &0.065 &0.061 &0.064 &0.052 &0.056 &0.053\\
&COM &0.044 &0.049 &0.056 &0.061 &0.063 &0.058 &0.046 &0.061 &0.053\\
&HC2&0.030&0.025&0.025&0.030&0.035&0.036&0.025&0.020&0.033\\ \hline \hline
\end{tabular}}
\end{center}
\end{table}

%\begin{figure}[tbph]
%\begin{center}
%\caption{\label{f2} Power of tests with different number of nonzero variables for Example 4.2.}
%\includegraphics[width=6 in]{mstwo.pdf}
%\end{center}
%\end{figure}

% \begin{figure}[!ht]
% \begin{center}
% \caption{\label{f2} Power vs. the number of variables with non-zero means for Example 4.2. The $x$-lab $m$ denotes the number of variables with non-zero means; the $y$-lab is the empirical power.}
% \includegraphics[width=6 in]{fig/twototal.pdf}
% \end{center}
% \end{figure}

\begin{figure}[h]
    \caption{Power vs. number of variables with non-zero means for Example 3. The $x$-lab $m$ denotes the number of variables with non-zero means; the $y$-lab is the empirical power.}
    \label{f2}
	\begin{center}
		\mbox{			    \includegraphics[width=2.15in]{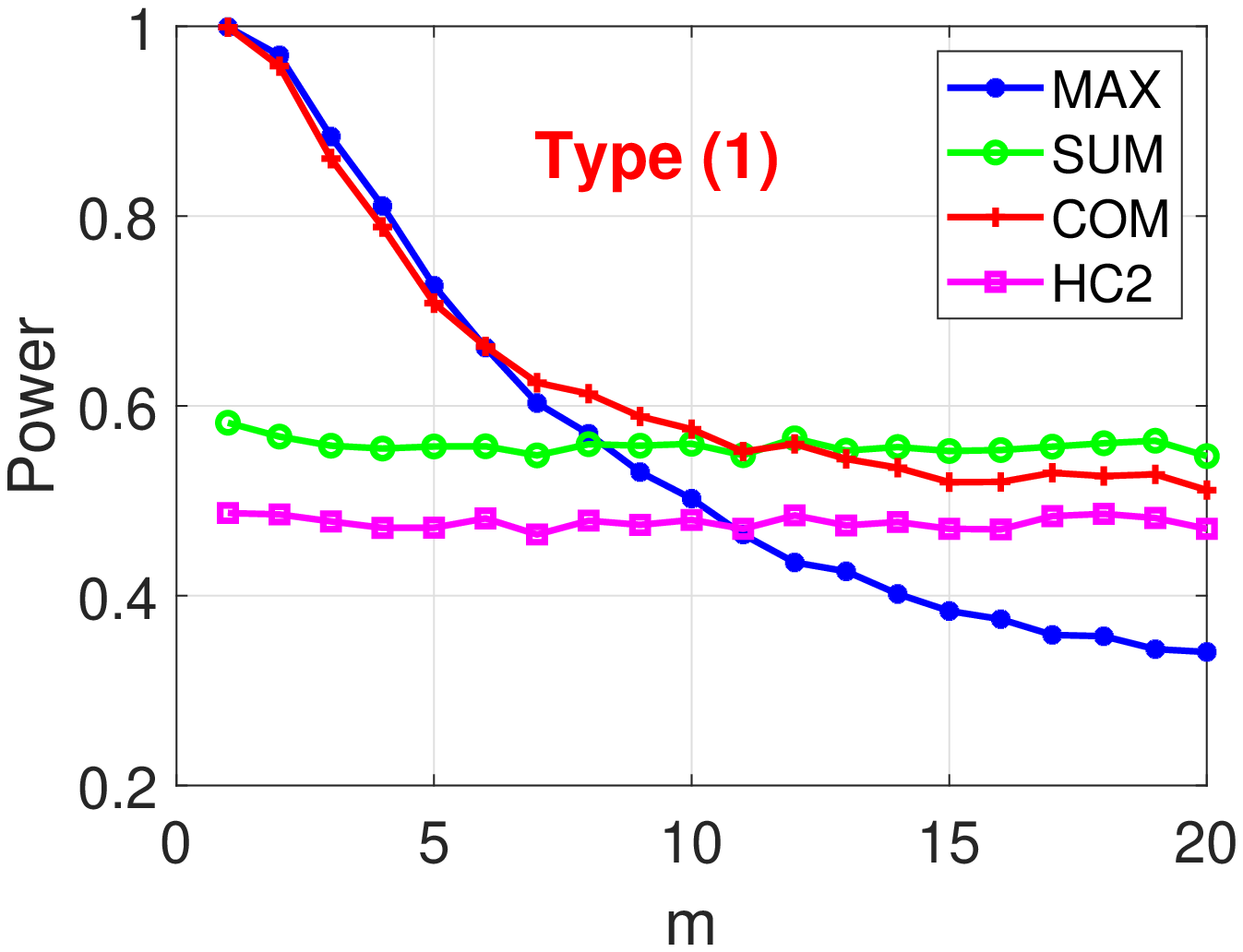}\hspace{-0.1in}
		\includegraphics[width=2.15in]{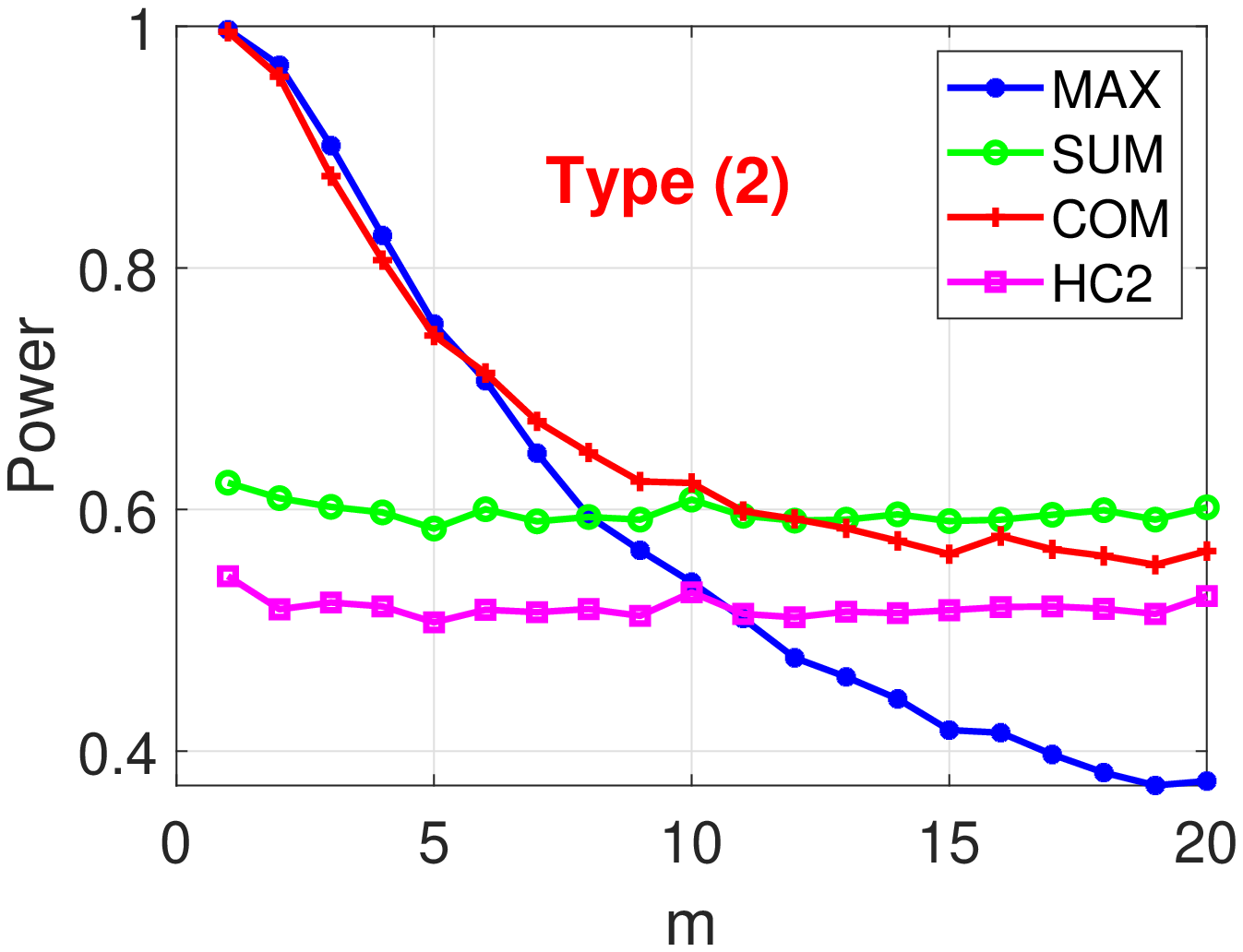}\hspace{-0.1in}
		\includegraphics[width=2.15in]{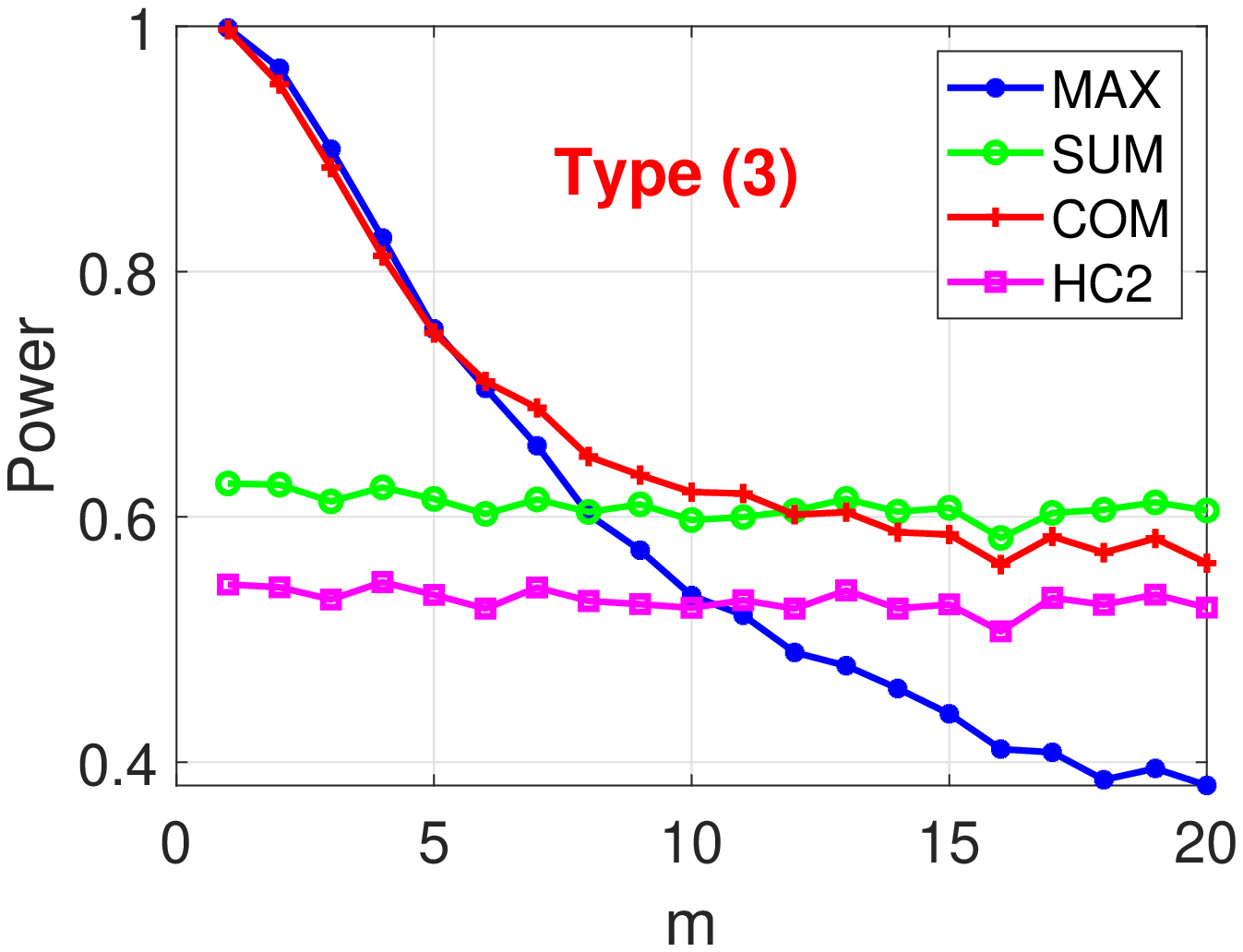}
		}
	\end{center}
\end{figure}

Table \ref{tt2} reports the empirical sizes of the compared tests. We see that all the tests control the empirical sizes in most cases except that the sizes of HC2 are a little smaller than the nominal level when $n=200$.

Figure \ref{f2} shows the power of each test, where we observe a similar pattern as in Example 1. The power of MAX declines as the number of variables with nonzero means  is increasing.  The power of SUM  and COM are always larger than that of HC2 in all cases. The proposed COM matches the power of MAX when the number of variables with nonzero means is small, and almost has same power as SUM when $m$ is large. This justifies the superiority of the proposed combo-type test in the two-sample testing problem, regardless of the sparsity of the data.

\section{Technical Proofs}

\subsection{Proof of Theorem \ref{theorem_1}}
\noindent\textbf{Proof of Theorem \ref{theorem_1}.} Let $\xi_1, \xi_2,\cdots$ be i.i.d. $N(0,1)$-distributed random variables. Let $\bd{\Sigma}^{1/2}$ be a non-negative definite matrix such that $\bd{\Sigma}^{1/2}\cdot \bd{\Sigma}^{1/2}=\bd{\Sigma}.$ Then $(Z_1, \cdots, Z_p)^T$ and  $\bd{\Sigma}^{1/2}(\xi_1, \cdots, \xi_p)^T$ have the same distribution. As a consequence, $Z_1^2+\cdots + Z_p^2$ has the same distribution as that of
\begin{align} \lbl{An_Ru}
(\xi_1, \cdots, \xi_p)\bd{\Sigma}^{1/2}\cdot\bd{\Sigma}^{1/2}(\xi_1, \cdots, \xi_p)^T
=(\xi_1, \cdots, \xi_p)\bd{\Sigma}(\xi_1, \cdots, \xi_p)^T.
\end{align}
%Recall the explanation from \eqref{deed_wong}.
Let $\lambda_{p, 1}, \lambda_{p,2}, \cdots, \lambda_{p,p}$ be the eigenvalues of $\bd{\Sigma}=\bd{\Sigma}_p$ and $\bd{O}$ be a $p\times p$ orthogonal matrix such that $\bd{\Sigma}_p=\bd{O}^T\,\mbox{diag}(\lambda_{p1}, \cdots, \lambda_{pp})\bd{O}$. In particular, since all of the diagonal entries of $\bd{\Sigma}$ are $1$, we have $\lambda_{p, 1}+ \cdots+ \lambda_{p,p}=p$. By the orthogonal invariance of normal distributions, $\bd{O}(\xi_1, \cdots, \xi_p)^T$ and $(\xi_1, \cdots, \xi_p)^T$ have the same distribution. By \eqref{An_Ru}, $Z_1^2+\cdots + Z_p^2$ is equal to
\beaa
\big[\bd{O}(\xi_1, \cdots, \xi_p)^T\big]^T\,\mbox{diag}(\lambda_{p, 1}, \cdots, \lambda_{p,p})\,\big[\bd{O}(\xi_1, \cdots, \xi_p)^T\big],
\eeaa
and hence has the same distribution as that of $\lambda_{p,1}\xi_1^2+\cdots +\lambda_{p, p}\xi_p^2$. It is easy to see $E(Z_1^2+\cdots + Z_p^2)=p$ and
\beaa
\mbox{Var}(Z_1^2+\cdots + Z_p^2)&=&\lambda_{p,1}^2\mbox{Var}(\xi_1^2)+\cdots +\lambda_{p, p}^2\mbox{Var}(\xi_p^2)\\
&=&2\lambda_{p,1}^2+\cdots +2\lambda_{p, p}^2\\
& = & 2\cdot \mbox{tr}(\bd{\Sigma}^2).
\eeaa
Easily, $m_k:=E(|\xi_1^2-1|^k)<\infty$ for any $k\geq 1$. Note that $\lambda_{p,1}\xi_1^2+\cdots +\lambda_{p, p}\xi_p^2$ is a sum of independent random variables.
Then,
\beaa
\frac{1}{[2\cdot \mbox{tr}(\bd{\Sigma}^2)]^{(2+\delta)/2}}
\sum_{i=1}^p E\big|\lambda_{p,i}\xi_i^2-E(\lambda_{p,i}\xi_i^2)\big|^{2+\delta}
&\leq & \frac{m_{2+\delta}}{[\mbox{tr}(\bd{\Sigma}^2)]^{(2+\delta)/2}}\cdot \sum_{i=1}^p\lambda_{p,i}^{2+\delta}\\
&=&m_{2+\delta}\cdot \frac{\mbox{tr}(\bd{\Sigma}^{2+\delta})}{[\mbox{tr}(\bd{\Sigma}^2)]^{(2+\delta)/2}},
\eeaa
which goes to zero by Assumption (\ref{condition_1}). Therefore, by the Lyapunov central limit theorem, $(\lambda_{p,1}\xi_1^2+\cdots +\lambda_{p, p}\xi_p^2-p)/\sqrt{2\mbox{tr}(\bd{\Sigma}^2)}$ converges weakly to $N(0,1)$ as $p\to \infty.$ This implies that $(Z_1^2+\cdots + Z_p^2-p)/\sqrt{2\mbox{tr}(\bd{\Sigma}^2)}$ converges weakly to $N(0,1)$ as $p\to \infty.$ \hfill$\square$

\subsection{Proof of Theorem \ref{theorem_2}}

For a graph $G$, we say vertices $i$ and $j$ are neighbors if there is an edge between them. For a set $A$, we write $|A|$ for its cardinality. We first prove some lemmas.

\begin{lemma}\lbl{jexbcsiyr} Let $G=(V, E)$ be an undirected graph with $n=|V|\geq 4$ vertices. Write $V=\{v_1, \cdots, v_n\}$. Assume each vertex in $V$ has at most $q$ neighbors. Let $G_t$ be the set of subgraphs of $G$ such that each subgraph has $t$ vertices and  at least one edge. The following are true.

(i) $|G_t| \leq qn^{t-1}$ for any $2\leq t \leq  n$.

(ii) Fix integer $t$ with $2\leq t \leq  n$. Let $G_t' \subset G_t$ such that each member of $G_t'$ is a clique, that is, any two vertices are neighbors. Then $|G_t'|\leq nq^{t-1}.$

The following conclusions are true for integer $t$ with $3\leq t \leq  n$.

(iii) For $j=2, \cdots, t-1$, let $H_j$ be the subset of $(i_1, \cdots, i_t)$ from $G_t$ satisfying the following: there exists a subgraph  $S$ of $\{i_1, \cdots, i_t\}$ with $|S|=j$ and without any edge such that any vertex from $\{i_1, \cdots, i_t\}\backslash S$ has at least two neighbors in $S$. Then $|H_j| \leq (qt)^{t-j+1}n^{j-1}$.

(iv) For $j=2, \cdots, t-1$, let $H_j'$ be the subset of $(i_1, \cdots, i_t)$ from $G_t$ satisfying the following: for any subgraph  $S$ of $\{i_1, \cdots, i_t\}$ with $|S|=j$ and without any edge, we know any vertex from $\{i_1, \cdots, i_t\}\backslash S$ has at least one neighbor in $S$. Then $|H_j'|\leq (qt)^{t-j}n^j.$

\end{lemma}
\noindent\textbf{Proof of Lemma \ref{jexbcsiyr}}.
(i). Choose one vertex from $V$ and choose one of its neighbors. The total number of ways to do this is $nq$. The total number of ways to fill the rest of $t-2$ vertices arbitrarily is no more than $n^{t-2}$. Hence, $|G_t|\leq nq\cdot n^{t-2}=qn^{t-1}.$

(ii). To form a clique from $G_t$, we first choose a vertex with $n$ ways. The next vertex has to be one of its $q$ neighbors, the third vertex has to be one of the neighbors of the first two vertices at the same time. Thus the number of choices for the third vertex is no more than $q$. This has to be true for the picks of the remaining vertices to form a clique. So $|G_t'|\leq nq^{t-1}$.

(iii).  Now we figure out the ways to get $(i_1, \cdots, i_t)\in H_t$. The number of ways to get $i_1$ is at most $n$. Once $i_1$ is chosen, another vertex $b_1$ from its neighbors has to be picked to be an element in $\{i_1, \cdots, i_t\}\backslash S$. Once $b_1$ is taken, since at leat two members from $(i_1, \cdots, i_t)$ are neighbors of $b_1$, a third vertex $i_2$ has to be in $S$. Keep in mind that $i_2$ has to be a neighbor of $b_1$. Thus, the total number of ways to pick these three vertices  is at most $n\cdot q\cdot  q.$ The rest of $j-2$ vertices in $S$ have at most $n^{j-2}$ choices to satisfy the requirement; the rest $t-j-1$ vertices from $\{i_1, \cdots, i_t\}\backslash S$ have to be the neighbors of the vertices in $S$, which amounts to at most $(qt)^{t-j-1}$ ways to fill the $t-j$ vertices. So $|H_t|\leq nq^2\cdot n^{j-2}\cdot (qt)^{t-j-1} =(qt)^{t-j+1}n^{j-1}.$

(iv). The choices of $S$ with $|S|=j$ is no more than $n^j$. Any of the rest $t-j$ vertices from $(i_1, \cdots, i_t)$ must be the neighbor of a vertex from the chosen $j$ vertices. The total number of neighbors is at most $qt$. This amounts to no more than $(qt)^{t-j}$ ways to achieve this. Hence $|H_j'| \leq (qt)^{t-j}n^j$. \hfill$\square$

\begin{lemma}\lbl{winter_fire} For $p\geq 1$, let $R=R_p$ be positive integers with $\lim_{p\to\infty}R_p/p=1$. Let $Z_{p1}, \cdots, Z_{pR}$ be $N(0,1)$-distributed random variables with covariance matrix $\bd{\Sigma}=\bd{\Sigma}_p=(\sigma_{ij})_{p\times p}.$ Assume $|\sigma_{ij}|\leq \delta_p$ for all $1\leq i<j \leq R$ and $p\geq 1$, where  $\{\delta_p;\, p\geq 1\}$ are constants satisfying $0<\delta_p=o(1/\log p).$ Given $x\in \mathbb{R}$, set $z=\big(2\log p - \log \log p+x\big)^{1/2}.$ Then, for any fixed $m\geq 1$, we have
\beaa\lbl{cskh}
\Big(\frac{\sqrt{\pi}p}{e^{-x/2}}\Big)^m\cdot P(|Z_{pi_1}|>z, \cdots, |Z_{pi_m}|>z)
\to 1
\eeaa
as $p\to\infty$ uniformly for all $1\leq i_1<\cdots < i_m\leq R.$
\end{lemma}
\noindent\textbf{Proof of Lemma \ref{winter_fire}}. To ease notation, we write ``$Z_{1}, \cdots, Z_{R}$" for ``$Z_{p1}, \cdots, Z_{pR}$" if there is no danger of confusion.
First, $z$ is well-defined as $p$ is large.   Note that $(Z_1, \cdots, Z_m)^T\sim N(\bd{0}, \bd{\Sigma}_m)$, where $\bd{\Sigma}_m=(\sigma_{ij})_{1\leq i, j \leq m}$. Recall the density function of $N(\bd{0}, \bd{\Sigma}_m)$ is given by $\frac{1}{(2\pi)^{m/2}\mbox{det}(\bd{\Sigma}_m)^{1/2}} \exp\big(-\frac{1}{2}x^T\bd{\Sigma}_m^{-1}x\big)$ for all $x:=(x_1, \cdots, x_m)^T\in \mathbb{R}^m$. It follows that
\begin{align}\lbl{doune}
 & P(Z_1>z, \cdots, Z_m>z) \nonumber\\
=&\frac{1}{(2\pi)^{m/2}\mbox{det}(\bd{\Sigma}_m)^{1/2}}\int_z^{\infty}\cdots\int_{z}^{\infty} \exp\Big(-\frac{1}{2}x^T\bd{\Sigma}_m^{-1}x\Big)\,dx_1\cdots dx_m.
\end{align}
For two non-negative definite matrices $\bd{A}$ and $\bd{B}$, we write $\bd{A}\leq \bd{B}$ if $\bd{B}-\bd{A}$ is also non-negative definite. We need to understand $\bd{\Sigma}_m^{-1}$ and $\mbox{det}(\bd{\Sigma}_m)$ on the right hand side of \eqref{doune}. First we claim that
\begin{align}\lbl{random_error}
(1-2m \delta_p)\bd{I}_m \leq \bd{\Sigma}_m^{-1}\leq (1+2m \delta_p)\bd{I}_m
\end{align}
 and that
\begin{align}\lbl{many_hello}
1-m!\delta_p\leq \mbox{det}(\bd{\Sigma}_m)^{-1/2}\leq 1+m!\delta_p
\end{align}
as $p$ is sufficiently large.

In fact, by the Gershgorin disc theorem (see, e.g., \cite{horn2012matrix}), all eigenvalues of $\bd{\Sigma}_m$ have to be in the set
\beaa
\bigcup_{1\leq i \leq m}\Big(\sigma_{ii}- \sum_{j\ne i}\sigma_{ij}, \sigma_{ii}+ \sum_{j\ne i}\sigma_{ij}\Big).
\eeaa
By assumption, $\sigma_{ii}=1$  and $|\sigma_{ij}|\leq \delta_p$ for all $1\leq i< j\leq R$. Thus, all of the eigenvalues of $\bd{\Sigma}_m$ are between $1-m\delta_p$ and $1+m\delta_p$. The two bounds are positive as $p$ is sufficiently large. On the other hand, $(1-m\delta_p)^{-1}\leq 1+2m\delta_p$ and $(1+m\delta_p)^{-1}\geq 1-2m\delta_p$ as $p$ is sufficiently large. The assertion \eqref{random_error} is obtained.

Second, $\mbox{det}(\bd{\Sigma}_m)$ is the sum of $m!$ terms. The term as the product of the diagonal entries of $\bd{\Sigma}_m$ is $1$; each of the remaining $m!-1$ terms is the product of $m$ entries from which at least one is an off-diagonal entry. Therefore, $|\mbox{det}(\bd{\Sigma}_m)-1|\leq m!\delta_p.$ This implies $1-m!\delta_p\leq \mbox{det}(\bd{\Sigma}_m)\leq 1+m!\delta_p$. Trivially, $(1+u)^{-1/2}=1-\frac{u}{2}(1+O(u))$ as $u\to 0$. The statement \eqref{many_hello} is confirmed.

The claim \eqref{random_error} implies that $(1-2m\delta_p)|x|^2\leq x^T\bd{\Sigma}_m^{-1}x\leq (1+2m\delta_p)|x|^2$ for each $x\in \mathbb{R}^m.$ The tail bound of Gaussian variable gives $P(N(0, 1)\geq t)\sim \frac{1}{\sqrt{2\pi}\, t}e^{-t^2/2}$ as $t\to\infty.$ Since $P(N(0, \sigma^2)\geq t)=P(N(0, 1)\geq t/\sigma)$ for any $\sigma>0$, for any $\epsilon \in (0, 1/2)$, there exists $t_0>0$ such that
\begin{align}\lbl{quiet_morning}
(1-\epsilon) \frac{\sigma}{\sqrt{2\pi}\, t}e^{-t^2/(2\sigma^2)} \leq P(N(0, \sigma^2)\geq t)\leq (1+\epsilon) \frac{\sigma}{\sqrt{2\pi}\, t}e^{-t^2/(2\sigma^2)}
\end{align}
for all $t\geq \sigma t_0.$ Thus, from \eqref{doune}-\eqref{quiet_morning}, we upper bound the probability of interest by
\beaa
&& P(Z_1>z, \cdots, Z_m>z) \\
&\leq &\frac{1+m!\delta_p}{(2\pi)^{m/2}}\int_z^{\infty}\cdots\int_{z}^{\infty} \exp\Big[-\frac{(1-2m \delta_p)}{2}(x_1^2+\cdots+x_m^2)\Big]\,dx_1\cdots dx_m\\
&=& (1+m!\delta_p)\sigma^m\cdot P\big(N(0, \sigma^2)\geq z\big)^m\\
& \leq & (1+m!\delta_p)\sigma^{2m}(1+\epsilon)^m\cdot \Big[\frac{1}{\sqrt{2\pi}\, z}e^{-(1-2m \delta_p)z^2/2}\Big]^m
\eeaa
as $z\geq \sigma t_0$, where $\sigma:=(1-2m \delta_p)^{-1/2}.$ Now
\beaa
\frac{1}{z}e^{-(1-2m \delta_p)z^2/2}
&=&e^{m \delta_pz^2}\cdot\frac{1}{z} e^{-z^2/2}\\
& = & e^{o(1)}\cdot \frac{1+o(1)}{\sqrt{2\log p}}\cdot \exp\Big[-\frac{1}{2}(2\log p - \log \log p+x)\Big]\\
&=& [1+o(1)]\cdot \frac{e^{-x/2}}{\sqrt{2} p}
\eeaa
as $p\to\infty$ since $\delta_p=o(1/\log p)$, where the last $o(1)$ depends on $m$ and $p$.  Consequently,
\beaa
 P(Z_1>z, \cdots, Z_m>z)
\leq  \Big(\frac{e^{-x/2}}{2\sqrt{\pi}p}\Big)^mC_m
\eeaa
where
\beaa
C_m:=(1+m!\delta_p)\sigma^{2m}(1+\epsilon)^m[1+o(1)]^m\leq (1+2\epsilon)^m
\eeaa
as $p$ is sufficiently large because $\delta_p\to 0$. In summary, for fixed $m\geq 1$,
\begin{align}\lbl{hahelj}
 P(Z_1>z, \cdots, Z_m>z)
\leq (1+2\epsilon)^m \Big(\frac{e^{-x/2}}{2\sqrt{\pi}p}\Big)^m
\end{align}
as $p$ is sufficiently large. Similarly, from \eqref{doune}-\eqref{quiet_morning}, a lower bound can be established as
\beaa
&& P(Z_1>z, \cdots, Z_m>z) \\
&\geq &\frac{1-m!\delta_p}{(2\pi)^{m/2}}\int_z^{\infty}\cdots\int_{z}^{\infty} \exp\Big(-\frac{(1+2m \delta_p)}{2}(x_1^2+\cdots+x_m^2)\Big)\,dx_1\cdots dx_m\\
&=& (1-m!\delta_p)\sigma_1^m\cdot P\big(N(0, \sigma_1^2)\geq z\big)^m\\
& \geq & (1-m!\delta_p)\sigma_1^{2m}(1-\epsilon)^m\cdot \Big[\frac{1}{\sqrt{2\pi}\, z}e^{-(1+2m \delta_p)z^2/2}\Big]^m
\eeaa
where $\sigma_1:=(1+2m \delta_p)^{-1/2}.$ By taking care of each term above as in the previous arguments, we can get
a reverse inequality of \eqref{hahelj} with ``$1+2\epsilon$" replaced by ``$1-2\epsilon$". Consequently, we know that
\begin{align}\lbl{qkuxnb}
\Big(\frac{e^{-x/2}}{2\sqrt{\pi}p}\Big)^m\cdot(1-2\epsilon)^{m}\leq P(Z_1>z, \cdots, Z_m>z)
\leq  \Big(\frac{e^{-x/2}}{2\sqrt{\pi}p}\Big)^m\cdot(1+2\epsilon)^{m}
\end{align}
as $p\geq p(m, \epsilon)$, where $p(m, \epsilon)\geq 1$ is a constant that depends on $m$ and $\epsilon$ only. Now we consider the decomposition
\begin{align}\lbl{shang_tou}
P(|Z_1|>z, \cdots, |Z_m|>z)=\sum P(\eta_1 Z_1>z, \cdots, \eta_m Z_m>z),
\end{align}
where the summation is over all the $2^m$ possible cases with $\eta_1=\pm 1, \cdots, \eta_m =\pm 1.$ Notice that  $|\mbox{Cov}(\eta_i Z_i, \eta_j Z_j)|=|\mbox{Cov}(Z_i, Z_j)|\leq \delta_p$, and the derivation of \eqref{qkuxnb} depends on $\delta_p$ rather than the exact values of  $\sigma_{ij}$'s. By \eqref{shang_tou}, we have
\beaa
\Big(\frac{e^{-x/2}}{\sqrt{\pi}p}\Big)^m\cdot(1-2\epsilon)^{m}\leq P(|Z_1|>z, \cdots, |Z_m|>z)
\leq  \Big(\frac{e^{-x/2}}{\sqrt{\pi}p}\Big)^m\cdot(1+2\epsilon)^{m}
\eeaa
as $p\geq p(m, \epsilon)$. Based on the same reasoning, for any  $\epsilon \in (0, 1/2)$, we have
\beaa\lbl{liquor}
\Big(\frac{e^{-x/2}}{\sqrt{\pi}p}\Big)^m\cdot(1-2\epsilon)^{m}\leq
P(|Z_{i_1}|>z, \cdots, |Z_{i_m}|>z)
\leq \Big(\frac{e^{-x/2}}{\sqrt{\pi}p}\Big)^m\cdot(1+2\epsilon)^{m}
\eeaa
as $p\geq p(m, \epsilon)$ uniformly for all $1\leq i_1<\cdots < i_m\leq R.$ Because the probability above does is independent of $\epsilon$, by letting $p \to \infty$ and $\epsilon\downarrow 0$, we have that
\beaa
\Big(\frac{\sqrt{\pi}p}{e^{-x/2}}\Big)^m\cdot P(|Z_{i_1}|>z, \cdots, |Z_{i_m}|>z)
\to 1
\eeaa
as $p\to\infty$ uniformly for all $1\leq i_1<\cdots < i_m\leq R$, which completes the proof.  \hfill$\square$

\medskip

For any $m\times m$ symmetric matrix $\bd{M}$, we use the notation $\|\bd{M}\|$ to denote its spectral norm, that is, $\|\bd{M}\|=\max\{\lambda_{max}(\bd{M}), -\lambda_{min}(\bd{M})\}$. Evidently, $\|\bd{M}\|^2\leq \mbox{ tr} (\bd{M}^T\bd{M})$. Also, $\|\bd{M}\|=\lambda_{max}(\bd{M})$ if $\bd{M}$ is non-negative definite. Obviously, for any symmetric matrix $\bd{M}$, if $\|\bd{M}\| \leq a$ then $x^T\bd{M}x\leq \lambda_{max}(\bd{M})x^Tx\leq ax^Tx$ for any $x\in \mathbb{R}^m$. The same argument applies to $-\bd{M}$. Thus, by symmetry we know that
\begin{align}\lbl{now_chat}
-ax^Tx \leq x^T\bd{M}x \leq ax^Tx
\end{align}
for any $x \in \mathbb{R}^m.$ The following lemma would be further needed.

\begin{lemma}\lbl{winter_cool}  Let $m\geq 2$ and $Z_1, \cdots, Z_m$ be $N(0,1)$-distributed random variables with positive definite covariance matrix $\bd{\Sigma}=(\sigma_{ij})_{m\times m}.$ For some $1>\varrho>\delta>0$, assume $|\sigma_{12}|\leq  \varrho$, and  $|\sigma_{ij}|\leq \delta$ for all $1\leq i<j \leq m$ but $(i, j)\ne (1, 2)$. Then, if $\delta \leq \frac{1}{8m^2}(1-\varrho)^3$, we have
\begin{align}\lbl{XGDD_street}
P(|Z_1|>z, \cdots, |Z_m|>z) \leq \frac{2^m}{z}\cdot  e^{-\alpha z^2/2}
\end{align}
for all $z>0$, where $\alpha=m-\frac{1}{4}(\varrho+3).$
%Moreover, the above inequality holds for $m=2$ and for any $\delta\in (0, 1)$.
\end{lemma}
\noindent\textbf{Proof of Lemma \ref{winter_cool}}. Let $\bd{1}=(1, \cdots, 1)^T\in \mathbb{R}^m$ and
\begin{align}\lbl{encouraging}
\bd{a}=(a_1, \cdots, a_m)^T=\frac{\bd{\Sigma}^{-1}\bd{1}}{\bd{1}^T\bd{\Sigma}^{-1}\bd{1}}.
%\in (0, \infty)^m.
\end{align}
Then $\bd{1}^T\bd{a}=1$. We claim that (which will be prove later)
\begin{align}\lbl{Hydrogen}
a_i\geq 0
\end{align}
for all $i=1,2,\cdots, m.$
Assuming this is true, then obviously,
\beaa
P(Z_1>z, \cdots, Z_m>z)\leq P((Z_1, \cdots, Z_m)\bd{a}\geq z).
\eeaa
Let $Y=(Z_1, \cdots, Z_m)\bd{a}$. Then $Y \sim N(0, \bd{a}^T\bd{\Sigma}\bd{a})$. By \eqref{encouraging}, $\bd{a}^T\bd{\Sigma}\bd{a}=(\bd{1}^T\bd{\Sigma}^{-1}\bd{1})^{-1}.$ Therefore,
\begin{align}\lbl{tusi}
P(Z_1>z, \cdots, Z_m>z) \leq & P\big(N(0, 1)\geq (\bd{1}^T\bd{\Sigma}^{-1}\bd{1})^{1/2}z\big) \nonumber\\
 \leq & \frac{1}{\sqrt{2\pi}}\cdot \frac{1}{(\bd{1}^T\bd{\Sigma}^{-1}\bd{1})^{1/2}z}\cdot e^{-(\bd{1}^T\bd{\Sigma}^{-1}\bd{1}) z^2/2}
\end{align}
for any $z>0$, where in the last step we use a well-known inequality of the Gaussian tail: $P(N(0, 1)\geq y)\leq \frac{1}{\sqrt{2\pi}\,y}e^{-y^2/2}$ for any $y>0$. Firstly, if $m=2$, then
\beaa
\bd{\Sigma}^{-1}=\frac{1}{1-\sigma_{12}^2}
\begin{pmatrix}
1 & -\sigma_{12}\\
-\sigma_{12} & 1
\end{pmatrix}
.
\eeaa
It is easy to check that $\bd{1}^T\bd{\Sigma}^{-1}\bd{1}=\frac{2}{1+\sigma_{12}}\geq \frac{2}{1+\varrho}\geq 2-\frac{1}{4}(\varrho+3)=\alpha.$ Notice $\alpha >1$. We have from \eqref{tusi} that
\begin{align}\lbl{Contract_holder}
P(Z_1>z,  Z_2>z) \leq \frac{1}{z}\cdot e^{-\alpha z^2/2}.
\end{align}
So the conclusion holds for $m=2$. From now on, we assume $m\geq 3$.

{\it Step 1: the proof of \eqref{Hydrogen}}. Define
\beaa
\bd{\Sigma}_{2}
=
\begin{pmatrix}
1& \sigma_{12}\\
\sigma_{12
} & 1
\end{pmatrix}
\ \ \mbox{and}\ \ \ \bd{A}=
\begin{pmatrix}
\bd{\Sigma}_{2}& \bd{0}\\
\bd{0}^T &  \bd{I}_{m-2}
\end{pmatrix}
\eeaa
where $\bd{0}$ is a $2\times (m-2)$ matrix whose entries are all equal to zero.
 Trivially, the eigenvalues of $\bd{\Sigma}_{2}$ are $1+\sigma_{12}$ and $1-\sigma_{12}$, respectively. Basic algebra gives
\begin{align}\lbl{too_no}
\bd{\Sigma}_{2}^{-1}=\frac{1}{1-\sigma_{12}^2}
\begin{pmatrix}
1 & -\sigma_{12}\\
-\sigma_{12} & 1
\end{pmatrix}
\ \ \mbox{and}\ \
\bd{A}^{-1}=
\begin{pmatrix}
\bd{\Sigma}_{2}^{-1} & \bd{0}\\
\bd{0} & \bd{I}_{m-2}
\end{pmatrix}
,
\end{align}
and the eigenvalues of $\bd{A}^{-1}$ are $1$ with $m-2$ folds, $\frac{1}{1+\sigma_{12}}$ and $\frac{1}{1-\sigma_{12}}$, respectively, which by assumption bounds the spectral norm as $\|\bd{A}^{-1}\|\leq \frac{1}{1-\varrho}.$ Also, $\|\bd{A}-\bd{\Sigma}\|^2 \leq \mbox{tr}[(\bd{A}-\bd{\Sigma})^T(\bd{A}-\bd{\Sigma})]\leq  m^2\delta^2$ since $|\sigma_{ij}|\leq \delta$ for all $1\leq i<j \leq m$ but $(i, j)\ne (1, 2)$. By the fact that $\bd{\Sigma}^{-1}-\bd{A}^{-1}=\bd{A}^{-1}(\bd{A}-\bd{\Sigma})\bd{\Sigma}^{-1}$, we obtain
\begin{align}\lbl{hurry}
\|\bd{\Sigma}^{-1}-\bd{A}^{-1}\| \leq & \|\bd{A}^{-1}\|\cdot \|\bd{A}-\bd{\Sigma}\|\cdot\|\bd{\Sigma}^{-1}\|
 \leq  \frac{m\delta}{1-\varrho}\cdot \|\bd{\Sigma}^{-1}\|.
\end{align}
In particular, this implies from the triangle inequality that
\beaa
\|\bd{\Sigma}^{-1}\| \leq \frac{1}{1-\varrho} + \frac{m\delta}{1-\varrho}\cdot \|\bd{\Sigma}^{-1}\|.
\eeaa
By assumption $\delta \leq \frac{1}{8m^2}(1-\varrho)^3$, we know $m\delta+\varrho<(1-\varrho) +\varrho=1$. By solving the inequality we obtain $\|\bd{\Sigma}^{-1}\| \leq (1-\varrho-m\delta)^{-1}\leq \frac{2}{1-\varrho}$. Substituting this to \eqref{hurry} we get
\begin{align}\lbl{dragon}
\|\bd{\Sigma}^{-1}-\bd{A}^{-1}\|  \leq  \frac{2m\delta}{(1-\varrho)^2}.
\end{align}
From \eqref{encouraging}, we know $a_i=\bd{e}_i^T\bd{\Sigma}^{-1}\bd{1}/\bd{1}^T\bd{\Sigma}^{-1}\bd{1}$ for $1\leq i \leq m$, where $\bd{e}_i=(0, \cdots, 1, \cdots,0)^T$ and the only ``$1$" appears in the $i$th position. As $\bd{\Sigma}$ is positive definite, $\bd{1}^T\bd{\Sigma}^{-1}\bd{1}>0.$ To show \eqref{Hydrogen}, it is enough to show $\tilde{a}_i:=\bd{e}_i^T\bd{\Sigma}^{-1}\bd{1}>0$ for each $i$.  Define $h_i=\bd{e}_i^T\bd{A}^{-1}\bd{1}$, which equals the $i$th row sum of $\bd{A}^{-1}$. We have
\begin{align}\lbl{cwkuvboit}
|\tilde{a}_i-h_i|=|\bd{e}_i^T(\bd{\Sigma}^{-1}-\bd{A}^{-1})\bd{1}|
\leq & \|\bd{e}_i\|\cdot \|(\bd{\Sigma}^{-1}-\bd{A}^{-1})\bd{1}\| \nonumber\\
 \leq & \|\bd{e}_i\|\cdot \|\bd{\Sigma}^{-1}-\bd{A}^{-1}\|\cdot \|\bd{1}\| \nonumber\\
 \leq & \frac{2m^2\delta}{(1-\varrho)^2}
\end{align}
induced by \eqref{dragon}. Therefore, $\tilde{a}_i\geq h_i-\frac{2m^2\delta}{(1-\varrho)^2}$. Now observe from \eqref{too_no} that
\beaa
h_1=h_2=\frac{1-\sigma_{12}}{1-\sigma_{12}^2}=\frac{1}{1+\sigma_{12}}\geq \frac{1}{1+\varrho},
\eeaa
and $h_i=1$ for $i=3, \cdots, m.$ Thus, \eqref{cwkuvboit} and condition $\delta \leq \frac{1}{8m^2}(1-\varrho)^3\leq \frac{1}{2m^2}\frac{(1-\varrho)^3}{1+\varrho}$ conclude that $\tilde{a}_i>0$ for each $i$, which implies \eqref{Hydrogen}.

{\it Step 2: the proof of \eqref{XGDD_street}}. From \eqref{now_chat} and \eqref{dragon},
\beaa
\bd{1}^T(\bd{\Sigma}^{-1}-\bd{A}^{-1})\bd{1} \geq -\frac{(2m\delta)\bd{1}^T\bd{1}}{(1-\varrho)^2}=-\frac{2m^2\delta}{(1-\varrho)^2}.
\eeaa
As a result, we have
\beaa
\bd{1}^T\bd{\Sigma}^{-1}\bd{1}
& \geq & \bd{1}^T\bd{A}^{-1}\bd{1}-\frac{2m^2\delta}{(1-\varrho^2)^2}\\
&=& \frac{2-2\sigma_{12}}{1-\sigma_{12}^2}+m-2-\frac{2m^2\delta}{(1-\varrho)^2}\\
& \geq & m-1+\frac{1-\varrho}{1+\varrho}-\frac{2m^2\delta}{(1-\varrho)^2}
\eeaa
by using the assumption $|\sigma_{12}|\leq \varrho$. By assumption $\delta \leq \frac{1}{8m^2}(1-\varrho)^3$, we further obtain
\beaa
\bd{1}^T\bd{\Sigma}^{-1}\bd{1}
\geq  m-1+\frac{1-\varrho}{2(1+\varrho)}.
\eeaa
In particular, $\bd{1}^T\bd{\Sigma}^{-1}\bd{1} \geq m-1\geq \frac{1}{4}m$ for $m\geq 3.$
Then, we can establish from \eqref{tusi} that
\begin{align}\lbl{one_duan}
P(Z_1>z, \cdots, Z_m>z) \leq
\frac{1}{z\sqrt{m}}\exp\Big\{-\frac{z^2}{2}\Big(m-1+\frac{1-\varrho}{2(1+\varrho)}\Big)\Big\},
\end{align}
under the assumption $\delta \leq \frac{1}{8m^2}(1-\varrho)^3$. In the above derivation, the true values of $\sigma_{ij}$'s are not used, instead their bounds $\varrho$ and $\delta$ are relevant. Therefore, \eqref{Contract_holder} and \eqref{one_duan} still hold if each ``$Z_i$" is replaced by ``$\eta_iZ_i$" with $\eta_i=\pm 1$. Trivially,  $-1+\frac{1-\varrho}{2(1+\varrho)}\geq -\frac{1}{4}(\varrho+3)$ for each $\varrho \in (0, 1)$. This combining with \eqref{one_duan} and \eqref{shang_tou} yields \eqref{XGDD_street}. \hfill$\square$

\medskip

To prove Theorem \ref{theorem_2}, we need a notation. Let $p\geq 2$ and $(\sigma_{ij})_{p\times p}$ be a non-negative definite matrix. For $\delta>0$ and a set $A\subset \{1,2,\cdots, m\}$ with $2\leq m\leq p$, define
\begin{align}\lbl{new_def}
\wp(A)=\max\Big\{|S|;\, S \subset A \ \mbox{and}\ \max_{i\in S, j\in S, i\ne j}|\sigma_{ij}|\leq \delta\Big\}.
\end{align}
Specifically, $\wp(A)$ takes possible values $0, 2 \cdots, |A|$, where we regard $|\varnothing|=0$. If $\wp(A)=0$, then $|\sigma_{ij}|> \delta$ for all $i\in A$ and $j\in A.$

\medskip

\noindent\textbf{Proof of Theorem \ref{theorem_2}}. For any $x\in \mathbb{R}$, write
\begin{align}\lbl{Chicago_Seattle}
z=\big(2\log p - \log \log p+x\big)^{1/2},
\end{align}
which is well defined as $p$ is sufficiently large. We will not mention this matter again since the conclusion is valid as $p\to\infty.$ It suffices to show
\begin{align}\lbl{asjasft}
\lim_{p\to \infty}P\Big(\max_{1\leq i \leq p}|Z_i|>z\Big)= 1-\exp\Big(-\frac{1}{\sqrt{\pi}}e^{-x/2}\Big)
\end{align}
as $p\to \infty.$ The proof will be divided into a few steps.

{\it Step 1: reducing ``$\{1\leq i \leq p\}$" in \eqref{asjasft} to a set of  friendly indices}.  First,
\begin{align}\lbl{money_road}
 P\big(|N(0, 1)|\geq z\big)
 \sim & \frac{2}{\sqrt{2\pi} z}e^{-z^2/2} \nonumber\\
 \sim & \frac{1}{\sqrt{\pi}}\cdot \frac{1}{\sqrt{\log p}}\exp\Big\{-\frac{1}{2}\big(2\log p - \log \log p+x\big)\Big\}  \nonumber\\
 \sim & \frac{1}{\sqrt{\pi}}\cdot \frac{e^{-x/2}}{p}
\end{align}
as $p\to\infty$, where for two sequences of real numbers $A_p$ and $H_p$, the notion $A_p \sim H_p$ means that $A_p/H_p\to 1$ as $p\to\infty.$ Immediately, a union bound implies
\beaa
P\Big(\max_{i\in C_p}|Z_i|>z\Big)
\leq  |C_p|\cdot P\big(|N(0, 1)|\geq z\big)\to 0,
\eeaa
as $p\to \infty$, where we recall the definition $C_p:=\{1\leq i \leq p;\, |B_{p,i}|\geq p^{\kappa}\}$ with $B_{p,i}=\big\{1\leq j \leq p;\, |\sigma_{ij}|\geq \delta_p\big\}$. Further denote $D_p:=\{1\leq i \leq p;\, |B_{p,i}|< p^{\kappa}\}$. By assumption, $|D_p|/p\to 1$ as $p\to\infty$. It follows that
\beaa
P\Big(\max_{i\in D_p}|Z_i|>z\Big)
&\leq & P\Big(\max_{1\leq i \leq p}|Z_i|>z\Big)\\
&\leq & P\Big(\max_{i\in D_p}|Z_i|>z\Big) + P\Big(\max_{i\in C_p}|Z_i|>z\Big).
\eeaa
Therefore, to prove \eqref{asjasft}, it is enough to show
\begin{align}\lbl{yellow_ear}
\lim_{p\to \infty}P\Big(\max_{i\in D_p}|Z_i|>z\Big)= 1- \exp\Big(-\frac{1}{\sqrt{\pi}}e^{-x/2}\Big),
\end{align}
as $p\to \infty$ asymptotically.

{\it Step 2: estimation of $P(\max_{i\in D_p}|Z_i|>z)$ via the inclusion-exclusion formula}. Set
\begin{align}\lbl{ksahc}
\alpha_t=\sum P(|Z_{i_1}|>z, \cdots, |Z_{i_t}|>z)
\end{align}
for $1\leq t \leq p$, where the sum runs over all $i_1\in D_p, \cdots, i_t\in D_p$ such that $i_1<\cdots < i_t$. Then,
\begin{align}\lbl{sing_you}
\sum_{t=1}^{2k}(-1)^{t-1}\alpha_t \leq P\Big(\max_{i\in D_p}|Z_i|>z\Big) \leq \sum_{t=1}^{2k+1}(-1)^{t-1}\alpha_t
\end{align}
for any $k\geq 1.$ We will prove next that
\begin{align}\lbl{di_da}
\lim_{p\to\infty}\alpha_t=\frac{1}{t!}\pi^{-t/2}e^{-tx/2}
\end{align}
for each $t\geq 1.$ Assuming this is true, let $p\to \infty$ in \eqref{sing_you}, we have
\begin{align*}
\sum_{t=1}^{2k}(-1)^{t-1}\frac{1}{t!}\Big(\frac{1}{\sqrt{\pi}}e^{-x/2}\Big)^t  \leq \liminf_{p\to\infty}P\Big(\max_{i\in D_p}|Z_i|>z\Big)\\
 \leq  \limsup_{p\to\infty}P\Big(\max_{i\in D_p}|Z_i|>z\Big)
&\leq & \sum_{t=1}^{2k+1}(-1)^{t-1}\frac{1}{t!}\Big(\frac{1}{\sqrt{\pi}}e^{-x/2}\Big)^t
\end{align*}
for each $k\geq 1$. By letting $k\to \infty$ and using the Taylor expansion of the function  $f(x)=1-e^{-x}$, we obtain \eqref{yellow_ear}. It remains to verify \eqref{di_da}. Evidently, by \eqref{money_road} and the assumption $|D_p|/p\to 1$, we immediately see \eqref{di_da} holds as $t=1.$ Now we prove \eqref{di_da} for any $t\geq 2$.

Recalling $D_p:=\{1\leq i \leq p;\, |B_{p,i}|< p^{\kappa}\}$, we write
\begin{align*}
\big\{(i_1, \cdots, i_t)\in (D_p)^t;\, i_1<\cdots < i_t\big\}=F_t\cup G_t,
\end{align*}
where
\begin{align}
F_t:=&\big\{(i_1, \cdots, i_t)\in (D_p)^t;\, i_1<\cdots < i_t\ \mbox{and}\ |\sigma_{i_r i_s}|\leq \delta_p\ \mbox{for all}\ 1\leq r< s\leq t\}; \nonumber\\
 G_t:=&\big\{(i_1, \cdots, i_t)\in (D_p)^t;\, i_1<\cdots < i_t\ \mbox{and}\ |\sigma_{i_r i_s}|> \delta_p\ \mbox{ for a pair}\ (r, s)\ \mbox{with}\ 1\leq r< s\leq t\big\}. \lbl{urine_problem}
\end{align}
Now, think $D_p$ as graph with $|D_p|$ vertices, with $|D_p|\leq p$ and $|D_p|/ p\to 1$ by assumption.  Any two different vertices from $D_p$, say, $i$ and $j$  are connected if $|\sigma_{i j}|> \delta_p.$ In this case we also say there is an edge between them. By the definition $D_p$, each vertex in the graph has at most $p^{\kappa}$ neighbors. Replacing ``$n$", ``$q$" and ``$t$" in Lemma \ref{jexbcsiyr}(i) with ``$|D_p|$", ``$p^{\kappa}$" and ``$t$", respectively, we have that $|G_t|\leq p^{t+\kappa-1}$ for each $2\leq t\leq p$. Therefore, $\binom{|D_p|}{t}\geq |F_t|\geq \binom{|D_p|}{t}-p^{t+\kappa-1}$. Since $D_p/p\to 1$  and $\kappa=\kappa_p\to 0$ as $p\to\infty$, we know
\begin{align}\lbl{hong_huo}
\lim_{p\to\infty}\frac{|F_t|}{p^t}=\frac{1}{t!}.
\end{align}
Decomposing \eqref{ksahc}, we see that
\begin{align*}
\alpha_t=\sum_{(i_1, \cdots, i_t)\in F_t} P(|Z_{i_1}|>z, \cdots, |Z_{i_t}|>z)+\sum_{(i_1, \cdots, i_t)\in G_t} P(|Z_{i_1}|>z, \cdots, |Z_{i_t}|>z).
\end{align*}
From Lemma \ref{winter_fire} and \eqref{hong_huo} we have
\begin{align*}
\sum_{(i_1, \cdots, i_t)\in F_t} P(|Z_{i_1}|>z, \cdots, |Z_{i_t}|>z)\to \frac{1}{t!} \Big(\frac{e^{-x/2}}{\sqrt{\pi}}\Big)^t=\frac{1}{t!}\pi^{-t/2}e^{-tx/2},
\end{align*}
as $p\to\infty.$ As a consequence, to derive \eqref{di_da}, we only need to show that
\begin{align}\lbl{wkcrby}
\sum_{(i_1, \cdots, i_t)\in G_t} P(|Z_{i_1}|>z, \cdots, |Z_{i_t}|>z) \to 0
\end{align}
as $p\to \infty$ asymptotically for each $t\geq 2.$

{\it Step 3: the proof of \eqref{wkcrby}}. If $t=2$, the sum of probabilities in \eqref{wkcrby} is bounded by
$|G_2|\cdot \max_{1\leq i<j\leq p}P(|Z_i|>z, |Z_j|>z )$. By Lemma \ref{jexbcsiyr}(i), $|G_2|\leq p^{\kappa+1}$. Since $|\sigma_{ij}|\leq \varrho$, by Lemma \ref{winter_cool}, \begin{align}\lbl{country_love}
P(|Z_i|>z, |Z_j|>z ) \leq \exp\big[-(5-\varrho)z^2/8\big]\leq \frac{(\log p)^C}{p^{(5-\varrho)/4}}
\end{align}
uniformly for all $1\leq i < j \leq p$ as $p$ is sufficiently large, where $C>0$ is a constant not depending on $p.$ We then know \eqref{wkcrby} holds. The remaining job is to prove  \eqref{wkcrby} for $t\geq 3.$

Take $\delta=\delta_p$ in \eqref{new_def} for the definition of $\wp(A)$ and compare it with $G_t$ from \eqref{urine_problem}. To proceed, we further classify $G_t$ into the following subsets
\begin{align*}
G_{t,j}=\big\{(i_1, \cdots, i_t)\in G_t;\, \wp(\{i_1, \cdots, i_t\})=j\big\},
\end{align*}
for $j=0, 2, \cdots, t-1.$ By the definition of $G_t$, we know that $G_t=\cup_{j=0, 2, \cdots, t-1} G_{t,j}$.  Since $t\geq 3$ is fixed, to show \eqref{wkcrby}, it suffices to prove
\begin{align}\lbl{what_tie}
\sum_{(i_1, \cdots, i_t)\in G_{t,j}} P(|Z_{i_1}|>z, \cdots, |Z_{i_t}|>z)  \to 0
\end{align}
for all $j\in\{0, 2, \cdots, t-1\}$.

Assume $(i_1, \cdots, i_t)\in G_{t,0}$, which implies $|\sigma_{i_r i_s}|> \delta_p$ for all  $1\leq r<s\leq t$ by \eqref{new_def}. Hence, the subgraph $\{i_1, \cdots, i_t\}\subset  G_t$ is a clique. Taking $n=|D_p|\leq p$, $t=t$ and $q=p^{\kappa}$ into Lemma \eqref{jexbcsiyr}(ii), we get $|G_{t,0}| \leq p^{1+\kappa(t-1)}\leq p^{1+t\kappa}$.  Thus,  we obtain
\begin{align}\lbl{Huang}
\eqref{what_tie}\leq p^{1+t\kappa}\cdot \max_{1\leq i< j\leq p}P(|Z_i|>z, |Z_j|>z)\leq p^{1+t\kappa}\cdot\frac{(\log p)^C}{p^{(5-\varrho)/4}}\to 0
\end{align}
as $p\to\infty$ by using \eqref{country_love}. So, \eqref{what_tie} holds with $t=0.$

Now we assume $(i_1, \cdots, i_t)\in G_{t,j}$ with $j\in \{2, \cdots, t-1\}$. By definition, there exits $S\subset \{i_1, \cdots, i_t\}$ such that $\max_{ i\in S, j\in S, i\ne j}|\sigma_{ij}|\leq \delta_p$ and for each $k\in \{i_1, \cdots, i_t\}\backslash S$, there exists $i\in S$ satisfying
$|\sigma_{ik}| > \delta_p$. Looking at the last statement we see two possibilities: (i) for each $k\in \{i_1, \cdots, i_t\}\backslash S$, there exist at leat two indices, say, $i\in S$, $j\in S$ with $i \ne j$ satisfying
$|\sigma_{ik}| > \delta_p$ and $|\sigma_{jk}| > \delta_p$; (ii) there exists $k\in \{i_1, \cdots, i_t\}\backslash S$ for which $|\sigma_{ik}| > \delta_p$ for an unique $i\in S$. However, for $(i_1, \cdots, i_t)\in G_{t,j}$, (i) and (ii) could happen at the same time for different $S$, say, (i) holds for $S_1$ and  (ii) holds for $S_2$ simultaneously. Thus, to differentiate the two cases, we consider the following two types of sets. Denote
\begin{align}
H_{t,j}=&\big\{(i_1, \cdots, i_t)\in G_{t,j};\, \ \mbox{there exist}\ S \subset \{i_1, \cdots, i_t\}\ \mbox{with}\ |S|=j\ \mbox{and} \nonumber\\
&\max_{ i\in S, j\in S, i\ne j}|\sigma_{ij}|\leq \delta_p\ \mbox{such that}
\ \mbox{for any}\ k \in \{i_1, \cdots, i_t\}\backslash S\ \mbox{there exist}\ r\in S, \nonumber\\
& s\in S, r \ne s\ \mbox{satisfying}\
  \min\{|\sigma_{kr}|,  |\sigma_{ks}|\}>\delta_p\big\}. \lbl{sdrtn}
\end{align}
Replacing ``$n$", ``$q$" and ``$t$" in Lemma \ref{jexbcsiyr}(iii) with ``$|D_p|$", ``$p^{\kappa}$" and ``$t$", respectively, we have that $|H_{t,j}|\leq t^t\cdot p^{j-1+ (t-j+1)\kappa}$ for each $t\geq 3$.
Analogously, set
\begin{align}\lbl{Take_ball}
H_{t,j}'=&\big\{(i_1, \cdots, i_t)\in G_{t,j};\, \ \mbox{for any}\ S \subset \{i_1, \cdots, i_t\}\ \mbox{with}\ |S|=j\ \mbox{and} \nonumber\\
&\max_{i\in S, j\in S, i\ne j}|\sigma_{ij}|\leq \delta_p\
\mbox{there exists}\ k \in \{i_1, \cdots, i_t\}\backslash S\ \mbox{such that}\ |\sigma_{kr}|>\delta_p\nonumber\\
& \mbox{for a unique}\ r \in S \}.
\end{align}
From Lemma \ref{jexbcsiyr}(iv) we see $|H_{t,j}'| \leq t^t\cdot p^{j+(t-j)\kappa}.$ It is easy to see $G_{t,j}=H_{t,j}\cup H_{t,j}'$. Therefore, to show \eqref{what_tie}, we only need to prove both
\begin{align}\lbl{what_tie_1}
\sum_{(i_1, \cdots, i_t)\in H_{t,j}} P(|Z_{i_1}|>z, \cdots, |Z_{i_t}|>z) \to 0
\end{align}
 and
\begin{align}\lbl{what_tie_2}
\sum_{(i_1, \cdots, i_t)\in H_{t,j}'} P(|Z_{i_1}|>z, \cdots, |Z_{i_t}|>z) \to 0
\end{align}
as $p\to\infty$ for $j=2, \cdots, t-1$. In fact, let $S$ be as in \eqref{sdrtn}, then using Lemma \ref{winter_fire}, the probability in \eqref{what_tie_1} is bounded by $P(\bigcap_{l\in S}\{|Z_{l}|>z\})\leq C\cdot p^{-j}$ uniformly for all $S$ as $p$ is sufficiently large, where $C$ is a constant independent of $p$. This leads to
\begin{align*}
\sum_{(i_1, \cdots, i_t)\in H_{t,j}} P(|Z_{i_1}|>z, \cdots, |Z_{i_t}|>z)
\leq & t^t\cdot p^{j-1+ (t-j+1)\kappa}\cdot \big(C\cdot p^{-j}\big)
 \leq  (Ct^t)\cdot p^{-1+t\kappa},
\end{align*}
as $p$ is sufficiently large. By the assumption that $\kappa=\kappa_p\to 0$, we arrive at \eqref{what_tie_1}.

Finally we validate \eqref{what_tie_2}. Recall the definition of $H_{t,j}'$. For  $(i_1, \cdots, i_t)\in H_{t,j}'$, pick $S \subset \{i_1, \cdots, i_t\}$ with $|S|=j$, $\max_{i\in S, j\in S, i\ne j}|\sigma_{ij}|\leq \delta_p$ and $k \in \{i_1, \cdots, i_t\}\backslash S$ such that $\delta_p<|\sigma_{kr}|\leq \varrho$ for a unique $r \in S.$  Then each probability in \eqref{what_tie_2} is bounded by
\begin{align*}
P\Big(|Z_{k}|>z, \bigcap_{l\in S}\{|Z_{l}|>z\}\Big),
\end{align*}
for $2\leq j \leq t-1$.  Taking $m=j+1$ and applying Lemma \ref{winter_cool}, the probability above is dominated by
\begin{align*}
\frac{2^{j+1}}{z}\cdot  \exp\Big\{-\frac{z^2}{2}\Big(j+\frac{1-\varrho}{4}\Big) \Big\}=O\Big(\frac{(\log p)^c}{p^{j+(1-\varrho)/4}}\Big),
\end{align*}
for some constant $c$. As stated earlier, $|H_{t,j}'| \leq t^t\cdot p^{j+(t-j)\kappa}.$   by union bound, since $\kappa=\kappa_p\to 0$, we see the sum from \eqref{what_tie_2} is of order $O(p^{-(1-\varrho)/8})$. Hence, \eqref{what_tie_2} holds. We have proved \eqref{what_tie} for any $j\in\{0, 2, \cdots, t-1\}$, which concludes the proof. \hfill$\square$

\subsection{Proof of Theorem \ref{theorem_3}}
The proof of Theorem \ref{theorem_3} is involved. A preparation with a few of lemmas is given below.
\begin{lemma}\lbl{Gu_xiao} Let $\bd{A}$ and $\bd{B}$ be nonnegative definite matrices. Then
\begin{align*}
\mbox{tr}\,(\bd{A}\bd{B})\leq \lambda_{max}(\bd{A})\,\mbox{tr}(\bd{B}).
\end{align*}
\end{lemma}
\noindent\textbf{Proof of Lemma \ref{Gu_xiao}}. Assume $\bd{A}$ and $\bd{B}$ are  $n\times n$ matrices. There is an orthogonal matrix $\bd{O}$ such that $\bd{A}=\bd{O}^T\bd{\Lambda}\bd{O}$, where $\bd{\Lambda}=\mbox{diag}(\lambda_1, \cdots, \lambda_n)$ and $\lambda_1\geq \cdots \geq \lambda_n\geq 0.$ Observe $\mbox{tr}\,(\bd{A}\bd{B})=\mbox{tr}\,[\bd{\Lambda}(\bd{O}\bd{B}\bd{O}^T)]$,  $\mbox{tr}\,(\bd{O}\bd{B}\bd{O}^T)=\mbox{tr}(\bd{B})$ and $\lambda_{max}(\bd{A})=\lambda_{max}(\bd{\Lambda})$. Thus, without loss of generality, we assume $\bd{A}=\bd{\Lambda}$. Write $\bd{B}=(b_{ij})$. Then, $b_{ii}\geq 0$ for each $i$ and
\begin{align*}
\mbox{tr}\,(\bd{A}\bd{B})=\sum_{i=1}^n \lambda_ib_{ii}\leq \lambda_1\cdot \sum_{i=1}^n b_{ii}
=\lambda_{max}(\bd{A})\,\mbox{tr}(\bd{B}).
\end{align*}
The proof is completed. \hfill$\square$.

The following is a well-known formula for the conditional distributions of multivariate normal distributions; see, for example, p. 12 from \cite{muirhead1982aspects}.

\begin{lemma}\lbl{sunny_coffee}
%Let $m>k\geq 1$ be integers.
Let $\bd{X} \sim N(\bm{\mu}, \bd{\Sigma})$ with $\bd{\Sigma}$ being invertible. Partition $\bd{X}, \bm{\mu}$ and $\bd{\Sigma}$ as
\begin{align}\lbl{similar_beer}
\bd{X}=
\begin{pmatrix}
\bd{X}_1\\
\bd{X}_2
\end{pmatrix}
,\ \ \ \
\bd{\bmu}=
\begin{pmatrix}
\bd{\bmu}_1\\
\bd{\bmu}_2
\end{pmatrix}
,\ \ \ \
\bd{\Sigma}=
\begin{pmatrix}
\bd{\Sigma}_{11} & \bd{\Sigma}_{12}\\
\bd{\Sigma}_{21} & \bd{\Sigma}_{22}
\end{pmatrix}
\end{align}
where $\bd{X}_2 \sim N(\bd{\bmu}_2, \bd{\Sigma}_{22})$. Set $\bd{\Sigma}_{22\cdot 1}=\bd{\Sigma}_{22}-\bd{\Sigma}_{21}\bd{\Sigma}_{11}^{-1}\bd{\Sigma}_{12}$. Then
$\bd{X}_2-\bd{\Sigma}_{21}\bd{\Sigma}_{11}^{-1}\bd{X}_1 \sim N(\bd{\bmu}_2-\bd{\Sigma}_{21}\bd{\Sigma}_{11}^{-1}\bd{\bmu}_1, \bd{\Sigma}_{22\cdot 1})$ and is independent of $\bd{X}_1$.
\end{lemma}

\begin{lemma}\lbl{Weyl} Let  $(Z_1, \cdots, Z_p)^T\sim N(\bd{0}, \bd{\Sigma})$. Under the notion of Lemma \ref{sunny_coffee}, for  $1\leq d<p$, write $\X_1=(Z_1, \cdots, Z_d)^T$ and $\X_2=(Z_{d+1}, \cdots, Z_p)^T$. Define  $\bd{U}_p=\X_2-\bms_{21}\bms_{11}^{-1}\X_1$ and  $\bd{V}_p=\bms_{21}\bms_{11}^{-1}\X_1$. Assume $\bd{\Sigma}=(\sigma_{ij})_{p\times p}$ with $\sum_{j=1}^p\sigma_{ij}^2\leq K_p$ for each $1\leq i \leq p$, where $K_p$ is a constant depending on $p$ only. Then there exists a constant $C>0$ free of $d, p$, $\bd{\Sigma}$ and $K_p$, such that the following holds:\\
(i) $Ee^{\theta\bd{U}_p^T\bd{V}_p}  \leq
\exp(dK_pJ_p\theta^2)$ for all $|\theta|\leq C/\lambda_{max}(\bd{\Sigma})$ where $J_p=\lambda_{max}(\bd{\Sigma})/\lambda_{min}(\bd{\Sigma}).$\\
(ii) $Ee^{\theta\|\bd{V}_p\|^2} \leq  \exp[2dK_p\theta/\lambda_{min}(\bd{\Sigma})]$ for all $0\leq \theta\leq C/\lambda_{max}(\bd{\Sigma})$.\\
(iii) $E\exp\big[\theta(Z_1^2+\cdots + Z_d^2)\big]\leq e^{2d\theta}$ for $0\leq \theta\leq C/d$.
\end{lemma}

\noindent\textbf{Proof of Lemma \ref{Weyl}}. Set $k=p-d$, so $\bd{\Sigma}_{11}$ is $d\times d$, $\bd{\Sigma}_{12}$ is $d\times k$ and $\bd{\Sigma}_{22}$ is $k\times k$. Let $\bd{\xi}=(\xi_1, \cdots, \xi_k)^T$ and $\bd{\eta}=(\eta_1, \cdots, \eta_k)^T$, where the $2k$ random variables $\xi_i$'s and $\eta_i$'s are  i.i.d. $N(0, 1)$-distributed. According to Lemma \ref{sunny_coffee},
\begin{align}\lbl{yu_qiuyu}
\bd{U}_p \overset{d}{=}(\bd{\Sigma}_{22\cdot 1})^{1/2}\bd{\xi}\ \ \
\mbox{and}\ \ \ \bd{V}_p \overset{d}{=}(\bd{\Sigma}_{21}\bd{\Sigma}_{11}^{-1}\bd{\Sigma}_{12})^{1/2}\bd{\eta},
\end{align}
and they are independent. From \eqref{yu_qiuyu} we see that   $\bd{U}_p^T\bd{V}_p\overset{d}{=}\bd{\xi}^T\hat{\bd{\Sigma}}_p\bd{\eta}$,  where
\begin{align*}
\hat{\bd{\Sigma}}_p = (\bd{\Sigma}_{22\cdot 1})^{1/2}\cdot (\bd{\Sigma}_{21}\bd{\Sigma}_{11}^{-1}\bd{\Sigma}_{12})^{1/2}.
\end{align*}
By the singular value decomposition theorem, we write $\hat{\bd{\Sigma}}_p=\bd{O}_1\,\mbox{diag}(\lambda_1, \cdots, \lambda_k)\,\bd{O}_2$ where $\lambda_1, \cdots, \lambda_k$ are the singular values of $\hat{\bd{\Sigma}}_p$, and $\bd{O}_1$ and $\bd{O}_2$ are orthogonal matrices. Review the well-known facts that
\begin{align}\lbl{old_cadre}
\mbox{tr}\,(\bd{A}\bd{B})=\mbox{tr}\,(\bd{B}\bd{A})\ \ \mbox{and}\ \ \mbox{tr}(\bd{C}\bd{D})\leq \mbox{tr}(\bd{C})\cdot \mbox{tr}(\bd{D})
\end{align}
for any matrices $\bd{A}$ and $\bd{B}$ and any non-negative definite matrices $\bd{C}$ and $\bd{D}$ (the second fact from \eqref{old_cadre} can also be thought as a consequence of Lemma \ref{Gu_xiao}). Note that
\begin{align}\lbl{chat_dophin}
\bd{\Sigma}_{22}=\bd{\Sigma}_{22\cdot 1}+\bd{\Sigma}_{21}\bd{\Sigma}_{11}^{-1}\bd{\Sigma}_{12}
\end{align}
and all three matrices are non-negative definite. This together with the Weyl interlacing inequality implies  $\lambda_{max}(\bd{\Sigma}_{22\cdot 1})\leq \lambda_{max}(\bd{\Sigma}_{22}) \leq \lambda_{max}(\bd{\Sigma})$. Consequently, we have from Lemma \ref{Gu_xiao} that
\begin{align*}
\lambda_1^2+ \cdots+ \lambda_k^2=\, \mbox{tr}\,(\hat{\bd{\Sigma}}_p\hat{\bd{\Sigma}}_p^T)
=&\mbox{tr}\,\big[\bd{\Sigma}_{22\cdot 1}\cdot (\bd{\Sigma}_{21}\bd{\Sigma}_{11}^{-1}\bd{\Sigma}_{12})\big]\\
 \leq & \lambda_{max}(\bd{\Sigma}_{22\cdot 1})\cdot\mbox{tr}\,(\bd{\Sigma}_{21}\bd{\Sigma}_{11}^{-1}\bd{\Sigma}_{12})\\
 \leq & \lambda_{max}(\bd{\Sigma})\cdot\mbox{tr}\,(\bd{\Sigma}_{21}\bd{\Sigma}_{11}^{-1}\bd{\Sigma}_{12}).
\end{align*}
 Furthermore, by Lemma \ref{Gu_xiao} again,
\begin{align}\lbl{hong_chi}
\mbox{tr}\,(\bd{\Sigma}_{21}\bd{\Sigma}_{11}^{-1}\bd{\Sigma}_{12})
=\mbox{tr}\,(\bd{\Sigma}_{11}^{-1}\bd{\Sigma}_{12}\bd{\Sigma}_{21})
\leq & \lambda_{max}(\bd{\Sigma}_{11}^{-1})\cdot \mbox{tr}(\bd{\Sigma}_{12}\bd{\Sigma}_{21})\nonumber\\
=& \frac{1}{\lambda_{min}(\bd{\Sigma}_{11})}\cdot \mbox{tr}(\bd{\Sigma}_{12}\bd{\Sigma}_{21})
\leq  \frac{d K_p}{\lambda_{min}(\bd{\Sigma})}.
\end{align}
In fact in the above we use the assertion $\lambda_{min}(\bd{\Sigma})\leq \lambda_{min}(\bd{\Sigma}_{11})$ by the Weyl interlacing inequality and the fact that
\begin{align*}
\mbox{tr}(\bd{\Sigma}_{12}\bd{\Sigma}_{21})=\sum_{i=1}^d\sum_{j=d+1}^p\sigma_{ij}^2\leq dK_p
\end{align*}
by assumption. Combing the above, we arrive at
\begin{align}\lbl{kscealvtu}
 \lambda_1^2+ \cdots+ \lambda_k^2\leq (dK_p)\cdot \frac{\lambda_{max}(\bd{\Sigma})}{\lambda_{min}(\bd{\Sigma})}.
\end{align}
Another fact we will use later on is that
\begin{align}\lbl{chi_le}
\Lambda_1:=\max\{\lambda_1, \cdots, \lambda_k\} \leq \lambda_{max}(\bd{\Sigma}).
\end{align}
In fact, recall that $\|\cdot\|$ denotes the spectral norm of a matrix. By definition,
\begin{align}\lbl{hei_a}
\Lambda_1
=& \|(\bd{\Sigma}_{22\cdot 1})^{1/2}\cdot (\bd{\Sigma}_{21}\bd{\Sigma}_{11}^{-1}\bd{\Sigma}_{12})^{1/2}\|\nonumber\\
 \leq &  \|\bd{\Sigma}_{22\cdot 1}\|^{1/2}\cdot \|\bd{\Sigma}_{21}\bd{\Sigma}_{11}^{-1}\bd{\Sigma}_{12}\|^{1/2}.
\end{align}
From \eqref{chat_dophin} we know that both norms in \eqref{hei_a} are bounded by  $\lambda_{max}(\bd{\Sigma}_{22})^{1/2}\leq \lambda_{max}(\bd{\Sigma})^{1/2}.$ So \eqref{chi_le} is obtained. With these preparation, we are ready to prove (i), (ii) and (iii).

(i)  Since $\bd{U}_p^T\bd{V}_p=\bd{\xi}^T\hat{\bd{\Sigma}}_p\bd{\eta}$ and $\hat{\bd{\Sigma}}_p=\bd{O}_1\,\mbox{diag}(\lambda_1, \cdots, \lambda_k)\,\bd{O}_2$, by the orthogonal invariant property of $N(\bd{0}, \bd{I}_k)$, we have that
\begin{align}\lbl{deng_dao}
\bd{U}_p^T\bd{V}_p \overset{d}{=} \sum_{i=1}^k\lambda_i\xi_i\eta_i.
\end{align}
Review the moment generating functions of Gaussian variables,
\begin{align}\lbl{apple_orange}
Ee^{\theta\xi_1}=e^{\theta^2/2}\ \ \ \mbox{and}\ \ \ Ee^{\theta\xi_1^2}=(1-2\theta)^{-1/2},\ \ \ \theta<\frac{1}{2}.
\end{align}
We can write
\begin{align*}
E\exp\big(\theta\bd{U}_p^T\bd{V}_p\big)=\prod_{i=1}^kEe^{\theta\lambda_i\xi_i\eta_i}
=\prod_{i=1}^kEe^{(\theta\lambda_i)^2\xi_i^2/2}=\exp\Big\{-\frac{1}{2}\sum_{i=1}^k\log [1-(\theta\lambda_i)^2]\Big\},
\end{align*}
for all $|\theta\lambda_i| <1$ with $i=1,\cdots, k$, that is, $|\theta|\leq \frac{1}{\Lambda_1}$.
Notice $-\frac{1}{2}\log (1-x)\sim \frac{1}{2}x$ as $x\to 0$, Thus, there exists $x_0\in (0, \frac{1}{2})$ such that $-\frac{1}{2}\log (1-x)\leq x$ for all $x \in (0, x_0^2)$. Then
\begin{align}\lbl{gang_tie}
E\exp\big(\theta\bd{U}_p^T\bd{V}_p\big) \leq \exp\Big(\theta^2\sum_{i=1}^k\lambda_i^2\Big) \leq
e^{dK_pJ_p\theta^2}
\end{align}
for all $\theta$ with $|\theta|\leq \frac{1}{\Lambda_1} \wedge \frac{x_0}{\Lambda_1}=\frac{x_0}{\Lambda_1}$
by \eqref{kscealvtu}, with $J_p=\lambda_{max}(\bd{\Sigma})/\lambda_{min}(\bd{\Sigma}).$ The inequality \eqref{gang_tie} is particularly true if $|\theta|\leq \frac{x_0}{\lambda_{max}(\bd{\Sigma})}$ by \eqref{chi_le}.

(ii) Let $\rho_1, \cdots, \rho_k$ be the eigenvalues of $\bd{\Sigma}_{21}\bd{\Sigma}_{11}^{-1}\bd{\Sigma}_{12}.$ From \eqref{chat_dophin} we have
\begin{align}\lbl{bu_yong}
\Lambda_2:=\max\{\rho_1, \cdots, \rho_k\} \leq \lambda_{max}(\bd{\Sigma}_{22})\leq \lambda_{max}(\bd{\Sigma}).
\end{align}
By \eqref{yu_qiuyu} and the orthogonal invariant property of normal distributions similar to \eqref{deng_dao},
\begin{align}\lbl{headache}
\|\bd{V}_p\|^2 \overset{d}{=} \bd{\eta}^T\big(\bd{\Sigma}_{21}\bd{\Sigma}_{11}^{-1}\bd{\Sigma}_{12}\big)
\bd{\eta}\overset{d}{=}\rho_1\eta_1^2+\cdots + \rho_k\eta_k^2.
\end{align}
As shown in \eqref{hong_chi},
\begin{align}\lbl{hong_chen}
\rho_1+\cdots + \rho_k \leq \frac{dK_p}{\lambda_{min}(\bd{\Sigma})}.
\end{align}
By \eqref{apple_orange} and \eqref{headache}, we have
\begin{align*}
Ee^{\theta\|\bd{V}_p\|^2}=\exp\Big[-\frac{1}{2}\sum_{i=1}^d\log (1-2\theta \rho_i)\Big]
\end{align*}
for all $\theta<\frac{1}{2\Lambda_2}$. Recalling $x_0$ defined earlier, we conclude
\begin{align*}
Ee^{\theta\|\bd{V}_p\|^2} \leq e^{2\theta(\rho_1+\cdots +\rho_k)} \leq \exp\Big(\frac{2dK_p\theta}{\lambda_{min}(\bd{\Sigma})}\Big)
\end{align*}
for all $0\leq \theta<\frac{1}{2\Lambda_2}\wedge\frac{x_0^2}{2\Lambda_2}=
\frac{x_0^2}{2\Lambda_2}$. By \eqref{bu_yong}, the above is particularly true provided $0\leq \theta<\frac{x_0^2}{2\lambda_{max}(\bd{\Sigma})}$, which proves the claim.

(iii) By the H\"{o}lder inequality and the fact that $Z_i \sim N(0, 1)$ for each $i$,
\begin{align*}
E\exp\big[\theta(Z_1^2+\cdots + Z_d^2)\big]\leq \Big[\big(Ee^{d\theta Z_1^2}\big)^{1/d}\Big]^d=(1-2d\theta)^{-1/2}
\end{align*}
for all $\theta< 1/(2d).$ For $x_0$ defined earlier, we obtain
\begin{align*}
E\exp\big[\theta(Z_1^2+\cdots + Z_d^2)\big]\leq e^{2d\theta}
\end{align*}
for  $0\leq \theta\leq \frac{x_0^2}{2d}$ as desired.   \hfill$\square$

\medskip

\begin{lemma}\lbl{sunset_street} Assume the same notations and conditions as in Lemma \ref{Weyl}.  Denote  $\Theta_p=\|\bd{V}_p\|^2+ 2\bd{U}_p^T\bd{V}_p +\sum_{i=1}^dZ_i^2$ and $\upsilon_p=[2\mbox{tr}(\bd{\Sigma}^2)]^{1/2}$. Suppose Assumption \eqref{assumption_A3} holds and  $C$ is the constant appearing in \eqref{assumption_A3}. Let $d\geq 1$ and $\epsilon>0$ be given. Then
\begin{align}\lbl{stone_shadow}
\epsilon_p:=\frac{(\log p)^C}{v_p\lambda_{min}(\bd{\Sigma})} \to 0\ \ \mbox{and}\ \ t=t_p:=\frac{C\epsilon}{8}\cdot\frac{\upsilon_p}{\lambda_{max}(\bd{\Sigma})\log p}\to \infty.
\end{align}
Furthermore,
\begin{align*}
P(|\Theta_p|\geq \epsilon\upsilon_p) \leq \frac{3}{p^{t}}+\exp\Big(-\frac{C\epsilon}{8d}\sqrt{p}\Big)
\end{align*}
for every  $p$ satisfying $p\geq \frac{256d^2}{\epsilon^2}$  and $\epsilon_p<\frac{\epsilon}{8(C+2)d}$.
\end{lemma}
\noindent\textbf{Proof of Lemma \ref{sunset_street}}. First we give an estimate for $\upsilon_p.$ Set $K_p=(\log p)^C$, where $C>0$ is the constant as given in Assumption \eqref{assumption_A3}. Evidently, $\mbox{tr}(\bd{\Sigma}^2)=\sum_{1\leq i, j \leq p}\sigma_{ij}^2$. Also $\sigma_{ii}=1$ and $\sum_{j=1}^p\sigma_{ij}^2\leq K_p$ for each $1\leq i \leq p$ by Assumption \eqref{assumption_A3}. It follows that
\begin{align}\lbl{liquor_care}
\sqrt{2p}=\Big(2\sum_{i=1}^p\sigma_{ii}^2\Big)^{1/2}\leq \upsilon_p\leq \sqrt{2p}\cdot (\log p)^C
\end{align}
for $p\geq 3$. Now, for any $\epsilon>0$, union bound gives
\begin{align}\lbl{Koch}
&P(|\Theta_p|\geq \epsilon\upsilon_p)\nonumber\\
\leq & P\Big(\|\bd{V}_p\|^2\geq \frac{1}{4}\epsilon\upsilon_p\Big) + P\Big(|\bd{U}_p^T\bd{V}_p|\geq \frac{1}{4}\epsilon\upsilon_p\Big) + P\Big(\sum_{i=1}^nZ_i^2\geq \frac{1}{4}\epsilon\upsilon_p\Big).
\end{align}
Let us bound them one by one. First, by the Markov inequality and Lemma \ref{Weyl}(ii),
\begin{align}\lbl{grass_no}
P\Big(\|\bd{V}_p\|^2\geq \frac{1}{4}\epsilon\upsilon_p\Big)
\leq & \exp\Big(-\frac{\theta\epsilon}{4}\upsilon_p\Big)\cdot Ee^{\theta\|\bd{V}_p\|^2} \nonumber\\
 \leq & \exp\Big[-\theta\Big(\frac{\epsilon}{4}\upsilon_p-
\frac{2dK_p}{\lambda_{min}(\bd{\Sigma})}\Big)\Big]
\end{align}
for any $0\leq \theta\leq C/\lambda_{max}(\bd{\Sigma})$ where $C>0$ is a constant free of $p$. By assumption, $K_p\leq (\log p)^C$ and  $p^{-1/2}(\log p)^C \ll \lambda_{min}(\bd{\Sigma})\leq \lambda_{max}(\bd{\Sigma})\ll \sqrt{p}/\log p$. Then,
\begin{align}\lbl{wolf_fox}
\frac{K_p}{\lambda_{min}(\bd{\Sigma})}=o(\upsilon_p)\ \ \mbox{and}\ \ \frac{\upsilon_p}{\lambda_{max}(\bd{\Sigma})}\gg \log p,
\end{align}
due to \eqref{liquor_care}. Thus, \eqref{stone_shadow} holds. Choosing $\theta= \frac{C}{\lambda_{max}(\bd{\Sigma})}$, we see from \eqref{grass_no} that
\begin{align}\lbl{desk}
P\Big(\|\bd{V}_p\|^2\geq \frac{1}{4}\epsilon\upsilon_p\Big)\leq \exp\Big[-\frac{C\epsilon}{8\lambda_{max}(\bd{\Sigma})}\upsilon_p\Big]=e^{-t\log p}=
\frac{1}{p^{t}}
\end{align}
for all $p$ satisfying
$\frac{\epsilon}{8}\upsilon_p >
\frac{2dK_p}{\lambda_{min}(\bd{\Sigma})}$, which is particularly true if
\begin{align}\lbl{birdy_sing}
\frac{K_p}{v_p\lambda_{min}(\bd{\Sigma})} \leq \epsilon_p < \frac{\epsilon}{8(C+2)d}.
\end{align}
Second,
\begin{align*}
P\Big(|\bd{U}_p^T\bd{V}_p|\geq \frac{1}{4}\epsilon\upsilon_p\Big) \leq P\Big(\bd{U}_p^T\bd{V}_p\geq \frac{1}{4}\epsilon\upsilon_p\Big)+P\Big(-\bd{U}_p^T\bd{V}_p\geq \frac{1}{4}\epsilon\upsilon_p\Big).
\end{align*}
By Lemma \ref{Weyl}(i) and a similar argument as \eqref{grass_no}, we have
\begin{align*}
P\Big(|\bd{U}_p^T\bd{V}_p|\geq \frac{1}{4}\epsilon\upsilon_p\Big) \leq 2\cdot\exp\Big[\theta\Big(-\frac{1}{4}\epsilon\upsilon_p+dK_pJ_p\theta\Big)\Big],
\end{align*}
for all $|\theta|\leq C/\lambda_{max}(\bd{\Sigma})$, where $J_p=\lambda_{max}(\bd{\Sigma})/\lambda_{min}(\bd{\Sigma}).$ It is trivial to see from the notation  $\bd{\Sigma}=(\sigma_{ij})_{p\times p}$ that  $\lambda_{max}(\bd{\Sigma})\geq \sigma_{11}=1$. Take  $\theta = C/\lambda_{max}(\bd{\Sigma})$  to obtain
\begin{align}\lbl{yi_yi}
P\Big(|\bd{U}_p^T\bd{V}_p|\geq \frac{1}{4}\epsilon\upsilon_p\Big)
 \leq &
2\cdot\exp\Big[
\frac{C}{\lambda_{max}(\bd{\Sigma})}\Big(-\frac{1}{4}\epsilon\upsilon_p+
\frac{CdK_p}{\lambda_{min}(\bd{\Sigma})}\Big)
\Big] \nonumber\\
 \leq & 2\cdot\exp\Big(-\frac{C\epsilon}{8\lambda_{max}(\bd{\Sigma})}\upsilon_p\Big)
 =  \frac{2}{p^{t}}
\end{align}
for all $p$ satisfying \eqref{birdy_sing}.
Finally, by the Markov inequality and by taking  $\theta= C/d$ from Lemma \ref{Weyl}(iii) we obtain
\begin{align*}
P\Big(\sum_{i=1}^dZ_i^2\geq \frac{1}{4}\epsilon\upsilon_p\Big) \leq \exp\Big(-\frac{C\epsilon}{4d}\upsilon_p+2C\Big) \leq \exp\Big(-\frac{C\epsilon}{8d}\sqrt{p}\Big)
\end{align*}
for every $p$ satisfying $2<\frac{\epsilon}{8d}\sqrt{p}$, or equivalently, $p\geq \frac{256d^2}{\epsilon^2}$. In the last step above we use the inequality $\upsilon_p\geq \sqrt{p}$ from  \eqref{liquor_care}. Combining this with \eqref{Koch}, \eqref{desk} and \eqref{yi_yi}, we conclude that
\begin{align*}
P(|\Theta_p|\geq \epsilon\upsilon_p) \leq \frac{3}{p^{t}}+\exp\Big(-\frac{C\epsilon}{8d}\sqrt{p}\Big)
\end{align*}
for every $p$ satisfying $\epsilon_p < \frac{\epsilon}{8(C+2)d}$ from \eqref{birdy_sing} and $p\geq \frac{256d^2}{\epsilon^2}$. \hfill$\square$

\medskip

We now introduce more general indexing. Let  $(Z_1, \cdots, Z_p)^T\sim N(\bd{0}, \bd{\Sigma})$. Assume $d$ is an integer with $1\leq d<p$. For any set $\Lambda=\{i_1, \cdots, i_d\}$ with $1\leq i_1<\cdots <i_d \leq p$, write $\X_{1,\Lambda}=(Z_{i_1}, \cdots, Z_{i_d})^T$. Let  $\X_{2,\Lambda}$ be the vector obtained with deleting $Z_{i_1}, \cdots, Z_{i_d}$ from $(Z_1, \cdots, Z_p)^T$, that is, $\X_{\Lambda, 2}=(Z_{j_1}, \cdots, Z_{j_{p-d}})^T$ where $j_1< \cdots <j_{p-d}$ and $\{j_1, \cdots, j_{p-d}\}=\{1, 2, \cdots, n\}\backslash\Lambda.$ Let $\bd{\Sigma}_{\Lambda}$ be the covariance matrix of
\begin{align*}
\bd{X}_{\Lambda}:=
\begin{pmatrix}
 \X_{1,\Lambda}\\
 \X_{2,\Lambda}
\end{pmatrix}
.
\end{align*}
Partition $\bd{\Sigma}_{\Lambda}$ similar to \eqref{similar_beer} such that
\begin{align*}
\bd{\Sigma}_{\Lambda}=
\begin{pmatrix}
\bd{\Sigma}_{11,\Lambda} & \bd{\Sigma}_{12,\Lambda}\\
\bd{\Sigma}_{21,\Lambda} & \bd{\Sigma}_{22,\Lambda}
\end{pmatrix}
.
\end{align*}
In particular, $\X_{1,\Lambda} \sim N(\bd{0}, \bd{\Sigma}_{11,\Lambda})$ and  $\X_{2,\Lambda} \sim N(\bd{0}, \bd{\Sigma}_{22,\Lambda})$. We have the following result.

\begin{lemma}\lbl{yyy_lll} Let  $\bd{X}=(Z_1, \cdots, Z_p)^T\sim N(\bd{0}, \bd{\Sigma})$. Assume $d$ is an integer with $1\leq d<p$. For any set $\Lambda=\{i_1, \cdots, i_d\}$ with $1\leq i_1<\cdots <i_d \leq p$, we define
 $\bd{U}_{p,\Lambda}=\X_{2,\Lambda}-\bms_{21,\Lambda}\bms_{11,\Lambda}^{-1}\X_{1,\Lambda}$ and  $\bd{V}_{p,\Lambda}=\bms_{21,\Lambda}\bms_{11,\Lambda}^{-1}\X_{1,\Lambda}$. Set  $\Theta_{p,\Lambda}=\|\bd{V}_{p,\Lambda}\|^2+ 2\bd{U}_{p,\Lambda}^T\bd{V}_{p,\Lambda} +\sum_{k=1}^dZ_{i_k}^2$ and $\upsilon_{p}=[2\mbox{tr}(\bd{\Sigma}^2)]^{1/2}$.
Then, under Assumption \eqref{assumption_A3}, for any $\epsilon>0$ there exists $t=t_p\to \infty$ such that
\begin{align*}
\max_{\Lambda}P(|\Theta_{p,\Lambda}|\geq \epsilon\upsilon_p) \leq \frac{1}{p^{t}}
\end{align*}
as $p$ is sufficiently large, where the maximum $\Lambda=\{i_1, \cdots, i_d\}$ runs over all possible indices $i_1, \cdots, i_d$ with $1\leq i_1<\cdots <i_d \leq p$.
\end{lemma}
\noindent\textbf{Proof of Lemma \ref{yyy_lll}}. View $\bd{X}_{\Lambda}$ as the vector after  exchanging some rows of $\bd{X}$. Then there is a permutation matrix $\bd{O}$ such that $\bd{X}_{\Lambda}=\bd{O}\bd{X}$. Therefore the covariance matrix of $\bd{X}_{\Lambda}$ is
$\bd{\Sigma}_{\Lambda}=E(\bd{O}\bd{X}(\bd{O}\bd{X})^T)=\bd{O}\bd{\Sigma}\bd{O}^T.$ Set $\upsilon_{p,\Lambda}=[2\mbox{tr}(\bd{\Sigma_{\Lambda}}^2)]^{1/2}$. Then
\begin{align}\lbl{waiters1}
\lambda_{max}(\bd{\Sigma}_{\Lambda})=\lambda_{max}(\bd{\Sigma}), \ \ \
\lambda_{min}(\bd{\Sigma}_{\Lambda})=\lambda_{min}(\bd{\Sigma})\ \ \ \mbox{and}\ \ \
\upsilon_{p,\Lambda}=\upsilon_p.
%\mbox{tr}(\bd{\Sigma_{\Lambda}}^2)=\mbox{tr}(\bd{\Sigma}^2).
\end{align}
Second, $\bd{O}\bd{\Sigma}\bd{O}^T$ is the matrix by exchanging some rows and then exchanging the corresponding columns. So the entries of $\bd{\Sigma}_{\Lambda}$ are the same as those of $\bd{\Sigma}$; the sum of squares of the entries of a row from $\bd{O}\bd{\Sigma}\bd{O}^T$ is the same as that of a row from $\bd{\Sigma}$, and vice versa. Write  $\bd{\Sigma}_{\Lambda}=(\sigma_{ij,\Lambda})_{p\times p}$. As a consequence,
\begin{align}\lbl{waiters2}
\max_{1\leq i<j \leq p}|\sigma_{ij,\Lambda}|=\max_{1\leq i<j \leq p}|\sigma_{ij}|,\ \ \mbox{and}\ \ \ \max_{1\leq i \leq p}\sum_{j=1}^p\sigma_{ij,\Lambda}^2=\max_{1\leq i \leq p}\sum_{j=1}^p\sigma_{ij}^2.
\end{align}
Notice that in Assumption \eqref{assumption_A3}, all conditions are imposed on the four quantities: $\lambda_{max}(\bd{\Sigma})$, $\lambda_{min}(\bd{\Sigma})$, $\max_{1\leq i<j \leq p}|\sigma_{ij}|$ and $\max_{1\leq i \leq p}\sum_{j=1}^p\sigma_{ij}^2$. As a result, by \eqref{waiters1} and \eqref{waiters2}, we see that \eqref{assumption_A3} still holds if ``$\bd{\Sigma}$" is replaced with ``$\bd{\Sigma}_{\Lambda}$". Review Lemma \ref{sunset_street}. Let $C$ be as in \eqref{assumption_A3}. Let $t=t_p$ be as in \eqref{stone_shadow}. By this display,
%Set $t'=t'_p=\frac{C\epsilon}{8d}\cdot\frac{\sqrt{p}}{\log p}$. Then
\begin{align*}
t'=t'_p:=\frac{C\epsilon}{8d}\cdot \min\Big\{\frac{\upsilon_p}{\lambda_{max}(\bd{\Sigma})\log p},\, \frac{\sqrt{p}}{\log p}\Big\} \to \infty
\end{align*}
as $p\to\infty$. Evidently, $t\geq t'$ and $\frac{C\epsilon}{8d}\frac{\sqrt{p}}{\log p}\geq t'$. Thus
\begin{align*}
\frac{3}{p^{t}}+\exp\Big(-\frac{C\epsilon}{8d}\sqrt{p}\Big)\leq \frac{4}{p^{t'}}\leq \frac{1}{p^{t'/2}}
\end{align*}
if $p^{t'/2}>4$. Taking $p_0\geq 3$ such that $p^{t'/2}>4$, $p\geq \frac{256d^2}{\epsilon^2}$  and $\epsilon_p<\frac{\epsilon}{8(C+2)d}$ for all $p\geq p_0$ and applying Lemma \ref{sunset_street}, we know that
\begin{align}\lbl{real_ha}
P(|\Theta_p|\geq \epsilon\upsilon_p) \leq \frac{1}{p^{t'/2}},
\end{align}
as $p\geq p_0$. Note that in the proof of \eqref{real_ha}, although the conclusion is on $\bd{\Sigma}$, only five quantities of $\bd{\Sigma}$ in \eqref{waiters1} and \eqref{waiters2} are required, and they are the same if ``$\bd{\Sigma}$" is replaced by ``$\bd{\Sigma}_{\Lambda}$" for different $\Lambda.$ Consequently, we induce from \eqref{real_ha} that
\begin{align*}
P(|\Theta_{p,\Lambda}|\geq \epsilon\upsilon_p) \leq \frac{1}{p^{t'/2}}.
\end{align*}
for any $p\geq p_0$ and any $\Lambda=\{i_1, \cdots, i_d\}$  with $1\leq i_1<\cdots <i_d \leq p$.  The desired conclusion then follows by writing $t'/2$ back to $t$. \hfill$\square$

\begin{lemma}\lbl{akxbcvgf} Assume $(Z_1, \cdots, Z_p)^T\sim N(\bd{0}, \bd{\Sigma})$ with $\bd{\Sigma}$ satisfying  \eqref{assumption_A3}. Set $S_p=Z_1^2+\cdots + Z_p^2$ and   $\upsilon_p=[2\mbox{tr}(\bd{\Sigma}^2)]^{1/2}$. For any $x\in \mathbb{R}$ and $y\in \mathbb{R}$, define $A_p=\{\frac{S_p-p}{\upsilon_p}\leq x\}$ and  $l_p= (2\log p -\log\log p+y)^{1/2}$  and  $B_{i}=\{|Z_i|>l_p\}.$ Then, for each $d\geq 1$,
\begin{align*}
\sum_{1\leq i_1<  \cdots < i_{d}\leq p}\big| P(A_pB_{i_1}\cdots B_{i_{d}}) - P(A_p)\cdot P(B_{i_1}\cdots B_{i_{d}}) \big|\to 0
\end{align*}
as $p\to\infty$.
\end{lemma}
\noindent\textbf{Proof of Lemma \ref{akxbcvgf}}. We prove the lemma in two steps.

{\it Step 1: appealing independence from normal distributions}. Note that $(Z_1, \cdots, Z_p)^T\sim N(\bd{0}, \bd{\Sigma})$. Take $\bd{X}_1=(Z_1, \cdots, Z_d)^T$ and $\bd{X}_2=(Z_{d+1}, \cdots, Z_p)^T$. Recall the notation in Lemma \ref{sunny_coffee}, which allows us to write
\begin{align*}
\bd{X}_2=\bd{U}_p + \bd{V}_p,
\end{align*}
where $\bd{U}_p =\bd{X}_2-\bd{\Sigma}_{21}\bd{\Sigma}_{11}^{-1}\bd{X}_1\sim N(\bd{0}, \bd{\Sigma}_{22\cdot 1})$ and $\bd{V}_p =\bd{\Sigma}_{21}\bd{\Sigma}_{11}^{-1}\bd{X}_1\sim N(\bd{0}, \bd{\Sigma}_{21}\bd{\Sigma}_{11}^{-1}\bd{\Sigma}_{12})$. Lemma \ref{sunny_coffee} says that
\begin{align}\lbl{old_man}
\bd{U}_p\ \mbox{and}\ \{Z_1, \cdots, Z_d\}\ \mbox{are independent}.
\end{align}
Further denote
\begin{align*}
S_p=\|\bd{X}_1\|^2+\|\bd{X}_2\|^2=\|\bd{U}_p\|^2+\|\bd{V}_p\|^2+ 2\bd{U}_p^T\bd{V}_p +\sum_{i=1}^dZ_i^2.
\end{align*}
We will show the last three terms on the right hand side  are negligible. Recall
\begin{align*}
\Theta_p=\|\bd{V}_p\|^2+ 2\bd{U}_p^T\bd{V}_p +\sum_{i=1}^dZ_i^2
\end{align*}
as defined in Lemma \ref{sunset_street}.
By Lemma \ref{yyy_lll} with $\bd{X}_{\Lambda}=\bd{X}$,  for any $d\geq 1$ and $\epsilon>0$, there exists $t=t_p>0$ with  $\lim_{p\to \infty}t_p=\infty$ and integer $p_0\geq 1$, such that
\begin{align}\lbl{when_what}
P(|\Theta_p|\geq \epsilon\upsilon_p)\leq \frac{1}{p^{t}}
\end{align}
as $p\geq p_0$. Now for clarity we re-write the definition of $A_p$  as
\begin{align*}
A_p(x)=\Big\{\frac{1}{\upsilon_p}(S_p-p)\leq x\Big\},\ \ x \in \mathbb{R},
\end{align*}
for $p\geq 1$. Since $S_p=\|\bd{U}_p\|^2+\Theta_p$, we see that
\begin{align*}
 P(A_p(x)B_1\cdots B_d)
\leq & P\Big(A_p(x)B_1\cdots B_d,\ \frac{|\Theta_p|}{\upsilon_p}<  \epsilon\Big) + \frac{1}{p^t}\\
 \leq & P\Big(\frac{1}{\upsilon_p}(\|\bd{U}_p\|^2-p)\leq x+\epsilon
,\ B_1\cdots B_d\Big) +\frac{1}{p^{t}}\\
 = & P\Big(\frac{1}{\upsilon_p}(\|\bd{U}_p\|^2-p)\leq x+\epsilon\Big)\cdot P\big(
B_1\cdots B_d\big) +\frac{1}{p^{t}},
\end{align*}
by the independence stated in \eqref{old_man}. Regarding the first probability, we have
\begin{align*}
 P\Big(\frac{1}{\upsilon_p}(\|\bd{U}_p\|^2-p)\leq x+\epsilon\Big)
 \leq & P\Big(\frac{1}{\upsilon_p}(\|\bd{U}_p\|^2-p)\leq x+\epsilon,\ \frac{|\Theta_p|}{\upsilon_p}<  \epsilon\Big) + \frac{1}{p^{t}} \\
 \leq & P\Big(\frac{1}{\upsilon_p}(\|\bd{U}_p\|^2+\Theta_p-p)\leq  x+2\epsilon\Big) + \frac{1}{p^{t}}\\
\leq & P\big(A_p(x+2\epsilon)\big) + \frac{1}{p^{t}}.
\end{align*}
Combine the two inequalities to get
\begin{align}\lbl{shakings}
 P(A_p(x)B_1\cdots B_d)
\leq P\big(A_p(x+2\epsilon)\big)\cdot P\big(
B_1\cdots B_d\big)  + \frac{2}{p^{t}}.
\end{align}
Similarly,
\begin{align*}
&P\Big(\frac{1}{\upsilon_p}(\|\bd{U}_p\|^2-p)\leq x-\epsilon,\
B_1\cdots B_d\Big)\\
\leq & P\Big(\frac{1}{\upsilon_p}(\|\bd{U}_p\|^2-p)\leq x-\epsilon,
B_1\cdots B_d, \frac{|\Theta_p|}{\upsilon_p}<  \epsilon\Big)  +\frac{1}{p^{t}}\\
 \leq & P\Big(\frac{1}{\upsilon_p}(S_p-p)\leq x,\ B_1\cdots B_d\Big) +\frac{1}{p^{t}}.
\end{align*}
By the independence from \eqref{old_man},
\begin{align*}
 P(A_p(x)B_1\cdots B_d) \geq
P\Big(\frac{1}{\upsilon_p}(\|\bd{U}_p\|^2-p)\leq x-\epsilon\Big)\cdot P(
B_1\cdots B_d)-\frac{1}{p^{t}}.
\end{align*}
Furthermore,
\begin{align*}
P\Big(\frac{1}{\upsilon_p}(S_p-p)\leq x-2\epsilon\Big)
 \leq & P\Big(\frac{1}{\upsilon_p}(S_p-p)\leq x-2\epsilon,\ \frac{|\Theta_p|}{\upsilon_p}<  \epsilon\Big) + \frac{1}{p^{t}} \\
 \leq & P\Big(\frac{1}{\upsilon_p}(\|\bd{U}_p\|^2-p)\leq x-\epsilon\Big) +\frac{1}{p^{t}}
\end{align*}
where the fact $S_p=\|\bd{U}_p\|^2+\Theta_p$ is used again.
Combining the above two inequalities we get
\begin{align*}
P(A_p(x)B_1\cdots B_d)
\geq  P(A_p(x-2\epsilon))\cdot P(B_1\cdots B_d)-\frac{2}{p^{t}}.
\end{align*}
This together with \eqref{shakings} implies that
\begin{align}\lbl{hei_po}
&\big|P(A_p(x)B_1\cdots B_d)-P(A_p(x))\cdot P(B_1\cdots B_d)\big|\nonumber\\
 \leq & \Delta_{p, \epsilon}\cdot  P(B_1\cdots B_d)+\frac{2}{p^{t}}
\end{align}
as $p\geq p_0$, where
\begin{align*}
\Delta_{p, \epsilon}:=&|P(A_p(x))-P(A_p(x+2\epsilon))| + |P(A_p(x))-P(A_p(x-2\epsilon))|\\
=& P(A_p(x+2\epsilon))-P(A_p(x-2\epsilon)),
\end{align*}
since $P(A_p(x))$ is increasing in $x\in \mathbb{R}.$
An important observation is that the derivation of \eqref{hei_po} is based on three key facts:  inequality \eqref{when_what}, the identity  $S_p=\|\bd{U}_p\|^2+\Theta_p$ and the fact $\bd{U}_p$ and $\{Z_1, \cdots, Z_d\}$ are independent from \eqref{old_man}.

Recall the notations in Lemma \ref{yyy_lll}. For any $1\leq i_1< i_2<\cdots <i_d\leq p$, denote   $\Lambda=\{i_1, \cdots, i_d\}$. Then, $\bd{X}_{2,\Lambda}=\bd{U}_{p,\Lambda} + \bd{V}_{p,\Lambda}$.
By Lemma \ref{sunny_coffee}, $\bd{U}_{p,\Lambda}$ and $\{Z_{i_1}, \cdots, Z_{i_d}\}$ are independent. In addition,
\begin{align*}
S_p=& \|\bd{X}_{1,\Lambda}\|^2+\|\bd{X}_{2,\Lambda}\|^2=\|\bd{U}_{p,\Lambda}\|^2+\|\bd{V}_{p,\Lambda}\|^2+ 2\bd{U}_{p,\Lambda}^T\bd{V}_{p,\Lambda} +\sum_{k=1}^dZ_{i_k}^2;\\
\Theta_{p,\Lambda}=& \|\bd{V}_{p,\Lambda}\|^2+ 2\bd{U}_{p,\Lambda}^T\bd{V}_{p,\Lambda} +\sum_{k=1}^dZ_{i_k}^2.
\end{align*}
Hence, we can write $S_p=\|\bd{U}_{p,\Lambda}\|^2+\Theta_{p,\Lambda}$. Based on Lemma \ref{yyy_lll},
\begin{align*}
\max_{\Lambda}P(|\Theta_{p,\Lambda}|\geq \epsilon\upsilon_p) \leq \frac{1}{p^{t}}
\end{align*}
when $p\geq p_0$. Consequently, the three  key facts aforementioned also hold for the corresponding quantities related to $\Lambda$. Thus, similar to the derivation of \eqref{hei_po},
we have
\begin{align*}
&\big|P(A_p(x)B_{i_1}\cdots B_{i_d})-P(A_p(x))\cdot P(B_{i_1}\cdots B_{i_d})\big|\\
 \leq & \Delta_{p, \epsilon}\cdot  P(B_{i_1}\cdots B_{i_d})+\frac{2}{p^{t}},
\end{align*}
as $p\geq p_0$. Taking the summation we get
\begin{align}\lbl{last_orange}
\zeta(p,d):=& \sum_{1\leq i_1< \cdots < i_{d}\leq p}\big|P(A_p(x) B_{i_1}\cdots B_{i_{d}}) - P(A_p(x))\cdot P(B_{i_1}\cdots B_{i_{d}})\big| \nonumber\\
 \leq & \sum_{1\leq i_1< \cdots < i_{d}\leq p}\Big[\Delta_{p, \epsilon}\cdot  P(B_{i_1}\cdots B_{i_{d}})+\frac{2}{p^{t}}\Big] \nonumber\\
 \leq & \Delta_{p, \epsilon}\cdot H(d, p)+ \binom{p}{d}\cdot \frac{2}{p^{t}},
\end{align}
where we denote
\begin{align*}
H(d, p):=\sum_{1\leq i_1< \cdots < i_{d}\leq p}P(B_{i_1}\cdots B_{i_{d}}).
\end{align*}
In the following we will show $\lim_{\epsilon\downarrow 0}\limsup_{p\to\infty}\Delta_{p, \epsilon}=0$ and $\limsup_{p\to\infty}H(d, p)<\infty$ for each $d\geq 1$. Assuming these are true, by using  $\binom{p}{d} \leq p^d$ and \eqref{last_orange}, for fixed $d\geq 1$, by sending $p\to\infty$ first and then sending $\epsilon\downarrow 0$, we obtain $\lim_{p\to \infty}\zeta(p,d)= 0$ for each $d\geq 1$. The proof is then completed.

{\it Step 2: the proofs of ``\,$\lim_{\epsilon\downarrow 0}\limsup_{p\to\infty}\Delta_{p, \epsilon}=0$" and ``\,$\limsup_{p\to\infty}H(d, p)<\infty$ for each $d\geq 1$"}.
First, as discussed below  \eqref{assumption_A3}, Assumption \eqref{assumption_A3} implies Assumption \eqref{condition_1}. Thus, Theorem \ref{theorem_1} holds and we have as $p\to\infty$,
\begin{align}\lbl{hunshui_bainian}
\frac{S_p-p}{\upsilon_p}\to N(0, 1)\ \mbox{weakly},
\end{align}
and hence
\begin{align}\lbl{864123}
\Delta_{p, \epsilon} \to  \Phi(x+2\epsilon)-\Phi(x-2\epsilon),
\end{align}
as $p\to\infty$, where $\Phi(x)=\frac{1}{\sqrt{2\pi}}\int_{-\infty}^xe^{-t^2/2}\,dt$.  This implies that  $\lim_{\epsilon\downarrow 0}\limsup_{p\to\infty}\Delta_{p, \epsilon}=0$.

Second, take $\delta_p=1/(\log p)^2.$ Recall
$B_{p,i}=\{1\leq j \leq p;\, |\sigma_{ij}|\geq \delta_p\}$ for $1\leq i \leq p$ defined in Theorem \ref{theorem_2}. By Assumption \eqref{assumption_A3}, we know $\max_{1\leq i \leq p}\sum_{j=1}^p\sigma_{ij}^2\leq (\log p)^C$ for all $p\geq 1$. Then
\begin{align*}
|B_{p,i}|\cdot\frac{1}{(\log p)^2}\leq  \sum_{j=1}^p\sigma_{ij}^2\leq (\log p)^C
\end{align*}
for each $i=1, \cdots, p$. This implies that $\max_{1\leq i \leq p}|B_{p,i}| \leq (\log p)^{C+2}$. Take $\kappa=\kappa_p=(C+3)(\log \log p)/\log p$ for $p\geq e^e$. Then, $\kappa_p\to 0$ and $(\log p)^{C+2}< p^{\kappa}$, which gives
\begin{align}\lbl{Lakers_Thunders}
C_p:=\{1\leq i \leq p;\, |B_{p,i}|\geq p^{\kappa}\}=\emptyset.
\end{align}
Hence, $D_p:=\{1\leq i \leq p;\, |B_{p,i}|< p^{\kappa}\}=\{1,2,\cdots, p\}.$ Recall \eqref{Chicago_Seattle}, \eqref{ksahc} and \eqref{di_da}. By noting that ``$H(t, p)$" here is exactly ``$\alpha_t$" there for each $t\geq 1$, we know
\begin{align}\lbl{Baoxian}
\lim_{p\to\infty}H(d, p)=\frac{1}{d!}\pi^{-d/2}e^{-dx/2},
\end{align}
for each $d\geq 1.$ The proof is finished.  \hfill $\square$

We are now in the position to prove Theorem \ref{theorem_3}.

\noindent\textbf{Proof of Theorem \ref{theorem_3}}. Again, since Assumption \eqref{assumption_A3} implies Assumption  \eqref{condition_1} and Assumption  \eqref{condition_2}, we know that Theorem \ref{theorem_1} and Theorem \ref{theorem_2} hold. Set $\upsilon_p=[2\mbox{tr}(\bd{\Sigma}^2)]^{1/2}$.
By Theorem \ref{theorem_1},
\begin{align}
P\Big(\frac{S_p-p}{\upsilon_p}\leq x\Big)=\Phi(x) \lbl{gan_doufu}
\end{align}
as $p\to \infty$ for any $x\in \mathbb{R}$, where  $\Phi(x)=\frac{1}{\sqrt{2\pi}}\int_{-\infty}^xe^{-t^2/2}\,dt$.
%Recall $\kappa$, $\delta_p$ and $C_d$ appeared in \eqref{Lakers_Thunders}.
From Theorem \ref{theorem_2}, we have
\begin{align}
P\big(\max_{1\leq i \leq p}\big\{Z_i^2\big\}-2\log p +\log\log p \leq y\big) \to
F(y)=\exp\Big\{-\frac{1}{\sqrt{\pi}}e^{-y/2}\Big\}\  \lbl{jian_jiao}
\end{align}
as $p\to \infty$ for any $y \in \mathbb{R}.$  To show asymptotic independence, it is enough to prove
\begin{align*}
\lim_{p\to \infty}P\Big(\frac{S_p-p}{\upsilon_p}\leq x,\ \max_{1\leq i \leq p}Z_i^2-2\log p +\log\log p\leq y\Big)= \Phi(x)\cdot F(y)
\end{align*}
for any $x\in \mathbb{R}$ and $y \in \mathbb{R}$. Define
\begin{align}\lbl{whale}
L_p=\max_{1\leq i \leq p}|Z_i|\ \ \mbox{and}\ \ l_p= (2\log p -\log\log p+y)^{1/2},
\end{align}
where the latter one  makes sense for sufficiently large $p$. Due to  \eqref{gan_doufu}, the above condition we want to prove is equivalent to
\begin{align}\lbl{wealth_no}
\lim_{p\to \infty}P\Big(\frac{S_p-p}{\upsilon_p}\leq x,\ L_p>l_p\Big)= \Phi(x)\cdot [1-F(y)],
\end{align}
for any $x\in \mathbb{R}$ and $y \in \mathbb{R}$.  Recalling the notation in Lemma \ref{akxbcvgf}, we have
\begin{align}
%&& \Lambda_p=\{1, \cdots,  p\}; \nonumber\\
%&&
A_p=\Big\{\frac{S_p-p}{\upsilon_p}\leq x\Big\}\ \ \ \mbox{and}\ \ \ B_{i}=\big\{|Z_i|>l_p\big\} \lbl{tea_red}
\end{align}
for $1\leq i\leq p$. We can then write
\begin{align}\lbl{abci}
P\Big(\frac{1}{\upsilon_p}(S_p-p)\leq x,\ L_p>l_p\Big)=P\Big(\bigcup_{i=1}^pA_pB_{i}\Big).
\end{align}
Here the notation $A_pB_i$ stands for $A_p\cap B_i$.  From the inclusion-exclusion principle,
\begin{align}
P\Big(\bigcup_{i=1}^pA_pB_{i}\Big)  \leq  &\sum_{1\leq i_1 \leq p}P(A_pB_{i_1})-\sum_{1\leq i_1< i_2\leq p}P(A_pB_{i_1}B_{i_2})\nonumber\\
& +\cdots+
\sum_{1\leq i_1<  \cdots < i_{2k+1}\leq p}P(A_pB_{i_1}\cdots B_{i_{2k+1}}),
\lbl{Upper_bound}
\end{align}
and
\begin{align}
P\Big(\bigcup_{i=1}^pA_pB_{i}\Big)  \geq & \sum_{1\leq i_1 \leq p}P(A_pB_{i_1})-\sum_{1\leq i_1< i_2\leq p}P(A_pB_{i_1}B_{i_2})\nonumber\\
&+\cdots-  \sum_{1\leq i_1<  \cdots < i_{2k}\leq  p}P(A_pB_{i_1}\cdots B_{i_{2k}}) \lbl{Lower_bound}
\end{align}
for any integer $k\geq 1$. As in the proof of Lemma \ref{akxbcvgf}, define
\begin{align*}
H(p, d)=\sum_{1\leq i_1<  \cdots < i_{d}\leq p}P(B_{i_1}\cdots B_{i_{d}})
\end{align*}
for $d\geq 1$. From \eqref{Baoxian} we know
\begin{align}\lbl{Maya1}
\lim_{d\to\infty}\limsup_{p\to\infty}H(p, d)=0.
\end{align}
Denote
\begin{align*}
\zeta(p,d)=\sum_{1\leq i_1<  \cdots < i_d\leq p}\big[P(A_pB_{i_1}\cdots B_{i_d}) - P(A_p)\cdot P(B_{i_1}\cdots B_{i_d})\big]
\end{align*}
By Lemma \ref{akxbcvgf}, we have
\begin{align}\lbl{back_campus}
\lim_{p\to\infty}\zeta(p,d)=0
\end{align}
for each $d\geq 1$. The assertion \eqref{Upper_bound} implies that
\begin{align}\lbl{639475}
P\Big(\bigcup_{i=1}^pA_pB_{i}\Big)
 \leq & P(A_p)\Big[\sum_{1\leq i_1 \leq p}P(B_{i_1})-\sum_{1\leq i_1< i_2\leq p}P(B_{i_1}B_{i_2})+\cdots-  \nonumber\\
& \sum_{1\leq i_1<  \cdots < i_{2k} \leq p}P(B_{i_1}\cdots B_{i_{2k}})\Big]+ \Big[\sum_{d=1}^{2k}\zeta(p,d)\Big] + H(p, 2k+1)  \nonumber\\
\leq & P(A_p)\cdot P\Big(\bigcup_{i=1}^pB_{i}\Big)+ \Big[\sum_{d=1}^{2k}\zeta(p,d)\Big] + H(p, 2k+1),
\end{align}
where the inclusion-exclusion formula is used again in the last inequality, that is,
\begin{align*}
P\Big(\bigcup_{i=1}^pB_{i}\Big) \geq & \sum_{1\leq i_1 \leq p}P(B_{i_1})-\sum_{1\leq i_1< i_2\leq p}P(B_{i_1}B_{i_2})+\cdots - \sum_{1\leq i_1<  \cdots < i_{2k}\leq p}P(B_{i_1}\cdots B_{i_{2k}}),
\end{align*}
for all $k\geq 1$.
By the definition of $l_p$ and \eqref{jian_jiao},
\begin{align*}
 P\Big(\bigcup_{i=1}^pB_{i}\Big)=P\big(L_p>l_p\big)=P\big(L_p^2-2\log p +\log\log p> y\big) \to 1-F(y)
\end{align*}
as $p\to\infty$. By \eqref{gan_doufu}, $P(A_p)\to \Phi(x)$ as $p\to\infty.$ From \eqref{abci}, \eqref{back_campus} and \eqref{639475}, by fixing $k$ first and sending $p\to \infty$ we obtain that
\begin{align*}
\limsup_{p\to\infty}P\Big(\frac{1}{\upsilon_p}(S_p-p)\leq x,\ L_p>l_p\Big)\leq \Phi(x)\cdot [1-F(y)] +\lim_{p\to\infty}H(p, 2k+1).
\end{align*}
Now, by letting $k\to \infty$ and using \eqref{Maya1}, we have
\begin{align}\lbl{vskdnti}
\limsup_{p\to\infty}P\Big(\frac{1}{\upsilon_p}(S_p-p)\leq x,\ L_p>l_p\Big)\leq \Phi(x)\cdot [1-F(y)].
\end{align}
We next prove the lower bound in a similar way. By applying the same argument to \eqref{Lower_bound}, we see that the counterpart of \eqref{639475} becomes
\begin{align*}
P\Big(\bigcup_{i=1}^pA_pB_{i}\Big)
 \geq & P(A_p)\Big[\sum_{1\leq i_1 \leq p}P(B_{i_1})-\sum_{1\leq i_1< i_2\leq p}P(B_{i_1}B_{i_2})+\cdots + \nonumber\\
& \sum_{1\leq i_1<  \cdots < i_{2k-1}\leq p}P(B_{i_1}\cdots B_{i_{2k-1}})\Big] + \Big[\sum_{d=1}^{2k-1}\zeta(p,d)\Big] - H(p, 2k)  \nonumber\\
\geq & P(A_p)\cdot P\Big(\bigcup_{i=1}^pB_{i}\Big) + \Big[\sum_{d=1}^{2k-1}\zeta(p,d)\Big] - H(p, 2k),
\end{align*}
where in the last step we use the inclusion-exclusion principle such that
\begin{align*}
P\Big(\bigcup_{i=1}^pB_{i}\Big) \leq & \sum_{1\leq i_1 \leq p}P(B_{i_1})-\sum_{1\leq i_1< i_2\leq p}P(B_{i_1}B_{i_2})+\cdots + \sum_{1\leq i_1<  \cdots < i_{2k-1}\leq p}P(B_{i_1}\cdots B_{i_{2k-1}})
\end{align*}
for all $k\geq 1$. Review \eqref{abci} and repeat the earlier procedure to see
\begin{align*}
\liminf_{p\to\infty}P\Big(\frac{1}{\upsilon_p}(S_p-p)\leq x,\ L_p>l_p\Big)\geq \Phi(x)\cdot [1-F(y)]
\end{align*}
with $p\to \infty$ and $k\to\infty$,
which, together with \eqref{vskdnti}, yields \eqref{wealth_no}. The proof is now complete. \hfill$\Box$

\subsection{Proof of Theorem \ref{thone} and Theorem S\ref{thtwo}}\lbl{jingshen}

\begin{lemma}\lbl{small_lemma} Let $\{(U, U_{p},\tilde{U}_p)\in \mathbb{R}^3;\, p\geq 1\}$ and  $\{(V, V_{p},\tilde{V}_p)\in \mathbb{R}^3;\, p\geq 1\}$ be two sequences of random variables with $U_p\to U$ and $V_p\to V$ in distribution as $p\to\infty.$ Assume $U$ and $V$ are continuous random variables and
\begin{align}\label{aixiaokey}
\tilde{U}_p=U_p+o_p(1)\ \ \ \mbox{and}\ \ \ \tilde{V}_p=V_p+o_p(1).
\end{align} If $U_p$ and $V_p$ are asymptotically independent, then $\tilde{U}_p$ and
$\tilde V_p$ are also asymptotically independent.
\end{lemma}
\noindent\textbf{Proof of Lemma \ref{small_lemma}}.  Define
\begin{align*}
\Omega_{p,\epsilon}=\Big\{\Big|U_p-\tilde{U}_p\Big|\leq \epsilon,\ \Big|V_p-\tilde{V}_p\Big|\leq \epsilon\Big\}
\end{align*}
for any $p\geq 1$ and $\epsilon>0$. By \eqref{aixiaokey},
\begin{align}\lbl{qianjinne}
\lim_{p\to\infty}P(\Omega_{p,\epsilon})=1
\end{align}
for any $\epsilon>0$. Fix $x\in \mathbb{R}$ and $y\in \mathbb{R}$. We note that
\begin{align}\lbl{xizao}
 P\Big(\tilde{U}_p\leq x, \tilde{V}_p\leq y\Big)
\leq &P\Big(\tilde{U}_p\leq x, \tilde{V}_p\leq y,\,  \Omega_{p,\epsilon}\Big) +P(\Omega_{p,\epsilon}^c) \nonumber\\
 \leq & P\Big(U_p\leq x+\epsilon, V_p\leq y+\epsilon\Big) +P(\Omega_{p,\epsilon}^c).
\end{align}
By the assumption on the asymptotic independence,
\begin{align}\label{jiancaojixiangle}
\lim_{p\to \infty}P(U_p\le s,V_p\le t)=P(U\le s)\cdot P(V\le t)
\end{align}
for any $s\in \mathbb{R}$ and $t\in \mathbb{R}$.
By letting $p\to \infty$ and then $\epsilon\downarrow 0$ in \eqref{xizao}, since $U$ and $V$ are continuous, we deduce from \eqref{qianjinne}  and \eqref{jiancaojixiangle} that
\begin{align}\lbl{love_ai}
\limsup_{p\to\infty} P\Big(\tilde{U}_p\leq x, \tilde{V}_p\leq y\Big) \leq P(U\leq x)\cdot P(V\leq y).
\end{align}
By switching the roles of ``$U_p \to U$ and $V_p\to V$" and ``$\tilde{U}_p$ and
$\tilde V_p$"in \eqref{xizao}, we have
\begin{align*}
P\Big(U_p\leq x, V_p\leq y\Big)
\leq & P\Big(\tilde{U}_p\leq x+\epsilon, \tilde{V}_p\leq y+\epsilon\Big) +P(\Omega_{p,\epsilon}^c)
\end{align*}
for any $x\in \mathbb{R}$,  $y\in \mathbb{R}$ and $\epsilon>0$.
Or, equivalently,
\begin{align*}
 P\Big(\tilde{U}_p\leq x, \tilde{V}_p\leq y\Big)
 \geq P\Big(U_p\leq x-\epsilon, V_p\leq y-\epsilon\Big)-P(\Omega_{p,\epsilon}^c).
\end{align*}
Similar to the derivation of \eqref{love_ai}, we get
\begin{align*}
\liminf_{p\to\infty} P\Big(\tilde{U}_p\leq x, \tilde{V}_p\leq y\Big) \geq P(U\leq x)\cdot P(V\leq y).
\end{align*}
This and \eqref{love_ai} lead to
\begin{align*}
\lim_{p\to \infty}P(\tilde U_p\le x,\tilde V_p\le y)=P(U\le x)\cdot P(V\le y),
\end{align*}
which shows the asymptotic independence between $\tilde U_p$ and $\tilde V_p$ as claimed. \hfill$\square$

\noindent{\bf Proof of Theorem \ref{thone}.} By Theorem 3.1 in \cite{srivastava2009test}, we get claim (i). We next prove claim (ii).

Recall $\X_1,\cdots,\X_n$ are i.i.d. $N(\bmu, \bms)$-distributed random vectors and  $\bar{\X}=\frac{1}{n}\sum_{i=1}^n\bd{X}_i=(\bar{\X}_1,\cdots,\bar{\X}_p)^T$. Note that $\sqrt{n}\bar{\X}\sim N(\bm 0,\bms)$ under the null hypothesis in  \eqref{one}. Write $\bms=(\sigma_{ij})_{p\times p}$, and let $\R=\D^{-1/2}\bms\D^{-1/2}=(\rho_{ij})_{1\le i,j\le p}$ denote the population correlation matrix, where  $\D$ is the diagonal matrix of $\bms$. Then  $Z_i:=\sqrt{n}\bar{\X}_i/\sqrt{\sigma_{ii}}\sim N(0, 1)$ for each $1\leq i \leq p$ and $\mbox{Cov}(Z_i, Z_j)=\sigma_{ij}/\sqrt{\sigma_{ii}\sigma_{jj}}=\rho_{ij}$ for $i\ne j.$ In other words, $(Z_1, \cdots, Z_p)^T \sim N(\bd{0}, \bd{R})$. By assumption, \eqref{condition_2} holds with ``$\bd{\Sigma}$'' replaced by ``$\bd{R}$''. Set $\tilde{T}_{max}^{(1)}=\max_{1\le i\le p}Z_i^2$. Since Assumption \eqref{assumption_A3} is stronger than Assumption \eqref{condition_2}, by Theorem \ref{theorem_2} and Assumption  \eqref{condition_2}, it holds that
\begin{align}\lbl{xiangshou}
&\tilde{T}_{max}^{(1)}-2\log p +\log\log p=\max_{1\leq i \leq p}Z_i^2-2\log p +\log\log p\ \mbox{
converges} \nonumber\\
& \mbox{weakly to a distribution with cdf}\ F(x)=\exp\big\{-e^{-x/2}/\sqrt{\pi}\big\},\ x\in \mathbb{R}.
\end{align}
Observe  that the distribution of $(Z_1, \cdots, Z_p)^T \sim N(\bd{0}, \bd{R})$ is free of $n$, hence the above limit holds for any $n=n_p$. Now,
%$\tilde{T}_{max}^{(1)}:=n\cdot\max_{1\le i\le p}\sigma_{ii}^{-1}\bar{\X}_i^2$. Note that
to prove (ii), we only need to show that $T_{max}^{(1)} - \tilde{T}_{max}^{(1)}=o_p(1)$. Indeed, we have
\begin{align}\lbl{haoya}
|T_{max}^{(1)}-\tilde{T}_{max}^{(1)}|
=& \Big|n\max_{1\le i\le p}\hat{\sigma}_{ii}^{-2}\bar{\X}_i^2-n\max_{1\le i\le p}\sigma_{ii}^{-1}\bar{\X}_i^2\Big| \nonumber\\
\le &\Big|n\max_{1\le i\le p}\sigma_{ii}^{-1}\bar{\X}_i^2\Big|\cdot \max_{1\le i\le p} |\sigma_{ii}\hat{\sigma}_{ii}^{-2}-1| \nonumber\\
=& \Big(\max_{1\le i\le p}|Z_i|\Big)^2\cdot \max_{1\le i\le p} |\sigma_{ii}\hat{\sigma}_{ii}^{-2}-1|.
\end{align}
 First, use the inequality $P(N(0, 1)\geq x)\leq e^{-x^2/2}$ for $x>0$ to see
\begin{align*}
P\big(\max_{1\le i\le p}|Z_i|\geq 2\sqrt{\log p}\big)\leq p \cdot P(|N(0, 1)|\geq 2\sqrt{\log p}) \leq \frac{2}{p}.
\end{align*}
Thus,
\begin{align}\lbl{xindeyitian}
(\max_{1\le i\le p}|Z_i|)^2=O_p(\log p).
\end{align}
Based on the explanation below \eqref{hero_grass}, we have
\begin{align}\lbl{nishuodeshi}
\sigma_{ii}^{-1}\hat{\sigma}_{ii}^{2} \sim \frac{1}{n}\chi^2(n-1)
\end{align}
for each $i$. Set $a_p=\alpha\sqrt{n^{-1}\log p}$ with the constant $\alpha$ to be determined. Then
\begin{align*}
P\big(\max_{1\le i\le p} |\sigma_{ii}\hat{\sigma}_{ii}^{-2}-1|\geq a_p\big)
\leq & p\cdot P\Big(\Big|\frac{n}{\chi^2(n-1)}-1\Big|\geq a_p\Big)\\
\leq & p\cdot P\Big(|\chi^2(n-1)-n|\geq \frac{na_p}{2}\Big) + p\cdot P\Big(\chi^2(n-1)< \frac{1}{2}n\Big)
\end{align*}
by considering $\chi^2(n-1)< \frac{1}{2}n$ or not. Recall the Chernoff bound and the moderate deviation for sum of i.i.d. random variables (see, for example, p.31 and p.109 from \cite{DZ1998}). There exists a constant $C>0$ such that
$P(\chi^2(n-1)< \frac{1}{2}n)\leq e^{-Cn}$ for all $n\geq 1$ and
\begin{align}\lbl{LDPDZ}
P\Big(|\chi^2(n-1)-n|\geq \frac{na_p}{2}\Big)\leq P\Big(\Big|\frac{\chi^2(m)-m}{\sqrt{m\log p}}\Big|\geq \frac{\alpha}{3}\Big) \leq \exp\Big(-\frac{1}{3}\cdot \frac{\alpha^2}{9}\log p\Big)
\end{align}
as $p$ is sufficiently large, where $m:=n-1$ and the fact $\log p=o(n)$ is used in the last step. Choose $\alpha=8$ to bound the above probability by $O(p^{-2})$. It follows that
\begin{align*}
P\big(\max_{1\le i\le p} |\sigma_{ii}\hat{\sigma}_{ii}^{-2}-1|\geq a_p\big) =p\cdot O(p^{-2}) + e^{\log p -Cn} \to 0,
\end{align*}
as $p\to \infty$. This says
\begin{align}\lbl{shuishou}
\max_{1\le i\le p} |\sigma_{ii}\hat{\sigma}_{ii}^{-2}-1|=o_p(\sqrt{n^{-1}\log p}).
\end{align}
This together with \eqref{haoya} and  \eqref{xindeyitian} implies $T_{max}^{(1)}-\tilde{T}_{max}^{(1)}=o_p((n^{-1}\log^3 p)^{1/2})\to 0$ as long as $\log p=o(n^{1/3})$. This confirms $T_{max}^{(1)}-\tilde{T}_{max}^{(1)}=o_p(1)$, and the proof of part (ii) is completed by using \eqref{xiangshou}.

Finally we prove part (iii).
According to the proof of Theorem 3.1 in \cite{srivastava2009test} or the proof of Theorem 1 in \cite{jiang2021mean}, we have
\begin{align}\lbl{taiyanga1}
T_{sum}^{(1)}=\tilde{T}_{sum}^{(1)}+o_p(1)=\frac{n\bar{\X}^T \D^{-1}\bar{\X}-p}{\sqrt{2\tr(\R^2)}}+o_p(1).
\end{align}
Also, using an conclusion from the proof of (ii) above,
\begin{align}\lbl{taiyanga2}
T_{max}^{(1)}=\tilde{T}_{max}^{(1)}+o_p(1)=\max_{1\leq i \leq p}\frac{\sqrt{n}\bar{\X}_i}{\sqrt{\sigma_{ii}}}+o_p(1).
\end{align}
Since $\sqrt{n}(\bar{\X}_1,\cdots,\bar{\X}_p)^T=\sqrt{n}\bar{\X}^T \sim N(\bd{0}, \bms)$, then $\sqrt{n}\D^{-1/2}\bar{\X}\sim N(\bd{0}, \bd{R})$ by using the notation $\R=\D^{-1/2}\bms\D^{-1/2}.$ Recall an earlier notation $Z_i=\sqrt{n}\bar{\X}_i/\sqrt{\sigma_{ii}}\sim N(0, 1)$ for each $1\leq i \leq p$. Obviously, $(Z_1, \cdots, Z_p)^T=\sqrt{n}\D^{-1/2}\bar{\X}\sim N(\bd{0}, \bd{R})$. We are able to rewrite \eqref{taiyanga1} and \eqref{taiyanga2} in terms of $Z$'s as
\begin{align*}
T_{sum}^{(1)}=\frac{Z_1^2+\cdots +Z_p^2-p}{\sqrt{2\tr(\R^2)}}+o_p(1)\ \ \mbox{and}\ \
T_{max}^{(1)}=\max_{1\leq i \leq p}Z_i^2+o_p(1).
\end{align*}
As aforementioned, Assumption \eqref{assumption_A3} is stronger than Assumption \eqref{condition_2}. We then conclude (iii) by Theorem \ref{theorem_3}, Lemma \ref{small_lemma} and the fact $(Z_1, \cdots, Z_p)^T \sim N(\bd{0}, \bd{R})$.
\hfill$\square$

\medskip

\noindent{\bf Verification of $\beta^{(1)}_{M}(\bmu,\alpha)\approx \alpha$.} Recall the simplified assumption that $\bms=(\sigma_{ij})_{p\times p}=\I_p$,   $\xi\in (1/2, 5/6]$ and $\delta=O(n^{-\xi})$. In this case,  $n^{1-2\xi}\to 0$. We also assume that $\log p=o(n^{\xi-\frac{1}{2}})$. Because of the condition $\xi\in (1/2, 5/6]$, we know $\log p=o(n^{1/3})$. As a consequence, the requirement on $p$ vs $n$ imposed in Theorem \ref{thone}(ii) is satisfied. Notice $\beta^{(1)}_{M}(\bmu,\alpha)$ is equal to
\begin{align*}
%&\beta^{(1)}_{M}(\bmu,\alpha)=
&P\left(T_{max}^{(1)}-2\log p+2\log\log p>q_{\alpha}\right)\\
=&P\left(n\max_{1\le i\le p}\frac{\bar{\X}_i^2}{\hat{\sigma}_{ii}^2}-2\log p+2\log\log p>q_{\alpha}\right)\\
=&P\left(n\max_{1\le i\le p}\frac{(\bar{\X}_i-\delta)^2+\delta^2+2\delta(\bar{\X}_i-\delta)}{\hat{\sigma}_{ii}^2}-2\log p+2\log\log p>q_{\alpha}\right)\\
\le &P\left(n\max_{1\le i\le p}\frac{(\bar{\X}_i-\delta)^2}{\hat{\sigma}_{ii}^2}+n\max_{1\le i\le p}\frac{\delta^2}{\hat{\sigma}_{ii}^2}+n\max_{1\le i\le p}\frac{2\delta|\bar{\X}_i-\delta|}{\hat{\sigma}_{ii}^2}-2\log p+2\log\log p>q_{\alpha}\right).
%\le &P\left(n\max_{1\le i\le p}\frac{(\bar{\X}_i-\delta)^2}{\hat{\sigma}_{ii}^2}+O_p(n^{1-2\xi})+O_p(n^{\frac{1}{2}-\xi}\log p)-2\log p+2\log\log (p)>q_{\alpha}\right)\\
%\le &P\left(n\max_{1\le i\le p}\frac{(\bar{\X}_i-\delta)^2}{\hat{\sigma}_{ii}^2}+o_p(1)-2\log p+2\log\log (p)>q_{\alpha}\right)\to\alpha.
\end{align*}
Since $\sigma_{ii}=1$ by assumption, we have from \eqref{shuishou} that $\max_{1\le i\le p}|\hat{\sigma}_{ii}^{-2}-\sigma_{ii}^{-1}|=O_p(\sqrt{n^{-1}\log p})$. In particular,  we have from the triangle inequality that
\begin{align}\lbl{anjing}
&&\max_{1\leq i \leq p}\hat{\sigma}_{ii}^{-1}\leq 1+\max_{1\leq i \leq p}\frac{|\hat{\sigma}_{ii}^{-2}-1|}{\hat{\sigma}_{ii}^{-1}+1}\leq 1+ \max_{1\leq i \leq p}|\hat{\sigma}_{ii}^{-2}-1|=1+O_p(\sqrt{n^{-1}\log p}).
\end{align}
From the fact  $\max_{1\le i\le p}|\hat{\sigma}_{ii}^{-2}-\sigma_{ii}^{-1}|=O_p(\sqrt{n^{-1}\log p})$, we see
\begin{align*}
n\max_{1\le i\le p}\frac{\delta^2}{\hat{\sigma}_{ii}^2}&\le n\max_{1\le i\le p}\frac{\delta^2}{{\sigma}_{ii}}+n\max_{1\le i\le p}{\delta^2}\left|\hat{\sigma}_{ii}^{-2}-{\sigma}_{ii}^{-1}\right|\\
&=O_p(n^{1-2\xi})+O_p(n^{1-2\xi}\sqrt{n^{-1}\log p})=O_p(n^{1-2\xi}).
\end{align*}
According to Theorem \ref{thone}(ii), we have $\max_{1\le i\le p}\frac{|\bar{\X}_i-\delta|}{\hat{\sigma}_{ii}}=O_p(\sqrt{(\log p)/n})$. This and \eqref{anjing} conclude that
\begin{align*}
n\max_{1\le i\le p}\frac{2\delta|\bar{\X}_i-\delta|}{\hat{\sigma}_{ii}^2}=O_p\big(n^{\frac{1}{2}-\xi}\sqrt{\log p}\big).
\end{align*}
Thus,
\begin{align*}
&\beta^{(1)}_{M}(\bmu,\alpha)\\
\le &P\left(n\max_{1\le i\le p}\frac{(\bar{\X}_i-\delta)^2}{\hat{\sigma}_{ii}^2}+O_p(n^{1-2\xi})+O_p(n^{\frac{1}{2}-\xi}\log p)-2\log p+2\log\log (p)>q_{\alpha}\right)\\
\le &P\left(n\max_{1\le i\le p}\frac{(\bar{\X}_i-\delta)^2}{\hat{\sigma}_{ii}^2}+o_p(1)-2\log p+2\log\log (p)>q_{\alpha}\right)
\end{align*}
which goes to $\alpha.$ The verification is completed. \hfill$\square$

\noindent{\bf Proof of Theorem \ref{thtwo}.} % The proof of Theorem \ref{thtwo} is similar to the proof of Theorem \ref{thone}, we omit it here.
The proof shares same spirit as Theorem \ref{thone}. Denote $n=n_1+n_2$. According to Section 5 in \cite{srivastava2008test} or Theorem 2 in \cite{jiang2021mean}, (i) holds. We prove (ii) next.

Under the normality assumption and the null hypothesis in  \eqref{ht}, we have $\bar{\X}_1-\bar{\X}_2\sim N(\bd{0},\frac{n_1+n_2}{n_1n_2}\bms)$. Let $\D=\mbox{diag}(\sigma_{11}, \cdots, \sigma_{pp})$ be the diagonal matrix of $\bms$. Recall $\R=\D^{-1/2}\bms\D^{-1/2}.$  Then
\begin{align}\lbl{tongxuezuo}
(Z_1, \cdots, Z_p)^T:=\Big(\frac{n_1n_2}{n_1+n_2}\Big)^{1/2}\D^{-1/2}(\bar{\X}_1-\bar{\X}_2)
\sim N(\bd{0}, \bd{R}).
\end{align}
According to Section 5 and the proof of Theorem 2.1 in \cite{srivastava2008test} or the proof of Theorem 2 in \cite{jiang2021mean}, we have
\begin{align}\lbl{dazhang}
T_{sum}^{(2)}
=&\frac{\frac{n_1n_2}{n_1+n_2}(\bar{\X}_1-\bar{\X}_2)^T \D^{-1}(\bar{\X}_1-\bar{\X}_2)-p}{\sqrt{2\tr(\R^2)}}+o_p(1) \nonumber\\
=& \frac{Z_1^2+\cdots +Z_p^2}{\sqrt{2\tr(\R^2)}}+o_p(1).
\end{align}
Let $\bar{\X}_{ji}$ be the $i$th coordinate of $\bar{\X}_j\in \mathbb{R}^p$ for $j=1,2$ and $1\leq i \leq p$. Then
\begin{align*}
\Big(\frac{n_1n_2}{n_1+n_2}\Big)^{1/2}\Big(\frac{\bar{\X}_{11}-\bar{\X}_{21}}{\sqrt{\sigma_{11}}}, \cdots, \frac{\bar{\X}_{1p}-\bar{\X}_{2p}}{\sqrt{\sigma_{pp}}}\Big)^T=(Z_1, \cdots, Z_p)^T.
\end{align*}
Recall
\begin{align*}
T^{(2)}_{max}=\frac{n_1n_2}{n_1+n_2}\max_{1\le i\le p}\frac{(\bar{\X}_{1i}-\bar{\X}_{2i})^2}{\hat{\sigma}_{ii}^2},
\end{align*}
where $\hat{\sigma}_{ii}^2$ is the $i$th diagonal element of $\hat{\S}$ in (\ref{duck_yazi}). Set
\begin{align*}
\tilde{T}_{max}^{(2)}=\frac{n_1n_2}{n_1+n_2}\max_{1\le i\le p}\frac{(\bar{\X}_{1i}-\bar{\X}_{2i})^2}{\sigma_{ii}}.
\end{align*}
Then $\tilde{T}_{max}^{(2)}=\max_{1\leq i \leq p}Z_i^2.$ Recalling the discussion below  \eqref{assumption_A3},  Assumption \eqref{assumption_A3} is stronger than Assumption \eqref{condition_2}. By Theorem \ref{theorem_2} and Assumption \eqref{condition_2}, we obtain
\begin{align}\lbl{xiangzi}
&\tilde{T}_{max}^{(2)}-2\log p +\log\log p=\max_{1\leq i \leq p}Z_i^2-2\log p +\log\log p\ \mbox{
converges} \nonumber\\
& \mbox{weakly to a distribution with cdf}\ F(x)=\exp\big\{-e^{-x/2}/\sqrt{\pi}\big\},\ x\in \mathbb{R}.
\end{align}
Thus, to prove (ii), it suffices to show that $T_{max}^{(2)}-\tilde{T}_{max}^{(2)}=o_p(1)$. By (\ref{duck_yazi}), $(n_1+n_2)\hat{\S}$ follows a Wishart distribution with parameter $n_1+n_2-2$ and covariance matrix $\bms$. Since $\hat{\sigma}_{ii}^2$ is the $i$th diagonal element of $\hat{\S}$, we know $(n_1+n_2)\hat{\sigma}_{ii}^2 \sim \sigma_{ii}\chi^2(n_1+n_2-2)$, or equivalently,
\begin{align*}
\hat{\sigma}_{ii}^2\sigma_{ii}^{-1} \sim \frac{\chi^2(n_1+n_2-2)}{n_1+n_2}
\end{align*}
for each $1\leq i \leq p.$ By the same argument as  between \eqref{nishuodeshi} and \eqref{shuishou}, we have from the above assertion that
$\max_{1\le i\le p}|\hat{\sigma}_{ii}^{-2}\sigma_{ii}-1|=O_p(\sqrt{n^{-1}\log p})$.
Notice \eqref{xiangzi} implies $\tilde{T}_{max}^{(2)}=O(\log p)$. By the triangle inequality of the maximum and the trivial inequality $\max_{1\leq i \leq p}|a_ib_i|\leq \max_{1\leq i \leq p}|a_i|\cdot \max_{1\leq i \leq p}|b_i|$ for any $\{a_i\}$ and $\{b_i\}$, we get
\begin{align}\lbl{nashahuo}
|T_{max}^{(2)}-\tilde{T}_{max}^{(2)}|=& \left|\frac{n_1n_2}{n_1+n_2}\max_{1\le i\le p}\hat{\sigma}_{ii}^{-2}(\bar{\X}_{1i}-\bar{\X}_{2i})^2-\frac{n_1n_2}{n_1+n_2}\max_{1\le i\le p}\sigma_{ii}^{-1}(\bar{\X}_{1i}-\bar{\X}_{2i})^2\right| \nonumber\\
\le &|\tilde{T}_{max}^{(2)}|\cdot
% \left|\frac{n_1n_2}{n_1+n_2}\max_{1\le i\le %p}\sigma_{ii}^{-2}(\bar{\X}_{1i}-\bar{\X}_{2i})^2\right|
\max_{1\le i\le p} |\hat{\sigma}_{ii}^{-2}\sigma_{ii}-1| \nonumber\\
=& O\big(n^{-1/2}(\log p)^{3/2}\big)\to 0.
\end{align}
Consequently, (ii) follows from \eqref{xiangzi} under the assumption $\log p=o(n^{1/3})$.

Now we prove (iii). Recall \eqref{dazhang}. By \eqref{nashahuo}, we see
\begin{align*}
\tilde{T}_{max}^{(2)}-2\log p+\log\log p=\max_{1\leq i \leq p}Z_i^2-2\log p +\log\log p +o_p(1).
\end{align*}
As discussed earlier, Assumption \eqref{assumption_A3} is stronger than Assumption \eqref{condition_2}. Then, under Assumption \eqref{assumption_A3} with ``$\bd{\Sigma}$'' replaced by ``$\bd{R}$'', we conclude (iii) from Theorem \ref{theorem_3}, \eqref{tongxuezuo} and Lemma \ref{small_lemma}.\hfill$\Box$

\subsection{Proof of Theorem \ref{thlm}}

%\subsection{Proof of Proposition ??}
To prove Theorem \ref{thlm}, we need a preparation. In fact,  an asymptotic ratio-consistent estimator of $\mbox{tr}\big(\bms_{b|a}^2\big)$ will be derived (the notation of ``$\bms_{b|a}$" is given in \eqref{youshayi}). It is stated in Proposition S\ref{sonfather}. We will develop a series of auxiliary results for this purpose.

Review the setting in Section \ref{regress_3}. In what follows, we assume the integers $n$, $p$ and $q$ satisfy $1\leq q <p$ and $q<n$.

\begin{lemma}\lbl{yizhixiaoniao} Let $\X_1, \cdots, \X_n$ be i.i.d. from the $p$-dimensional distribution $N(\bd{0}, \bd{\Sigma}).$ Write
\begin{align*}
\X_i
=\begin{pmatrix}
\bd{X}_{ia}\\
\bd{X}_{ib}
\end{pmatrix}
\ \ \ \mbox{and} \ \ \
\bd{\Sigma}=
\begin{pmatrix}
\bd{\Sigma}_{aa} & \bd{\Sigma}_{ab}\\
\bd{\Sigma}_{ba} & \bd{\Sigma}_{bb}
\end{pmatrix}
\end{align*}
for each $i=1, \cdots, n$, where $\bd{X}_{ia}$ is a $q$-dimensional vector with distribution $N(\bd{0}, \bd{\Sigma}_{aa})$. Then $\bd{\Sigma}_{bb\cdot a}:=\bd{\Sigma}_{bb}-\bd{\Sigma}_{ba}\bd{\Sigma}_{aa}^{-1}\bd{\Sigma}_{ab}$ is a $(p-q)\times (p-q)$ matrix. Recall $\hat{\bms}_{b|a}$  in \eqref{hondulasi}. We then have
\begin{align*}
\hat{\bms}_{b|a}\overset{d}{=}
   \frac{1}{n}\bd{\Sigma}_{bb\cdot a}^{1/2}\cdot \bd{W}\bd{W}^T\bd{\Sigma}_{bb\cdot a}^{1/2}
\end{align*}
where $\bd{W}$ is a $(p-q)\times (n-q)$ matrix and the entries are i.i.d. $N(0, 1).$
\end{lemma}
\noindent\textbf{Proof of Lemma \ref{yizhixiaoniao}}. Recall the notation between \eqref{h1j} and \eqref{hondulasi},
\begin{align*}
&& \X_{a}=(\X_{1a}, \cdots, \X_{na})^T,\ \ \H_a=\mX_a(\mX_a^{T} \mX_a)^{-1}\mX_a^{T},\ \ \X_{b}=(\X_{1b}, \cdots, \X_{nb})^T,\\
&&  \wX_b=(\bd{I}_n-\H_a)\X_{b},\ \ \hat{\bms}_{b|a}=n^{-1}\wX_b^{T} \wX_b,
\end{align*}
where  $\X_{a}$ is $n\times q$, $\H_a$ is $n\times n$, both $\X_{b}$ and $\wX_b$ are $n\times (p-q)$ and $\hat{\bms}_{b|a}$ is $(p-q)\times (p-q)$. $\hat{\bms}_{b|a}$ is defined as
\begin{align}\lbl{qianfang}
\hat{\bms}_{b|a}=\frac{1}{n}\wX_b^{T} \wX_b=\frac{1}{n}\X_{b}^T(\bd{I}_n-\H_a)\X_{b}.
\end{align}
Then, by Lemma \ref{sunny_coffee}, the $(p-q)$-dimensional random vector $\bd{X}_{ib}-\bd{\Sigma}_{ba}\bd{\Sigma}_{aa}^{-1}\bd{X}_{ia} \sim N(\bd{0}, \bd{\Sigma}_{bb\cdot a})$ and is independent of $\bd{X}_{ia}$.
% , where $\bd{\Sigma}_{bb\cdot a}:=\bd{\Sigma}_{bb}-\bd{\Sigma}_{ba}\bd{\Sigma}_{aa}^{-1}\bd{\Sigma}_{ab}$
% is a $(p-q)\times (p-q)$ matrix.
It follows that the conditional distribution of $\bd{X}_{ib}$ given $\bd{X}_{ia}$ is characterized by
\begin{align*}
\ml{L}(\bd{X}_{ib}|\bd{X}_{ia})= N(\bd{\Sigma}_{ba}\bd{\Sigma}_{aa}^{-1}\bd{X}_{ia}, \bd{\Sigma}_{bb\cdot a}).
\end{align*}
Therefore, we have from the definition $\bms_{b|a}=E\mbox{Cov}(\bd{X}_{1b}|\bd{X}_{1a})$ that
\begin{align}\lbl{youshayi}
\bms_{b|a}=\bd{\Sigma}_{bb\cdot a}.
\end{align}
Write
\begin{align*}
\X_{b}^T=\big(\bd{X}_{1b}-\bd{\Sigma}_{ba}\bd{\Sigma}_{aa}^{-1}\bd{X}_{1a}, \cdots, \bd{X}_{nb}-\bd{\Sigma}_{ba}\bd{\Sigma}_{aa}^{-1}\bd{X}_{na}\big) + \big(\bd{\Sigma}_{ba}\bd{\Sigma}_{aa}^{-1}\bd{X}_{1a}, \cdots, \bd{\Sigma}_{ba}\bd{\Sigma}_{aa}^{-1}\bd{X}_{na}\big).
\end{align*}
Notice the last two vectors are both normal and they are independent since that $\X_1, \cdots, \X_n$ are i.i.d. from the $p$-dimensional population  $N(\bd{0}, \bd{\Sigma})$. Moreover, we can also write
\begin{align}\lbl{haoshiduo}
\X_{b}^T=(\bd{V}_1,\cdots, \bd{V}_n)  + \bd{\Sigma}_{ba}\bd{\Sigma}_{aa}^{-1}\bd{X}_a^T
\end{align}
where $\bd{V}_1, \cdots, \bd{V}_n$ are i.i.d. $(p-q)$-dimensional random vectors with distribution $N(\bd{0}, \bd{\Sigma}_{bb\cdot a})$ also independent of $\bd{X}_a$. By definition \eqref{qianfang},
\begin{align}\lbl{xiao_guo}
\hat{\bms}_{b|a}
= &\frac{1}{n}\X_{b}^T(\bd{I}_n-\H_a)\X_{b} \nonumber\\
 = & \frac{1}{n}\big[(\bd{V}_1,\cdots, \bd{V}_n)  + \bd{\Sigma}_{ba}\bd{\Sigma}_{aa}^{-1}\bd{X}_a^T\big](\bd{I}_n-\H_a)
\big[(\bd{V}_1,\cdots, \bd{V}_n)  + \bd{\Sigma}_{ba}\bd{\Sigma}_{aa}^{-1}\bd{X}_a^T\big]^T\nonumber\\
 = & \frac{1}{n}(\bd{V}_1,\cdots, \bd{V}_n)(\bd{I}_n-\H_a)(\bd{V}_1,\cdots, \bd{V}_n)^T,
\end{align}
since $\bd{X}_a^T(\bd{I}_n-\H_a)=\bd{0}$. Let $\{\xi_{ij};\, 1\leq i \leq p-q, 1\leq j \leq n \}$ be i.i.d. $N(0, 1)$-distributed random variables independent of $\bd{X}_a$.  Without loss of generality, assume  $\bd{V}_j=\bd{\Sigma}_{bb\cdot a}^{1/2}\cdot (\xi_{ij})_{(p-q)\times 1}$ for each $j$. Therefore,
\begin{align}\lbl{hanshane}
(\bd{V}_1,\cdots, \bd{V}_n)=\bd{\Sigma}_{bb\cdot a}^{1/2}\cdot (\xi_{ij})_{(p-q)\times n}.
\end{align}
% where $\{\xi_{ij};\, 1\leq i \leq p-q, 1\leq j \leq n \}$ are i.i.d. $N(0, 1)$-distributed random variables.
Since the $n\times n$ idempotent matrix $\H_a$ has rank $q$, we know $\bd{I}_n-\H_a$ has  rank $n-q$. As a function of $\X_{a}$, $\bd{I}_n-\H_a$ is independent of  $(\bd{V}_1,\cdots, \bd{V}_n)$. As a result, there exists an $n\times n$  random orthogonal matrix $\bd{\Gamma}$ independent of $\{\xi_{ij};\, 1\leq i \leq p-q, 1\leq j \leq n \}$ such that
\begin{align*}
 \bd{I}_n-\H_a=\bd{\Gamma}
\begin{pmatrix}
\bd{I}_{(n-q)\times (n-q)} & \bd{0}\\
\bd{0} & \bd{0}_{q\times q}
\end{pmatrix}
\bd{\Gamma}^T
,
\end{align*}
where the three $\bd{0}$ above  are matrices of entries $0$ with proper size. By the orthogonal invariance of i.i.d. standard normals, $(\xi_{ij})_{(p-q)\times n}\bd{\Gamma}$ has the same distribution as that of $(\xi_{ij})_{(p-q)\times n}$. From \eqref{xiao_guo} and \eqref{hanshane}, we have
\begin{align*}
\hat{\bms}_{b|a}\overset{d}{=}&\frac{1}{n}\bd{\Sigma}_{bb\cdot a}^{1/2}\cdot (\xi_{ij})_{(p-q)\times n}\begin{pmatrix}
\bd{I}_{(n-q)\times (n-q)} & \bd{0}\\
\bd{0}  \bd{0}_{q\times q}
\end{pmatrix}
(\xi_{ij})_{(p-q)\times n}^T\bd{\Sigma}_{bb\cdot a}^{1/2}\\
 = & \frac{1}{n}\bd{\Sigma}_{bb\cdot a}^{1/2}\cdot \bd{W}\bd{W}^T\bd{\Sigma}_{bb\cdot a}^{1/2}
\end{align*}
where $\bd{W}=(\xi_{ij})_{(p-q)\times (n-q)}$. \hfill$\square$

\begin{lemma}\lbl{quwanle} Recall the notations in Lemma \ref{yizhixiaoniao} and \eqref{youshayi}. Let $\lambda_1, \cdots \lambda_{p-q}$ be the eigenvalues of $\bms_{b|a}=\bd{\Sigma}_{bb\cdot a}$. Let $\{\bd{w}_i;\, 1\leq i \leq p-q\}$ be i.i.d. $(n-q)$-dimensional vectors whose entries are i.i.d. $N(0,1).$ Then $\mbox{tr}\big(\hat{\bms}_{b|a}\big) \overset{d}{=}\frac{1}{n}\sum_{i=1}^{p-q}\lambda_i
\|\bd{w}_i\|^2$
and
\begin{align*}
\mbox{tr}\big(\hat{\bms}_{b|a}^2\big) \overset{d}{=}
\frac{1}{n^2}\sum_{i=1}^{p-q}\lambda_i^2(\bd{w}_i^T\bd{w}_i)^2 + \frac{2}{n^2}\sum_{1\leq i< j\leq p-q}\lambda_i\lambda_j(\bd{w}_i^T\bd{w}_j)^2.
\end{align*}
In particular,
\begin{align}\lbl{buhaoyisi}
E\mbox{tr}\big(\hat{\bms}_{b|a}^2\big)
=\frac{n-q}{n^2}\cdot\big[\mbox{tr}(
\bms_{b|a})\big]^2 + \frac{(n-q)(n-q+1)}{n^2}\mbox{tr}\big(
\bms_{b|a}^2\big).
\end{align}
\end{lemma}
\noindent\textbf{Proof of Lemma \ref{quwanle}}. Let $\bd{W}$ be a $(p-q)\times (n-q)$ matrix whose entries are i.i.d. $N(0, 1).$ Immediately, $E\hat{\bms}_{b|a}= \frac{1}{n}\bd{\Sigma}_{bb\cdot a}^{1/2}\cdot E(\bd{W}\bd{W}^T)\bd{\Sigma}_{bb\cdot a}^{1/2}=\frac{n-q}{n}\cdot\bd{\Sigma}_{bb\cdot a}$. Using the identity  $\mbox{tr}(\bd{A}\bd{B})=\mbox{tr}(\bd{B}\bd{A})$ for any matrices $\bd{A}$ and $\bd{B}$, we know that
\begin{align}\lbl{Havard_xiaoyou}
\mbox{tr}\big(\hat{\bms}_{b|a}\big)\overset{d}{=}&\frac{1}{n}\mbox{tr}\big(\bd{\Sigma}_{bb\cdot a}^{1/2}\cdot \bd{W}\bd{W}^T\bd{\Sigma}_{bb\cdot a}^{1/2}\big)\nonumber\\
=& \frac{1}{n}\mbox{tr}\big(\bd{W}^T\bd{\Sigma}_{bb\cdot a}\bd{W}\big)
\end{align}
and
\begin{align}\lbl{stan1}
\mbox{tr}\big(\hat{\bms}_{b|a}^2\big)\overset{d}{=}&\frac{1}{n^2}\mbox{tr}\big[\bd{\Sigma}_{bb\cdot a}^{1/2}\cdot \bd{W}\bd{W}^T\bd{\Sigma}_{bb\cdot a}\bd{W}\bd{W}^T\bd{\Sigma}_{bb\cdot a}^{1/2}\big]\nonumber\\
=& \frac{1}{n^2}\mbox{tr}\big[(\bd{W}^T\bd{\Sigma}_{bb\cdot a}\bd{W})^2\big].
\end{align}
Since $\lambda_1, \cdots, \lambda_{p-q}$ are the eigenvalues of $\bd{\Sigma}_{bb\cdot a}$, we are able to decompose $\bd{\Sigma}_{bb\cdot a}=\bd{\Gamma}_1^T\bd{\Lambda}\bd{\Gamma}_1$, where $\bd{\Gamma}_1$ is an orthogonal matrix and $\bd{\Lambda}=\mbox{diag}(\lambda_1, \cdots, \lambda_{p-q})$. By the orthogonal invariance of i.i.d. $N(0, 1)$-entries, we get
\begin{align*}%\lbl{zhishi}
\mbox{tr}\big(\hat{\bms}_{b|a}\big) \overset{d}{=}
\frac{1}{n}\mbox{tr}\big[(\bd{W}^T\bd{\Lambda}\bd{W})\big]\ \ \mbox{and}\ \ \mbox{tr}\big(\hat{\bms}_{b|a}^2\big) \overset{d}{=}
\frac{1}{n^2}\mbox{tr}\big[(\bd{W}^T\bd{\Lambda}\bd{W})^2\big]. \nonumber
\end{align*}
Furthermore write $\bd{W}^T=(\bd{w}_1, \cdots, \bd{w}_{p-q})$. Then $\bd{w}_1, \cdots, \bd{w}_{p-q}$ are i.i.d. $(n-q)$-dimensional vectors of distribution $N(\bd{0}, \bd{I}_{n-q})$. Hence,
\begin{align*}%\lbl{haiyouni}
\bd{W}^T\bd{\Lambda}\bd{W}=\sum_{i=1}^{p-q}\lambda_i\bd{w}_i\bd{w}_i^T, \nonumber
\end{align*}
which gives
\begin{align}\lbl{tin_beizi}
\mbox{tr}\big(\hat{\bms}_{b|a}\big) \overset{d}{=} \frac{1}{n}\sum_{i=1}^{p-q}\lambda_i
\|\bd{w}_i\|^2.
\end{align}
Additionally, we have
\begin{align*}
(\bd{W}^T\bd{\Lambda}\bd{W})^2=\sum_{i=1}^{p-q}\lambda_i^2(\bd{w}_i\bd{w}_i^T)^2 + 2\sum_{1\leq i< j\leq p-q}\lambda_i\lambda_j(\bd{w}_i\bd{w}_i^T)(\bd{w}_j\bd{w}_j^T).
\end{align*}
Observe
\begin{align*}
\mbox{tr}[(\bd{w}_i\bd{w}_i^T)^2]=\mbox{tr}[(\bd{w}_i\bd{w}_i^T)\bd{w}_i\bd{w}_i^T]
=\mbox{tr}[(\bd{w}_i^T\bd{w}_i\bd{w}_i^T\bd{w}_i]=(\bd{w}_i^T\bd{w}_i)^2.
\end{align*}
Similarly, $\mbox{tr}[(\bd{w}_i\bd{w}_i^T)(\bd{w}_j\bd{w}_j^T)]=(\bd{w}_i^T\bd{w}_j)^2$. Then we end up with
\begin{align*}
\mbox{tr}\big[(\bd{W}^T\bd{\Lambda}\bd{W})^2\big] = \sum_{i=1}^{p-q}\lambda_i^2(\bd{w}_i^T\bd{w}_i)^2 + 2\sum_{1\leq i< j\leq p-q}\lambda_i\lambda_j(\bd{w}_i^T\bd{w}_j)^2.
\end{align*}
It follows from \eqref{stan1} that
\begin{align}\lbl{bangmang}
\mbox{tr}\big(\hat{\bms}_{b|a}^2\big)\overset{d}{=}&
\frac{1}{n^2}\sum_{i=1}^{p-q}\lambda_i^2(\bd{w}_i^T\bd{w}_i)^2 + \frac{2}{n^2}\sum_{1\leq i< j\leq p-q}\lambda_i\lambda_j(\bd{w}_i^T\bd{w}_j)^2.
\end{align}
Since $\bd{w}_1, \cdots, \bd{w}_{p-q}$ are i.i.d. $N(\bd{0}, \bd{I}_{n-q})$, we know $\bd{w}_i^T\bd{w}_i\sim \chi^2(n-q)$ and $\bd{w}_i^T\bd{w}_j \overset{d}{=} \|\bd{w}_i\|\cdot \eta$, where $\eta \sim N(0, 1)$ and is independent of $\|\bd{w}_i\|$. Recall $E\chi^2(m)=m$ and  $\mbox{Var}(\chi^2(m))=2m$ for any integer $m\geq 1$. Thus, by independence we have $E(\|\bd{w}_i\|^2\cdot \eta^2)=n-q.$ From \eqref{bangmang}, we obtain
\begin{align*}
E\mbox{tr}\big(\hat{\bms}_{b|a}^2\big)=\frac{1}{n^2}\cdot \big[2(n-q)+(n-q)^2\big]\sum_{i=1}^{p-q}\lambda_i^2
+ \frac{2}{n^2}\cdot (n-q)\sum_{1\leq i< j\leq p-q}\lambda_i\lambda_j.
\end{align*}
Using the identity
\begin{align*}
2\sum_{1\leq i< j\leq p-q}\lambda_i\lambda_j=\Big(\sum_{i=1}^{p-q}\lambda_i\Big)^2-\sum_{i=1}^{p-q}\lambda_i^2,
\end{align*}
we arrive at
\begin{align*}
E\mbox{tr}\big(\hat{\bms}_{b|a}^2\big)
=\frac{n-q}{n^2}\cdot\big[\mbox{tr}(
\bms_{b|a})\big]^2 + \frac{(n-q)(n-q+1)}{n^2}\mbox{tr}\big(
\bms_{b|a}^2\big)
\end{align*}
by using \eqref{youshayi}. This completes the proof of the lemma. \hfill$\square$

\begin{lemma}\lbl{tianneji1}
%Set $r=n-q$.
Let $r\geq 1$ and $\bd{w}_1$ and $\bd{w}_2$ be i.i.d. $r$-dimensional random vectors with distribution $N(\bd{0}, \bd{I}_r)$. Then the following are true.

(i) $\mbox{Var}((\bd{w}_1^T\bd{w}_1)^2)=8r(r+2)(r+3).$

(ii) $\mbox{Var}((\bd{w}_1^T\bd{w}_2)^2)=2r(r+3).$

(iii) $\mbox{Cov}((\bd{w}_1^T\bd{w}_2)^2, (\bd{w}_1^T\bd{w}_3)^2)=2r$.

(iv) $\mbox{Cov}((\bd{w}_1^T\bd{w}_1)^2, (\bd{w}_1^T\bd{w}_2)^2)=4r(r+2).$
\end{lemma}
\noindent\textbf{Proof of Lemma \ref{tianneji1}}. It is well-known that
\begin{align}\lbl{xiao_1}
E[\chi^2(r)^m]=r(r+2)\cdots (r+2m-2)
\end{align}
for all integers $m\geq 1$. We will use this formula to prove the results.

(i) Since $\bd{w}_1^T\bd{w}_1\sim \chi^2(r)$, we have
\begin{align}\lbl{nipinpin}
E(\|\bd{w}_1\|^4)=E(\bd{w}_1^T\bd{w}_1)^2=r(r+2)
\end{align}
and $E(\bd{w}_1^T\bd{w}_1)^4=r(r+2)(r+4)(r+6)$. This leads to
\begin{align*}
\mbox{Var}((\bd{w}_1^T\bd{w}_1)^2)
=&r(r+2)(r+4)(r+6)-[r(r+2)]^2\\
=&8r(r+2)(r+3).
\end{align*}

(ii) By Proposition 7.3 from \cite{eaton1983multivariate} or Theorem 1.5.6 from  \cite{muirhead1982aspects}, it holds that
\begin{align}\lbl{zhulidepy}
\|\bd{w}_i\|\ \ \mbox{and}\ \ \bd{e}_i:=\frac{\bd{w}_i}{\|\bd{w}_i\|}\ \ \mbox{are independent}
\end{align}
for $i=1, 2$. Therefore,  $(\bd{w}_1^T\bd{w}_2)^2=\|\bd{w}_1\|^2\cdot(\bd{e}_1^T\bd{w}_2)^2\overset{d}{=} \|\bd{w}_1\|^2\cdot \eta^2$, where    $\eta \sim N(0, 1)$  and $\eta$ is independent of $\|\bd{w}_1\|$. Consequently, by \eqref{nipinpin},
\begin{align*}
\mbox{Var}((\bd{w}_1^T\bd{w}_2)^2)
=& E(\|\bd{w}_1\|^4\cdot \eta^4)-\big[E(\|\bd{w}_1\|^2\cdot \eta^2)\big]^2\\
 = & r(r+2)\cdot 3-r^2\\
= & 2r(r+3).
\end{align*}

(iii) By \eqref{zhulidepy}, we have
\begin{align*}
(\bd{w}_1^T\bd{w}_2)^2 (\bd{w}_1^T\bd{w}_3)^2=\|\bd{w}_1\|^4\cdot (\bd{e}_1^T\bd{w}_2)^2\cdot (\bd{e}_1^T\bd{w}_3)^2.
\end{align*}
Easily, be checking their covariance, we know $\bd{e}_1^T\bd{w}_2$ and $\bd{e}_1^T\bd{w}_3$ are i.i.d. $N(0,1)$. This implies that $\|\bd{w}_1\|^4$,  $(\bd{e}_1^T\bd{w}_2)^2$ and  $(\bd{e}_1^T\bd{w}_3)^2$ are independent. Thus, from \eqref{nipinpin} we obtain
\begin{align*}
\mbox{Cov}((\bd{w}_1^T\bd{w}_2)^2, (\bd{w}_1^T\bd{w}_3)^2)
= & E\|\bd{w}_1\|^4\cdot 1\cdot 1-E(\bd{w}_1^T\bd{w}_2)^2\cdot E(\bd{w}_1^T\bd{w}_3)^2\\
= & r(r+2)-r^2\\
=&2r.
\end{align*}

(iv) Write $(\bd{w}_1^T\bd{w}_1)^2 (\bd{w}_1^T\bd{w}_2)^2=\|\bd{w}_1\|^6\cdot (\bd{e}_1^T\bd{w}_2)^2$. From the independence between $\|\bd{w}_1\|$ and $\bd{e}_1^T\bd{w}_2$, we conclude that
\begin{align*}
& \mbox{Cov}((\bd{w}_1^T\bd{w}_1)^2, (\bd{w}_1^T\bd{w}_2)^2)\\
=& E\|\bd{w}_1\|^6\cdot E(\bd{e}_1^T\bd{w}_2)^2-E(\bd{w}_1^T\bd{w}_1)^2\cdot E(\bd{w}_1^T\bd{w}_2)^2\\
 = & r(r+2)(r+4)- r(r+2)\cdot r\\
 = & 4r(r+2).
\end{align*}
Thus, the verification is completed. \hfill$\square$

\begin{lemma}\lbl{jianjicao}  Let $\hat{\bms}_{b|a}$ and $\bms_{b|a}$ be as in \eqref{hondulasi} and \eqref{youshayi}, respectively. Then
\begin{align*}
\mbox{Var}\big(\mbox{tr}\big(\hat{\bms}_{b|a}^2\big)\big)
\leq \frac{1024(n-q)(n+p-2q)^2}{n^4}\cdot \mbox{tr}(\bms_{b|a}^4).
%\frac{44(r+3)^3}{n^4}\cdot \mbox{tr}\big(\bms_{b|a}^4\big).
\end{align*}
\end{lemma}
\noindent\textbf{Proof of Lemma \ref{jianjicao}}. Set $r=n-q$ and $s=p-q$. Let $\lambda_1, \cdots \lambda_{s}$ be the eigenvalues of $\bms_{b|a}=\bd{\Sigma}_{bb\cdot a}$. Standard computation gives
\begin{align}\lbl{shuxue_sense}
\mbox{Var}(\gamma_1+\cdots +\gamma_s)=\sum_{i=1}^s\mbox{Var}(\gamma_i)+2\sum_{1\leq i<j\leq s}\mbox{Cov}(\gamma_i, \gamma_j),
\end{align}
for any random variables $\gamma_1,\cdots,\gamma_s$.   Then, from Lemma \ref{quwanle}
%Back to \eqref{bangmang}
we see
\begin{align}\lbl{shangz}
n^4\cdot \mbox{Var}\big(\mbox{tr}\big(\hat{\bms}_{b|a}^2\big)\big)
=&
\sum_{i=1}^{s}\lambda_i^4\,\mbox{Var}\big((\bd{w}_i^T\bd{w}_i)^2\big) +
4\sum_{1\leq i< j\leq s}\lambda_i^2\lambda_j^2\,\mbox{Var}\big((\bd{w}_i^T\bd{w}_j)^2\big) \nonumber\\
& +8\sum_{1\leq i< j\leq s}\sum_{(k, l)\in A_{i,j}}\lambda_i\lambda_j\lambda_k\lambda_l\,\mbox{Cov}\big((\bd{w}_i^T\bd{w}_j)^2, (\bd{w}_k^T\bd{w}_l)^2\big)\nonumber\\
 & + 4\sum_{1\leq i< j\leq s}\sum_{k\in \{i, j\}}\lambda_i\lambda_j\lambda_k^2\,
\mbox{Cov}\big((\bd{w}_i^T\bd{w}_j)^2, (\bd{w}_k^T\bd{w}_k)^2\big)\nonumber\\
:=& C_1+C_2+C_3+C_4,
\end{align}
where $A_{i,j}=\{(k, l)\ne (i,j);\, 1\leq k <l \leq s,\ \{k,l\}\cap\{i, j\}\ne\emptyset\}$. By Lemma \ref{tianneji1}(i),
\begin{align}\lbl{C1co}
C_1=8r(r+2)(r+3)\sum_{i=1}^{s}\lambda_i^4\leq 8(r+3)^3\cdot \mbox{tr}(\bms_{b|a}^4).
\end{align}
By Lemma \ref{tianneji1}(ii),
\begin{align}\lbl{cu_lu_x}
C_2 = & 4r(r+3)\cdot 2\sum_{1\leq i< j\leq s}\lambda_i^2\lambda_j^2 \nonumber\\
\leq& 4(r+3)^2\Big(\sum_{1\leq i\leq s}\lambda_i^2\Big)^2\nonumber\\
= & 4(r+3)^2\,\big[\mbox{tr}(\bms_{b|a}^2)\big]^2.
\end{align}
By Lemma \ref{tianneji1}(iii),
\begin{align}\lbl{copC3}
C_3 = & 2r\cdot 8\sum_{1\leq i< j\leq s}\sum_{(k, l)\in A_{i,j}}\lambda_i\lambda_j\lambda_k\lambda_l\nonumber\\
 \leq & 16r\sum_{1\leq i, j, k\leq s}\lambda_i\lambda_j\lambda_k^2\nonumber\\
 = & 16r\cdot \big[\mbox{tr}(\bms_{b|a})\big]^2\cdot \mbox{tr}(\bms_{b|a}^2).
\end{align}
By Lemma \ref{tianneji1}(iv),
\begin{align}\lbl{qiutianle}
C_4 \leq & 4r(r+2)\cdot 4\sum_{1\leq i< j\leq s}\sum_{k\in \{i, j\}}\lambda_i\lambda_j\lambda_k^2 \nonumber\\
 \leq & 16(r+2)^2\sum_{1\leq i, j\leq s}\lambda_i\lambda_j^3\nonumber\\
 = & 16(r+2)^2\big[\mbox{tr}(\bms_{b|a})\big]\cdot \mbox{tr}(\bms_{b|a}^3).
\end{align}
Let $I$ be uniformly distributed over $\{1,\cdots, s\}.$ By H\"{o}lder's inequality, $(E\lambda_I^{\alpha})^{1/\alpha} \leq (E\lambda_I^{\beta})^{1/\beta}$ for any $0<\alpha<\beta$. This says that
\begin{align*}
\Big(\frac{\lambda_1^{\alpha}+\cdots + \lambda_s^{\alpha}}{s}\Big)^{1/\alpha}\leq \Big(\frac{\lambda_1^{\beta}+\cdots + \lambda_s^{\beta}}{s}\Big)^{1/\beta}.
\end{align*}
By taking $\alpha=1, 2, 3$, respectively, and $\beta=4$, we have
\begin{align}\lbl{aiain}
\mbox{tr}(\bms_{b|a}^i)\leq s^{1-(i/4)}\cdot[\mbox{tr}(\bms_{b|a}^4)]^{i/4}
\end{align}
for $i=1,2,3.$ Consequently,
\begin{align*}
& C_2\leq 4(r+3)^2s\cdot \mbox{tr}(\bms_{b|a}^4),\ \ \ C_3\leq 16rs^2\cdot \mbox{tr}(\bms_{b|a}^4),\\
&  C_4\leq 16(r+2)^2s\cdot \mbox{tr}(\bms_{b|a}^4).
\end{align*}
Combing the above with \eqref{shangz}, we get
\begin{align*}
\mbox{Var}\big(\mbox{tr}\big(\hat{\bms}_{b|a}^2\big)\big)
\leq & \frac{8(r+3)^3+4(r+3)^2s+16rs^2+16(r+2)^2s}{n^4}\cdot \mbox{tr}(\bms_{b|a}^4)\\
\leq & \frac{16(r+3)(r+s+3)^2}{n^4}\cdot \mbox{tr}(\bms_{b|a}^4).
\end{align*}
Using the fact $3\leq 3r$ to see $16(r+3)(r+s+3)^2\leq 4^5r(r+s)^2$ and plugging in $r=n-q$ and $s=p-q$, we obtain
\begin{align*}
\mbox{Var}\big(\mbox{tr}\big(\hat{\bms}_{b|a}^2\big)\big)\leq \frac{4^5(n-q)(n+p-2q)^2}{n^4}\cdot \mbox{tr}(\bms_{b|a}^4),
\end{align*}
which concludes the proof. \hfill$\square$

\begin{lemma}\lbl{jianjicaoji} Let $\hat{\bms}_{b|a}$ and $\bms_{b|a}$ be as in \eqref{hondulasi} and \eqref{youshayi}, respectively. Then, there exists a constant $K>0$ not depending on $p, q, n$ or $\bms_{b|a}$ such that
\begin{align*}
\mbox{Var}\Big(\big(\mbox{tr}\big(\hat{\bms}_{b|a}\big)\big)^2\Big) \leq  \frac{K(n-q)^2(p-q)^2}{n^4}\mbox{tr}\big(\bms_{b|a}^4\big).
\end{align*}
\end{lemma}
\noindent\textbf{Proof of Lemma \ref{jianjicaoji}}. Set $r=n-q$ and $s=p-q$. Let $\lambda_1, \cdots \lambda_s$ be the eigenvalues of the $s\times s$ matrix  $\bms_{b|a}$. Let $\xi_1, \cdots, \xi_s$ be i.i.d. random variables with distribution $\chi^2(r)$, with $E\xi_1=r$ and  $\mbox{Var}(\xi_1)=2r$.  From \eqref{tin_beizi},
\begin{align}\lbl{diaole}
n^4\cdot\mbox{Var}\big(\big(\mbox{tr}\big(\hat{\bms}_{b|a}\big)\big)^2\big)=
\mbox{Var}\Big(\Big(\sum_{i=1}^{s}\lambda_i\xi_i\Big)^2\Big).
\end{align}
Let $U$ be a random variable with mean $\mu$. Write $U^2=(U-\mu)^2+2\mu U-\mu^2$. Trivially, $\mbox{Var}(W_1+W_2)\leq 2\,\mbox{Var}(W_1)+ 2\,\mbox{Var}(W_2)$ for any random variables $W_1$ and $W_2$. Then
\begin{align*}
\mbox{Var}(U^2)
\leq & 2\,\mbox{Var}((U-\mu)^2) + 8\mu^2\,\mbox{Var}(U)\\
 \leq & 2E(U-\mu)^4 + 8\mu^2\,\mbox{Var}(U).
\end{align*}
Now consider $U$ to be $U=\sum_{i=1}^{s}\lambda_i\xi_i$, with $\mu=EU=r\,\mbox{tr}(\bms_{b|a})$. Use \eqref{aiain} to see
\begin{align}\lbl{Akron}
\mu^2\cdot \mbox{Var}(U)=2\mu^2\sum_{i=1}^s\lambda_i^2=2r^2\,[\mbox{tr}(\bms_{b|a})]^2\cdot\mbox{tr}\big(\bms_{b|a}^2\big) \leq 2r^2s^2\cdot \mbox{tr}(\bms_{b|a}^4).
\end{align}
By the Marcinkiewtz-Zygmund inequality [see, e.g., Theorem 2 on p. 386 from \cite{chow1997probability}] and the Cauchy-Schwartz inequality, we obtain
\begin{align*}
E(U-\mu)^4=&E\Big[\sum_{i=1}^{s}\lambda_i(\xi_i-r)\Big]^4\\
\leq & K_1\cdot E\Big[\sum_{i=1}^{s}\lambda_i^2(\xi_i-r)^2\Big]^2\\
 \leq & K_1\cdot \Big(\sum_{i=1}^{s}\lambda_i^4\Big)\cdot E\sum_{i=1}^{s}(\xi_i-r)^4,
\end{align*}
where $K_1$ is a numerical constant. Write $\xi_1=\sum_{j=1}^r\eta_i$, where $\eta_1, \cdots, \eta_r$ are i.i.d. $\chi^2(1)$-distributed random variables. By Corollary 2 on p. 387 from \cite{chow1997probability}, $E[\sum_{j=1}^r(\eta_i-1)]^4 \leq K_2r^2$, where $K_2$ is also a numerical constant. That is, $E(\xi_1-r)^4 \leq K_2r^2$. In summary, we get $E(U-\mu)^4 \leq (K_1K_2)r^2s\cdot \mbox{tr}\big(\bms_{b|a}^4\big)$ and hence
\begin{align*}
n^4\cdot\mbox{Var}\big(\big(\mbox{tr}\big(\hat{\bms}_{b|a}\big)\big)^2\big)=
\mbox{Var}(U^2)\leq (2K_1K_2)r^2s\cdot \mbox{tr}\big(\bms_{b|a}^4\big) + 16r^2s^2\cdot \mbox{tr}(\bms_{b|a}^4)
\end{align*}
from \eqref{diaole} and \eqref{Akron}. The conclusion then follows by using the fact $r^2s\leq r^2s^2$. \hfill$\square$

\begin{prop}\lbl{sonfather} Let $\hat{\bms}_{b|a}$ and $\bms_{b|a}$ be as in \eqref{hondulasi} and \eqref{youshayi}. Set $r=n-q$ and
\begin{align*}
\widehat{\mbox{tr}\big(\bms_{b|a}^2\big)}=\frac{n^2}{r(r+1)}\cdot
\Big[\mbox{tr}\big(\hat{\bms}_{b|a}^2\big)-
\frac{1}{r}\big(\mbox{tr}\big(\hat{\bms}_{b|a}\big)\big)^2\Big].
\end{align*}
Assume \eqref{con_(C3)} holds. If $q=o(n)$, $q=o(p)$ and $p=o(n^{3})$ then $\widehat{\mbox{tr}\big(\bms_{b|a}^2\big)}/\mbox{tr}\big(\bms_{b|a}^2\big)\to 1$ in\\ probability as $p\to \infty$. Hence, $\widehat{\mbox{tr}\big(\bms_{b|a}^2\big)}$ is an asymptotic ratio-consistent estimator of $\mbox{tr}\big(\bms_{b|a}^2\big)$.
\end{prop}
\noindent\textbf{Proof of Proposition \ref{sonfather}}. Since  $\mbox{Var}(W_1+W_2)\leq 2\,\mbox{Var}(W_1)+ 2\,\mbox{Var}(W_2)$ for any random variables $W_1$ and $W_2$, by Lemma \ref{jianjicao} and Lemma \ref{jianjicaoji}, there exists a constant $K>0$ independent of $n, p, q$ or $\bms_{b|a}$ such that
\begin{align}\lbl{wahaha}
&\mbox{Var}\Big(\mbox{tr}\big(\hat{\bms}_{b|a}^2\big)-
\frac{1}{r}\big(\mbox{tr}\big(\hat{\bms}_{b|a}\big)\big)^2\Big) \nonumber\\
\leq & K\cdot \Big[\frac{(n-q)(n+p-2q)^2}{n^4} + \frac{(n-q)^2(p-q)^2}{n^4r^2}\Big]\cdot \mbox{tr}\big(\bms_{b|a}^4\big) \nonumber\\
\leq & K\cdot\Big[\frac{(n+p)^2}{n^3}+\frac{p^2}{n^4}\Big]\cdot \mbox{tr}(\bms_{b|a}^4).
%& = & O\Big(\frac{r^3}{n^4}\big)\cdot \mbox{tr}(\bms_{b|a}^4).
\end{align}
On the other hand, by \eqref{buhaoyisi},
\begin{align}\lbl{dawanfan}
& E\Big[\mbox{tr}\big(\hat{\bms}_{b|a}^2\big)-
\frac{1}{r}\big(\mbox{tr}\big(\hat{\bms}_{b|a}\big)\big)^2\Big] \nonumber\\
= &\frac{r}{n^2}\cdot\big[\mbox{tr}(
\bms_{b|a})\big]^2 + \frac{r(r+1)}{n^2}\mbox{tr}\big(\bms_{b|a}^2\big)-\frac{1}{r}
E\big[\big(\mbox{tr}\big(\hat{\bms}_{b|a}\big)\big)^2\big].
\end{align}
Define $s=p-q$. Let $\lambda_1, \cdots \lambda_s$ be the eigenvalues of the $s\times s$ matrix  $\bms_{b|a}$. By \eqref{tin_beizi},
\begin{align*}
n^2\cdot E\big[\big(\mbox{tr}\big(\hat{\bms}_{b|a}\big)\big)^2\big]
=&E\Big(\sum_{i=1}^{s}\lambda_i\|\bd{w}_i\|^2\Big)^2\\
= & \sum_{i=1}^{s}\lambda_i^2E\big(\|\bd{w}_i\|^4)+
2\sum_{1\leq i< j\leq s}\lambda_i\lambda_j E\big(\|\bd{w}_i\|^2\|\bd{w}_j\|^2\big),
\end{align*}
where  $\|\bd{w}_1\|^2, \cdots, \|\bd{w}_{s}\|^2$ are i.i.d. $\chi^2(r)$-distributed random variables.
From \eqref{nipinpin}, $E(\|\bd{w}_1\|^4)=r(r+2)$. Thus, we have from independence that
\begin{align*}
n^2\cdot E\big[\big(\mbox{tr}\big(\hat{\bms}_{b|a}\big)\big)^2\big]
=& r(r+2)\cdot \mbox{tr}\big(\bms_{b|a}^2\big)+
r^2\cdot 2\sum_{1\leq i< j\leq s}\lambda_i\lambda_j\\
=& r(r+2)\cdot \mbox{tr}\big(\bms_{b|a}^2\big)+
r^2\big\{\big[\mbox{tr}\big(\bms_{b|a}\big)\big]^2-\mbox{tr}\big(\bms_{b|a}^2\big)\big\}\\
 = & 2r\cdot \mbox{tr}\big(\bms_{b|a}^2\big)+r^2\cdot \big[\mbox{tr}\big(\bms_{b|a}\big)\big]^2,
\end{align*}
or equivalently,
\begin{align*}
\frac{1}{r}
E\big[\big(\mbox{tr}\big(\hat{\bms}_{b|a}\big)\big)^2\big]=\frac{2}{n^2}\cdot  \mbox{tr}\big(\bms_{b|a}^2\big) + \frac{r}{n^2}\cdot \big[\mbox{tr}\big(\bms_{b|a}\big)\big]^2.
\end{align*}
Combine this with \eqref{dawanfan} to get
\begin{align}\lbl{yilian}
EF_p=E\Big[\mbox{tr}\big(\hat{\bms}_{b|a}^2\big)-
\frac{1}{r}\big(\mbox{tr}\big(\hat{\bms}_{b|a}\big)\big)^2\Big] =\frac{r^2+r-2}{n^2}\cdot \mbox{tr}\big(\bms_{b|a}^2\big),
\end{align}
where $F_p=\mbox{tr}\big(\hat{\bms}_{b|a}^2\big)-
\frac{1}{r}\big[\mbox{tr}\big(\hat{\bms}_{b|a}\big)\big]^2$. Under Assumption \eqref{con_(C3)},
$\tau^{-1}<\lambda_{min}(\bms_{b|a})\
\leq\lambda_{max}(\bms_{b|a})<\tau$ for some constant $\tau>1$. This then implies
\begin{align*}
\mbox{tr}(\bms_{b|a}^4)=\lambda_1^4+\cdots + \lambda_{s}^4\leq s\tau^4\ \ \mbox{and}\ \  \big[\mbox{tr}\big(\bms_{b|a}^2\big)\big]^2=(\lambda_1^2+\cdots+\lambda_{s}^2)^2 \geq s^2\tau^{-4}.
\end{align*}
We deduce from \eqref{wahaha} and \eqref{yilian} that
\begin{align*}
\mbox{Var}\Big(\frac{F_p}{EF_p}\Big)
\leq & K\cdot \frac{n(n+p)^2+p^2}{(r^2+r-2)^2}\cdot \frac{\mbox{tr}(\bms_{b|a}^4)}{\big[\mbox{tr}\big(\bms_{b|a}^2\big)\big]^2}\\
\leq & (3K)\tau^8\cdot \frac{n^3+np^2}{(r^2+r-2)^2}\cdot \frac{1}{s},
\end{align*}
where the fact $n(n+p)^2+p^2 \leq 3(n^3+np^2)$ is used in the second inequality.
Under the conditions $q/n\to 0$, $q/p\to 0$ and $p=o(n^{3})$, we know $r/n\to 1$ and $s/p\to 1$, hence the last term is of order $\frac{1}{np}+\frac{p}{n^3} \to 0$. This leads to $\frac{F_p}{EF_p}\to 1$ in probability and we conclude from \eqref{yilian} that
\begin{align*}
\Big[\mbox{tr}\big(\hat{\bms}_{b|a}^2\big)-
\frac{1}{r}\big(\mbox{tr}\big(\hat{\bms}_{b|a}\big)\big)^2\Big]\cdot \Big[\frac{r(r+1)}{n^2}\cdot \mbox{tr}\big(\bms_{b|a}^2\big)\Big]^{-1}\to 1
\end{align*}
in probability. The proof is complete. \hfill$\square$

\begin{lemma}\lbl{Aux_Gaussian} Let $\bd{y}_1, \cdots, \bd{y}_m$ be i.i.d. $p$-dimensional random vectors with distribution $N(\bd{0}, \bd{B})$, where $\bd{B}$ is a $p\times p$ non-negative definite matrix.  Let $\bd{x}\in \mathbb{R}^m$ be a non-zero random vector independent of $\bd{y}_1, \cdots, \bd{y}_m$. Write $\bd{x}_1=\bd{x}/\|\bd{x}\|$ and
the $m\times p$ matrix $(\bd{y}_1 \cdots \bd{y}_m)^T=(\bd{z}_1  \cdots \bd{z}_p)$. Then $(\bd{x}_1^T\bd{z}_1 \cdots  \bd{x}_1^T\bd{z}_p)^T \sim N(\bd{0}, \bd{B})$ and is independent of $\|\bd{x}\|$.
\end{lemma}

\noindent\textbf{Proof of Lemma \ref{Aux_Gaussian}}.   Write $\bd{B}=(b_{ij})_{p\times p}$ and $(\bd{y}_1 \cdots \bd{y}_m)^T=(y_{ij})_{m\times p}$.  Use the fact that the rows of the matrix are i.i.d. to see $E(y_{ki}y_{lj})=0$ if $k\ne l$ and $E(y_{ki}y_{lj})=b_{ij}$ if $k=l$. Thus,
\begin{align}\lbl{second_chance}
E(\bd{z}_i\bd{z}_j^T)=\big(E(y_{ki}y_{lj})\big)_{1\leq k, l\leq p}=b_{ij} \bd{I}_{p}.
\end{align}
Obviously, $\bd{z}_i \sim N(0, b_{ii}\bd{I}_m)$ for each $i$. Since $\bd{x}_1$ is a unit random vector, then conditional on $\bd{x}$, we know $\bd{x}_1^T\bd{z}_1, \cdots,  \bd{x}_1^T\bd{z}_p$ are jointly Gaussian random variables with $\bd{x}_1^T\bd{z}_i \sim N(0, b_{ii})$ for each $1\leq i \leq p$.
% mean $0$ and variance $1$ {\color{red} $diag(\bm B)$}.
Let us check their covariance matrix. In fact, conditional on $\bd{x}$,
\begin{align*}
E\big[(\bd{x}_1^T\bd{z}_i)(\bd{x}_1^T\bd{z}_j)\big]=E\big(\bd{x}_1^T\bd{z}_i\bd{z}_j^T\bd{x}_1\big)
=\bd{x}_1^TE\big(\bd{z}_i\bd{z}_j^T\big)\bd{x}_1=b_{ij},
\end{align*}
by \eqref{second_chance} and the fact $\bd{x}_1^T\bd{x}_1=1$. In summary, conditional on $\bd{x}$, the random vector  $(\bd{x}_1^T\bd{z}_1 \cdots  \bd{x}_1^T\bd{z}_p)^T \sim N(\bd{0}, \bd{B})$. Since $N(\bd{0}, \bd{B})$ is free of $\bd{x}_1$, this implies, unconditionally, it is also true that $(\bd{x}_1^T\bd{z}_1 \cdots  \bd{x}_1^T\bd{z}_p)^T \sim N(\bd{0}, \bd{B})$. Finally, for any set $F\subset \mathbb{R}^p$ and   $G\subset [0, \infty)$,
\begin{align*}
 P\big((\bd{x}_1^T\bd{z}_1 \cdots  \bd{x}_1^T\bd{z}_p)^T \in F,\, \|\bd{x}\| \in G\big)
 = & E\big[P\big((\bd{x}_1^T\bd{z}_1 \cdots  \bd{x}_1^T\bd{z}_p)^T \in F\big|\bd{x}\big)\cdot I(\|\bd{x}\| \in G)\big]\\
=& E\big[P\big(N(\bd{0}, \bd{B}) \in F\big)\cdot I(\|\bd{x}\| \in G)\big]\\
=& P\big(N(\bd{0}, \bd{B}) \in F\big)\cdot P(\|\bd{x}\| \in G).
\end{align*}
This shows that $(\bd{x}_1^T\bd{z}_1 \cdots  \bd{x}_1^T\bd{z}_p)^T$ and $\|\bd{x}\|$ are independent.    \hfill$\square$

\begin{lemma}[\cite{bai2010spectral}]\lbl{BaiS}
Let $\mathbf{A}=\left(a_{i j}\right)$ be an $n \times n$ non-random matrix and $\mathbf{X}=$ $\left(x_{1}, \cdots, x_{n}\right)^{T}$ be a random vector of independent entries. Assume that $\mathrm{E} x_{i}=0$, $E\left|x_{i}\right|^{2}=1$ and $E\left|x_{j}\right|^{\ell} \leq \nu_{\ell}$. Then, for any $p \geq 1$,
$$
E\left|\mathbf{X}^T \mathbf{A X}-\operatorname{tr} \mathbf{A}\right|^{p} \leq C_{p}\left(\left(\nu_{4} \operatorname{tr}\left(\mathbf{A} \mathbf{A}^T\right)\right)^{p / 2}+\nu_{2 p} \operatorname{tr}\left(\mathbf{A} \mathbf{A}^T\right)^{p / 2}\right),\nonumber
$$
where $C_{p}$ is a constant depending on $p$ only.
\end{lemma}

\begin{lemma}\lbl{Matrix_moment} Assume $\bme=(\epsilon_1, \cdots, \epsilon_n)^T\in \mathbb{R}^n$ satisfies that $\epsilon_1, \cdots, \epsilon_n$ are i.i.d. with $E\epsilon_1=0$,  $E\epsilon_1^2=\sigma^2$ and $E|\epsilon_1|^{2k}<\infty$ for an integer $k\geq 2$. Let $\bd{H}$ be an  $n\times n$ symmetric, random matrix  satisfying $\bd{H}^2=\bd{H}$ and $\mbox{rank}\,(\bd{H})=q$. Assume $\bd{u} \sim N(\bm 0, \bd{I}_n)$ and that $\bd{u}, \bd{H}$ and $\bme$ are independent. Then

(i) $\bme^T(\bd{I}_n-\H)\bme=(n-q)\sigma^2 +O_p(\sqrt{n})$;

(ii) $E(\bd{u}^T\H\bme)^{2k} \leq C(k, \sigma)\cdot q^{k}$, where $C(k, \sigma)$ is a constant depending on $k, \sigma$ only.
\end{lemma}

\noindent\textbf{Proof of Lemma \ref{Matrix_moment}}. It is easy to see $E(\bme^T\bd{A}\bme)=\sigma^2\,\mbox{tr}(A)$ for any matrix $\bd{A}$.

(i) Obviously, the conditional mean $E[\bme^T(\bd{I}_n-\H)\bme|\H]=(n-q)\sigma^2.$ Take another expectation to see $E[\bme^T(\bd{I}_n-\H)\bme)]=(n-q)\sigma^2.$ By Lemma \ref{BaiS},
\begin{align*}
E\big[\big(\bme^T(\bd{I}_n-\H)\bme-(n-q)\sigma^2\big)^2\big|\H\big]
\leq &
C\cdot E\epsilon_1^4\cdot \mbox{tr}(\bd{I}_n-\H)\\
= & C\cdot E\epsilon_1^4\cdot(n-q),
\end{align*}
where $C>0$ is a constant free of $n, q, k, \sigma$. By taking another expectation, we get $\mbox{Var}(\bme^T(\bd{I}_n-\H)\bme) \leq Cn$. Thus, by the Chebyshev inequality,
\begin{align*}
P\big(|\bme^T(\bd{I}_n-\H)\bme-(n-q)\sigma^2|\geq A\sqrt{n}\big)\leq \frac{\mbox{Var}(\bme^T(\bd{I}_n-\H)\bme)}{A^2}\leq \frac{C}{A^2}.
\end{align*}
This implies $\bme^T(\bd{I}_n-\H)\bme=(n-q)\sigma^2 +O_p(\sqrt{n})$.

(ii) First,  $E(\bme^T\H\bme)=\sigma^2\cdot\mbox{tr}(\H)=\sigma^2q$. Trivially,
\begin{align*}
E\big(\bme^T\H\bme\big)^k
\leq & 2^k\cdot \big[E\big|\bme^T\H\bme-\sigma^2q\big|^{k}+ (\sigma^2q)^{k}\big].
\end{align*}
 From Lemma \ref{BaiS} and the fact $\H^l=\H$ for any $l=1,2,\cdots$, we see
\begin{align*}
E\big|\bme^T\H\bme-\sigma^2q\big|^{k}
\leq & C_k\cdot \big[(E\epsilon_1^4)^{k/2}+
E\epsilon_1^{2k}\big]\cdot \big[(\mbox{tr}(\H))^{k/2}+ \mbox{tr}(\H)\big]\\
\leq & C_kq^{k/2},
\end{align*}
where $C_k$ is a constant depending on $k$ only. Therefore,
\begin{align}\lbl{jiaohuandaishu}
E\big(\bme^T\H\bme\big)^k \leq C(k, \sigma)\cdot q^{k},
\end{align}
 where $C(k, \sigma)$ is a constant depending on $k, \sigma$ only. On the other hand, noting that the $n$ entries of $\bd{u}$ are i.i.d. $N(0, 1)$, by conditioning on $\H$ and $\bme$, the random variable $\bd{u}^T\H\bme$ has the distribution of $G\cdot \|\H\bme\|$, where $G\sim N(0, 1)$. Or, equivalently,
\begin{align*}
%\lbl{nishuoshane1}
\mbox{given}\ \{\H, \bme\},\ \mbox{random variable}\ \bd{u}^T\H\bme\  \mbox{has distribution} \ G\cdot \big(\bme^T\H\bme\big)^{1/2}.
\end{align*}
In particular, this implies that
\begin{align*}
%\lbl{GuanLN1}
E(\bd{u}^T\H\bme)^{2k} = E\big(G^{2k}\big)\cdot E\big(\bme^T\H\bme\big)^k.
\end{align*}
By combining this, \eqref{jiaohuandaishu} and the fact $E(G^{2k})=(2k-1)!!$, we obtain
\begin{align*}
E(\bd{u}^T\H\bme)^{2k} \leq (2k-1!!\cdot C(k, \sigma)\cdot q^{k}.
\end{align*}
The proof is completed. \hfill$\square$

\medskip

Recall random error vector $\bme=(\varepsilon_1, \cdots, \varepsilon_n)^T\in \mathbb{R}^n$ in the linear regression model from Section \ref{regress_3}.  The components  $\{\varepsilon_i;\, 1\leq i \leq n\}$ are assumed to be i.i.d. random variables.
\begin{prop}\lbl{Extreme_for_3} Let $T^{(3)}_{max}$ be defined as in \eqref{3rdmax}. Assume  \eqref{con_(C2)} and \eqref{con_(C3)} are true and \eqref{assumption_A3} holds with ``$\bd{\Sigma}$'' replaced by ``$\bms_{b|a}$''. Suppose  $p=o(n^3)$, $q=o(p)$,  $q\leq n^{\delta}$ for some $\delta \in (0,1)$ and $E(|\epsilon_1|^{\ell}) < \infty$ with  $\ell=14(1-\delta)^{-1}$.
Then, under $H_0$ from \eqref{h1}, $T^{(3)}_{max}-2\log(p-q)+\log\log(p-q)$ converges weakly to a distribution with cdf $F(x)=\exp\{-\frac{1}{\sqrt{\pi}}
\exp(-\frac{x}{2})\}$.
\end{prop}
\noindent\textbf {Proof of Proposition \ref{Extreme_for_3}}.
Recall the notation between \eqref{h1j} and \eqref{hondulasi}. In particular, $\H_a=\mX_a(\mX_a^{T} \mX_a)^{-1}\mX_a^{T}$. Under the null hypothesis in  \eqref{h1}, $\bd{Y}=\X_{a}\bb_a+\bme$, where $\bme=(\varepsilon_1, \cdots, \varepsilon_n)^T\in \mathbb{R}^n$ and the random errors $\{\varepsilon_i;\, 1\leq i \leq n\}$ are i.i.d. with $E\varepsilon_i=0$ and $\mbox{Var}(\varepsilon_i)=\sigma^2$ for each $i$. Also, $\bme$  is assumed to be  independent of $\{\X_i;\, 1\leq i \leq n\}$. Use $\X_{a}^T(\I_n-\H_a)=\bd{0}$ and $(\I_n-\H_a)\X_{a}=\bd{0}$ to see
\begin{align}\lbl{xieguangkun}
\mY^{T}(\I_n-\H_a)\mY
=&(\X_{a}\bb_a+\bme)^T(\I_n-\H_a) (\X_{a}\bb_a+\bme) \nonumber\\
=&\bme^T(\I_n-\H_a) \bme.
\end{align}
Recalling \eqref{hondulasi}, we denote
\begin{align}\lbl{nishuowo}
\X_{b}=(\X_{1b}, \cdots, \X_{nb})^T\ \ \ \mbox{and}\ \ \ \wX_b=(\wX_{q+1}, \cdots, \wX_p)=(\bd{I}_n-\H_a)\X_{b},
\end{align}
where $\X_{b}$ and $\wX_b$ are $n\times (p-q)$ matrices. Write $\X_{b}=(\bd{w}_{q+1}, \cdots, \bd{w}_p)$. Then the last assertion from \eqref{nishuowo} says that $\wX_j=(\bd{I}_n-\H_a)\bd{w}_j$ for each $q+1\leq j \leq p$. This leads to
\begin{align*}
\mY^{T} \wX_j\wX_j^{T} \mY
=&(\X_{a}\bb_a+\bme)^T(\I_n-\H_a)\bd{w}_j\bd{w}_j^T(\I_n-\H_a) (\X_{a}\bb_a+\bme)\\
=& \bme^T(\I_n-\H_a)\bd{w}_j \bd{w}_j^T(\I_n-\H_a) \bme\\
=& \bme^T\wX_j\wX_j^T \bme
\end{align*}
via the facts $\X_{a}^T(\I_n-\H_a)=\bd{0}$ and $(\I_n-\H_a)\X_{a}=\bd{0}$ again. Note that $(\wX_j^{T} \wX_j)^{-1}$ is a scalar. By the definition of $T^{(3)}_{max}$ in \eqref{3rdmax}, we derive
\begin{align*}
T^{(3)}_{max}
%=(n-q)\max_{q+1\le j\le p} \tilde{F}_j
=& \max_{q+1\le j\le p}\frac{\mY^{T} \wX_j(\wX_j^{T} \wX_j)^{-1}\wX_j^{T} \mY
}{\mY^{T}(\I_n-\H_a)\mY/(n-q)}\\
= & \frac{(n-q)\sigma^2}{\bme^T(\I_n-\H_a) \bme}\cdot \max_{q+1\le j\le p}\big\{n^{-1}\sigma^{-2}\bme^T \wX_j(n^{-1}\wX_j^{T} \wX_j)^{-1}\wX_j^{T} \bme\big\}.
\end{align*}
Set
\begin{align*}
W_p=\frac{(n-q)\sigma^2}{\bme^T(\I_n-\H_a)\bme}.
\end{align*}
By the triangle inequality,
\begin{align}\lbl{daishujihe}
& \big|T^{(3)}_{max} - W_p\cdot \max_{q+1\le j\le p}\big\{n^{-1}\sigma^{-2}\bme^{T} \wX_j\wX_j^{T} \bme\big\}\big| \nonumber\\
\leq & W_p\cdot \max_{q+1\le j\le p}\big\{n^{-1}\sigma^{-2}\bme^{T} \wX_j\wX_j^{T} \bme\cdot  \big|1-(n^{-1}\wX_j^{T} \wX_j)^{-1}\big|\big\}.
\end{align}
Define
\begin{align*}
& H_{31}=\max_{q+1\le j\le p}\big\{n^{-1}\sigma^{-2}\bme^{T} \wX_j\wX_j^{T} \bme\big\};\\
& H_{32}=\max_{q+1\le j\le p}\big\{n^{-1}\sigma^{-2}\bme^{T} \wX_j\wX_j^{T} \bme \cdot\big|1-(n^{-1}\wX_j^{T} \wX_j)^{-1}\big|\big\}.
\end{align*}
By Lemma \ref{Matrix_moment}(i), we know $1/W_p= 1 + O_p(n^{-1/2})$, which implies
\begin{align}\lbl{alias1}
& W_p= 1 + O_p\Big(\frac{1}{\sqrt{n}}\Big).
\end{align}
We will show that, as $p\to\infty$,
\begin{align}
& H_{31}':=H_{31}-2\log(p-q)+\log\log(p-q)\to  \mbox{a distribution with cdf}\
 e^{-e^{-x/2}/\sqrt{\pi}};\lbl{alias2}\ \ \ \ \ \ \  \\
&  H_{32} \to 0 \ \mbox{in probability}.\lbl{alias3}
\end{align}
Assuming they are true, using assumption $p=o(n^3)$ and \eqref{daishujihe} we have
\begin{align}\lbl{woyaode}
T^{(3)}_{max}
=& \Big[1 + O_p\Big(\frac{1}{\sqrt{n}}\Big)\Big]\cdot \big[H_{31}'+2\log(p-q)-\log\log(p-q) +o_p(1)\big] \nonumber\\
=& H_{31}'+2\log(p-q)-\log\log(p-q) +o_p(1).
\end{align}
This and \eqref{alias2} show that $T^{(3)}_{max}-2\log(p-q)+\log\log(p-q)$ converges weakly to a distribution with cdf $F(x)=\exp\{-\frac{1}{\sqrt{\pi}}
\exp(-\frac{x}{2})\}$, $x \in \mathbb{R}$. It remains to prove \eqref{alias2}  and \eqref{alias3}, which will be done in two steps as follows.

{\it Step 1: the proof of \eqref{alias2}}.  By \eqref{haoshiduo} and \eqref{nishuowo}, $\X_{b}^T=(\bd{V}_1,\cdots, \bd{V}_n)  + \bd{\Sigma}_{ba}\bd{\Sigma}_{aa}^{-1}\bd{X}_a^T$,
where $\bd{V}_1, \cdots, \bd{V}_n$ are i.i.d. $(p-q)$-dimensional random vectors with distribution $N(\bd{0}, \bd{\Sigma}_{bb\cdot a})$ and they are also independent of $\bd{X}_a$. In particular, $\bd{V}_1,\cdots, \bd{V}_n$ are independent of $\H_a$, a function of $\bd{X}_a$.  As $(\bd{I}_n-\H_a)\X_{a}=\bd{0}$, we see that
\begin{align*}
(\wX_{q+1}, \cdots, \wX_p)
=& (\bd{I}_n-\H_a)\X_{b}\\
=& (\bd{I}_n-\H_a)\big[\bd{X}_a\bd{\Sigma}_{aa}^{-1}\bd{\Sigma}_{ba}^T+(\bd{V}_1,\cdots, \bd{V}_n)^T\big]\\
=& (\bd{I}_n-\H_a)(\bd{V}_1,\cdots, \bd{V}_n)^T.
\end{align*}
Write $(\bd{V}_1,\cdots, \bd{V}_n)^T=(\bd{u}_{q+1}, \cdots, \bd{u}_p)$. In other words, each $\bd{u}_{q+1}, \cdots, \bd{u}_p\in\mathbb R^n$ is a  column of $(\bd{V}_1,\cdots, \bd{V}_n)^T$.  Immediately we have
\begin{align}\lbl{jiutai}
\wX_{j}=(\bd{I}_n-\H_a)\bd{u}_j\ \ \ \mbox{and}\ \ \ \bd{u}_j \sim N(\bd{0}, \bd{I}_n)
\end{align}
for each $j=q+1, \cdots, p.$ This means $\wX_j^{T} \bme=\bd{u}_j^T\bme-\bd{u}_j^T\H_a\bme$, which then leads to
\begin{align}\lbl{chaowan}
(\wX_j^{T} \bme)^2
=&\big(\bd{u}_j^T\bme\big)^2+ (\bd{u}_j^T\H_a\bme)^2-2 (\bme^T\bd{u}_j)\bd{u}_j^T\H_a\bme
\end{align}
by using the fact $\bd{u}_j^T\bme=\bme^T\bd{u}_j \in \mathbb{R}.$ Obviously, it holds that
\begin{align}\lbl{zhuanquan}
& \big|\max_{q+1\leq j \leq p}\big\{n^{-1}(\wX_j^{T} \bme)^2\big\}- \max_{q+1\leq j \leq p}\big\{n^{-1}\big(\bd{u}_j^T\bme\big)^2\big\}\big| \nonumber\\
 \leq & \max_{q+1\leq j \leq p}\big\{n^{-1}(\bd{u}_j^T\H_a\bme)^2\big\} +2\cdot\max_{q+1\leq j \leq p}\big\{n^{-1}\big|(\bme^T\bd{u}_j)\bd{u}_j^T\H_a\bme\big|\big\}.
\end{align}
A key observation is that the three random quantities $\bd{u}_j$, $\H_a$ and $\bme$ are independent.
By the last assertion from \eqref{jiutai} and Lemma \ref{Matrix_moment}(ii), we have
\begin{align}\lbl{zhenshiniya}
E(\bd{u}_j^T\H_a\bme)^{2k} \leq C(k, \sigma)\cdot q^{k}
\end{align}
for any $q+1\leq j \leq n$, where $k=[\frac{6}{1-\delta}]+1$ and $C(k, \sigma)$ is a constant depending on $k, \sigma$ only.
We next show
\begin{align}\lbl{jiu}
\frac{1}{n}\cdot\max_{q+1\leq j \leq p}(\bd{u}_j^T\H_a\bme)^2 \to 0\ \ \ \ \mbox{and}\ \ \ \
\frac{1}{n}\cdot\max_{q+1\leq j \leq p}\big|(\bme^T\bd{u}_j)\bd{u}_j^T\H_a\bme\big| \to 0
\end{align}
in probability as $p\to \infty$. In fact, for any $\beta>0$,
\begin{align*}
P\Big(\frac{1}{n}\cdot\max_{q+1\leq j \leq p}(\bd{u}_j^T\H_a\bme)^2\geq \beta\Big)
\leq & p\cdot \max_{q+1\leq j \leq p}P\Big((\bd{u}_j^T\H_a\bme)^2\geq n\beta\Big)\\
 \leq & p\cdot \max_{q+1\leq j \leq p}
\frac{E(\bd{u}_j^T\H_a\bme)^{2k}}{ (n\beta)^{k}}.
% \\
% & = & \beta^{-k}E(G^{2k})\cdot\frac{p}{n^k}\cdot E\big(\bme^T\H_a\bme\big)^k
\end{align*}
Therefore, from assumption $q\leq n^{\delta}$ for some $\delta \in (0,1)$,  we have
\begin{align*}
P\Big(\frac{1}{n}\cdot\max_{q+1\leq j \leq p}(\bd{u}_j^T\H_a\bme)^2\geq \beta\Big)
\leq C(\beta, k, \sigma)\cdot \frac{pq^{k}}{n^{k}}\leq C(\beta, k, \sigma)\cdot \frac{p}{n^{k(1-\delta)}},
\end{align*}
where $C(\beta, k, \sigma)$ is a constant depending on $\beta, k$ and $\sigma$. Since $k(1-\delta)> 6$, we get the first limit of \eqref{jiu} by using the assumption $p=o(n^3)$. For the second limit, by setting $\bme_1=\bme/\|\bme\|$ we have
\begin{align*}
\frac{1}{n}\cdot\max_{q+1\leq j \leq p}\big|(\bme^T\bd{u}_j)\bd{u}_j^T\H_a\bme\big| =\frac{\|\bme\|}{\sqrt{n}}\cdot\frac{1}{\sqrt{n}}\cdot \max_{q+1\leq j \leq p}\big|(\bme_1^T\bd{u}_j)\bd{u}_j^T\H_a\bme\big|.
\end{align*}
By the law of large numbers, $\|\bme\|/\sqrt{n} \to \sigma$ in probability. Thus, to get the second limit of \eqref{jiu}, it suffices to prove
\begin{align}\lbl{zhenhuanzhuan}
\frac{1}{\sqrt{n}}\cdot \max_{q+1\leq j \leq p}\big|(\bme_1^T\bd{u}_j)\bd{u}_j^T\H_a\bme\big|
\to 0
\end{align}
in probability. Similar to an earlier argument, we have from the fact $\bme_1^T\bd{u}_j \sim N(0, 1)$ that
\begin{align*}
P\Big(\frac{1}{\sqrt{n}}\cdot \max_{q+1\leq j \leq p}\big|(\bme_1^T\bd{u}_j)\bd{u}_j^T\H_a\bme\big|\geq \beta\Big)
\leq & p\cdot \max_{q+1\leq j \leq p}P\Big(|\bme_1^T\bd{u}_j|\cdot|\bd{u}_j^T\H_a\bme|\geq \sqrt{n}\beta\Big)\\
 \leq & \frac{p}{(\sqrt{n}\beta)^k}\cdot\max_{q+1\leq j \leq p} E\big[|\bme_1^T\bd{u}_j|^k\cdot|\bd{u}_j^T\H_a\bme|^k\big]\\
 \leq & C(k, \beta)\cdot \frac{p}{n^{k/2}}\cdot \big[E(\bd{u}_1^T\H_a\bme)^{2k}\big]^{1/2}
\end{align*}
by the Cauchy-Schwartz inequality and $E(\bd{u}_j^T\H_a\bme)^{2k}=E(\bd{u}_1^T\H_a\bme)^{2k}$, where $C(k, \beta)$ is a constant depending on $k$ and $\beta$ only. Use \eqref{zhenshiniya} to see
\begin{align*}
P\Big(\frac{1}{\sqrt{n}}\cdot \max_{q+1\leq j \leq p}\big|(\bme_1^T\bd{u}_j)\bd{u}_j^T\H_a\bme\big|\geq \beta\Big)
\leq C(k, \beta, \sigma)\cdot \frac{pq^{k/2}}{n^{k/2}}=O\Big(\frac{p}{n^{k(1-\delta)/2}}\Big),
\end{align*}
by the assumption $q\leq n^{\delta}$ again. Since $p=o(n^3)$ and $k(1-\delta)>6$, we get \eqref{zhenhuanzhuan} and then the second limit of \eqref{jiu}.

Now we study $\max_{q+1\leq j \leq p}\big\{n^{-1}\big(\bd{u}_j^T\bme\big)^2\big\}$ in  \eqref{zhuanquan}. Since $\epsilon_1,  \cdots, \epsilon_n$ are i.i.d. with mean zero and variance $\sigma^2$. By assumption, $E(|\epsilon_1|^{\ell})< \infty$ with  $\ell=14(1-\delta)^{-1}$. This concludes  $E(\epsilon_1^{14})< \infty$. Write $\|\bme\|^2=\epsilon_1^2 +\cdots +\epsilon_n^2$. By the central limit theorem, $n^{-1}\sigma^{-2}\|\bme\|^2=1+O_p(n^{-1/2})$. Use $\bme_1=\bme/\|\bme\|$ to see
\begin{align}\lbl{meiyong}
\max_{q+1\leq j \leq p}\big\{n^{-1}\sigma^{-2}\big(\bd{u}_j^T\bme\big)^2\big\}
=&\big(n^{-1}\sigma^{-2}\|\bme\|^2\big)\cdot \max_{q+1\leq j \leq p}\big(\bd{u}_j^T\bme_1\big)^2 \nonumber\\
=&\big[1+O_p(n^{-1/2})\big]\cdot \max_{q+1\leq j \leq p}\big(\bd{u}_j^T\bme_1\big)^2.
\end{align}
Review $\bd{V}_1, \cdots, \bd{V}_n$ are i.i.d. $(p-q)$-dimensional random vectors with distribution $N(\bd{0}, \bd{\Sigma}_{bb\cdot a})$ and $(\bd{V}_1,\cdots, \bd{V}_n)^T=(\bd{u}_{q+1}, \cdots, \bd{u}_p)$.
By Lemma \ref{Aux_Gaussian},
\begin{align}\lbl{hazainali}
(\bd{u}_{q+1}^T\bme_1, \cdots, \bd{u}_{p}^T\bme_1)^T \sim N(\bd{0}, \bd{\Sigma}_{bb\cdot a}).
\end{align}
%for any random vector $\bme$ independent of $\{(\bd{u}_{q+1}^T, \cdots, \bd{u}_{p}\}$.
Based on assumption, \eqref{assumption_A3}  holds with ``$\bd{\Sigma}$'' replaced by ``$\bms_{b|a}$'', we know Assumption \eqref{condition_2} also holds by the discussion below \eqref{assumption_A3}. We then have from Theorem \ref{theorem_2} that
\begin{align*}
P\big(\max_{q+1\leq j \leq p}(\bd{u}_j^T\bme_1)^2-2\log (p-q) +\log\log (p-q) \leq x\big) \to \exp\Big\{-\frac{1}{\sqrt{\pi}}e^{-x/2}\Big\}
\end{align*}
as $p\to\infty$. Denote $U_p=\max_{q+1\leq j \leq p}(\bd{u}_j^T\bme_1)^2-2\log (p-q) +\log\log (p-q) $. Then the above says $U_p$ converges weakly to the Gumbel distribution with cdf $F(x)=\exp\{-\frac{1}{\sqrt{\pi}}e^{-x/2}\}$. It then follows from \eqref{meiyong} that
\begin{align*}
\max_{q+1\leq j \leq p}\big\{n^{-1}\sigma^{-2}\big(\bd{u}_j^T\bme\big)^2\big\}
=&\big[1+O_p(n^{-1/2})\big]\cdot \big[U_p+2\log (p-q) -\log\log (p-q)\big]\\
=& U_p+2\log (p-q) -\log\log (p-q) +O_p(n^{-1/2}\log p).
\end{align*}
Obviously, the last term goes to zero since  $p=o(n^3)$ by assumption. Use the Slutsky lemma to see that
\begin{align*}
\max_{q+1\leq j \leq p}\big\{n^{-1}\big(\bd{u}_j^T\bme\big)^2\big\}-2\log (p-q) +\log\log (p-q)
\end{align*}
converges weakly to the Gumbel distribution with cdf $F(x)=\exp\{-\frac{1}{\sqrt{\pi}}e^{-x/2}\}$.
We then obtain \eqref{alias2} by \eqref{zhuanquan} and \eqref{jiu}.

{\it Step 2: the proof of \eqref{alias3}}. Easily, by definition,
\begin{align*}
|H_{32}| \leq |H_{31}| \cdot \max_{q+1\le j\le p} \big|1-(n^{-1}\wX_j^{T} \wX_j)^{-1}\big|.
\end{align*}
By \eqref{alias2}, $H_{31}=O(\log p)$. Thus, to prove \eqref{alias3}, it suffices to
show
\begin{align}\lbl{chongzi}
H_{32}':=\max_{q+1\le j\le p} \big|1-(n^{-1}\wX_j^{T} \wX_j)^{-1}\big|=o_p\Big(\frac{1}{\log p}\Big).
\end{align}
Now we prove this assertion. From assumption $q\leq n^{\delta}$ for some $\delta \in (0,1)$, we choose $\delta'\in (0, \min\{1/2, 1-\delta\}).$ This indicates that
\begin{align}\lbl{xihuanba}
\max\Big\{\frac{1}{2}, \delta\Big\}<1-\delta' < 1.
\end{align}
Note that, for $x>0$ and $s\in (0, 1)$ satisfying $|1-x^{-1}|\geq s$, one of the two inequalities $x^{-1}\geq 1+s$ and $x^{-1}\leq 1-s$ must hold. Equivalently, $x\leq \frac{1}{1+s}$ or $x\geq \frac{1}{1-s}$. Both of them imply that $|x-1|\geq \frac{s}{1+s}\geq \frac{1}{2}s$. Consequently,
\begin{align*}
P\Big(H_{32}'\geq n^{-\delta'}\Big)
\leq &  p\cdot \max_{q+1\le j\le p} P\Big(\big|1-(n^{-1}\wX_j^{T} \wX_j)^{-1}\big|\geq n^{-\delta'}\Big) \\
\leq & p\cdot \max_{q+1\le j\le p} P\Big(\Big|\frac{1}{n}\wX_j^{T} \wX_j-1\Big|\geq \frac{1}{2}n^{-\delta'}\Big).
\end{align*}
Recalling \eqref{jiutai} and that the matrix   $(\bd{I}_n-\H_a)^2=\bd{I}_n-\H_a$ has rank  $n-q$, we know $\wX_j^{T} \wX_j \sim \chi^2(n-q).$ Denote $n'=n-q$. Observe that $|\frac{y}{n}-1|\geq (1/2)n^{-\delta'}$ implies $|y-n| \geq (1/2)n^{1-\delta'}$, which implies $|y-n'| \geq (1/2)n^{1-\delta'}-q$. Use \eqref{xihuanba} and the assumption $q\leq n^{\delta}$ to see $|y-n'|/\sqrt{n'}\geq [(1/2)n^{1-\delta'}-q]/\sqrt{n'} \sim (1/2)n^{(1/2)-\delta'}.$ By definition, $0<(1/2)-\delta'<1/2$. Similar to \eqref{LDPDZ}, we have
\begin{align*}
P\big(H_{32}'\geq n^{-\delta'}\big)
\leq & p \cdot P\Big(\Big|\frac{1}{\sqrt{n'}}(\chi^2(n')-n')\Big|\geq \frac{1}{3}n^{(1/2)-\delta'}\Big) \\
 \leq & p \cdot \exp \big(C\cdot n^{1-2\delta'}\big),
\end{align*}
which is equal to $o(1)$ by assumption $p=o(n^3)$. This says $H_{32}'= o_p(n^{-\delta'})$, which implies \eqref{chongzi},  and hence \eqref{alias3} holds as aforementioned. \hfill$\square$

\medskip

Recall the error vector  $\bme=(\varepsilon_1, \cdots, \varepsilon_n)^T\in \mathbb{R}^n$ in the linear regression from Section \ref{regress_3}. We assume  $\{\varepsilon_i;\, 1\leq i \leq n\}$ are i.i.d. random variables.

\begin{prop}\lbl{CLT_for_3} Let $T^{(3)}_{sum}$ be defined as in \eqref{Stat_3}.  Assume  \eqref{con_(C2)} and \eqref{con_(C3)} hold and \eqref{assumption_A3} also holds with ``$\bd{\Sigma}$'' replaced by ``$\bms_{b|a}$''. Suppose $p=o(n^3)$, $q=o(p)$, $q\leq n^{\delta}$ for some $\delta \in (0,1)$ and $E(|\epsilon_1|^{\ell}) < \infty$ with  $\ell=14(1-\delta)^{-1}$.
Under $H_0$ from \eqref{h1} we have $T^{(3)}_{sum}$ converges to $N(0, 1)$ in distribution  as $p\to\infty$.
\end{prop}
\noindent\textbf {Proof of Proposition \ref{CLT_for_3}}. Recall the notation from Section \ref{regress_3}. In particular,
\begin{align*}
&\hat{\bms}_{b|a}=n^{-1}\wX_b^{T} \wX_b,\ \ \ \ \hat{\sigma}^2= (n-q)^{-1}\hat{\bme}^{T}\hat{\bme},\\
&\widehat{\tr(\bms^2_{b|a})}=\frac{n^2}{(n+1-q)(n-q)}\Big\{\tr\big(\hat{\bms}_{b|a}^2\big)-\frac{1}{n-q}
\tr^2(\hat{\bms}_{b|a})\Big\}
\end{align*}
and
\begin{align}\lbl{zhuyao3}
T^{(3)}_{sum}=\frac{n^{-1}\hat{\bm\varepsilon}^{T} \X_b\X_b^{T} \hat{\bme} -n^{-1}(n-q)(p-q)\hat{\sigma}^{2}}{\sqrt{2\hat{\sigma}^{4}\,\widehat{\mbox{tr}\big(\bms_{b|a}^2\big)}}}.
\end{align}
Next we will first derive a workable form for the main ingredient $\hat{\bm\varepsilon}^{T} \X_b\X_b^{T} \hat{\bme}$ above.

Recall $\hat{\bme}=(\I_n-\H_a)\Y$. Under the null hypothesis in  \eqref{h1}, $\bd{Y}=\X_{a}\bb_a+\bme$, where $\bme=(\varepsilon_1, \cdots, \varepsilon_n)^T\in \mathbb{R}^n$, and $\{\varepsilon_i;\, 1\leq i \leq n\}$ are i.i.d. random variables  with $E\varepsilon_1=0$ and $\mbox{Var}(\varepsilon_1)=\sigma^2$. By assumption, $\{\varepsilon_i;\, 1\leq i \leq n\}$ are also independent of $\{\X_i;\, 1\leq i \leq n\}$.   Recalling \eqref{nishuowo}, we denote
\begin{align*}
\X_{b}=(\X_{1b}, \cdots, \X_{nb})^T\ \ \ \mbox{and}\ \ \ \wX_b=(\wX_{q+1}, \cdots, \wX_p)=(\bd{I}_n-\H_a)\X_{b}.
\end{align*}
Since $\H_a=\mX_a(\mX_a^{T} \mX_a)^{-1}\mX_a^{T}$, both  $\X_{a}^T(\I_n-\H_a)=\bd{0}$ and $(\I_n-\H_a)\X_{a}=\bd{0}$. Thus,  $\hat{\bme}=(\I_n-\H_a)\bme$. By definition,
\begin{align}\lbl{huiguo}
\hat{\sigma}^2= \frac{1}{n-q}\hat{\bme}^{T} \hat{\bme}=\frac{1}{n-q}\bme^T(\I_n-\H_a)\bme
\end{align}
and
\begin{align*}
\hat{\bm\varepsilon}^{T} \X_b\X_b^{T} \hat{\bme}
=& \bme^T(\I_n-\H_a)\X_b\X_b^{T}(\I_n-\H_a)\bme\\
= & \|(\wX_{q+1}, \cdots, \wX_p)^T\bme\|^2\\
= & \sum_{j=q+1}^p(\wX_{j}^T\bme)^2.
\end{align*}
It is easy to verify that $E(\bme^T(\I_n-\H_a)\bme)=\sigma^2\mbox{tr}(\I_n-\H_a)=(n-q)\sigma^2$. Consequently,
\begin{align}\lbl{xiaopingguo}
E\hat{\sigma}^2=\sigma^2.
\end{align}
According to \eqref{jiutai}, $\wX_{j}=(\bd{I}_n-\H_a)\bd{u}_j$ and $\bd{u}_j \sim N(\bd{0}, \bd{I}_n)$. Hence,
\begin{align}\lbl{Haerxi}
(\wX_{j}^T\bme)^2=[\bd{u}_j^T(\bd{I}_n-\H_a)\bme]^2=\|(\bd{I}_n-\H_a)\bme\|^2\cdot (\bd{u}_j^T\bd{e}_2)^2,
\end{align}
where $\bd{e}_2:=(\bd{I}_n-\H_a)\bme/\|(\bd{I}_n-\H_a)\bme\|$. Eventually, we arrive at an ideal form to work with, that is,
\begin{align}\lbl{zouzheqiao}
\frac{1}{n}\hat{\bm\varepsilon}^{T} \X_b\X_b^{T} \hat{\bme}=\frac{1}{n}\bme^T(\bd{I}_n-\H_a)\bme\cdot \sum_{j=q+1}^{p} (\bd{u}_j^T\bd{e}_2)^2.
\end{align}

Now we start to prove the central limit theorem.
The assumption  $E(|\epsilon_1|^{\ell}) < \infty$ with  $\ell=14(1-\delta)^{-1}$ implies that $E(|\epsilon_1|^{2k}) < \infty$ with  $k=[\frac{6}{1-\delta}]+1$. It follows from  Lemma \ref{Matrix_moment}(i) that
\begin{align}\lbl{jiangchongsuan}
\bme^T(\bd{I}_n-\H_a)\bme=(n-q)\sigma^2 +O_p(\sqrt{n}).
\end{align}
This and \eqref{huiguo} imply that
\begin{align}\lbl{zuguo}
\hat{\sigma}^2=\sigma^2 +O_p(n^{-1/2}).
\end{align}
By assumption $q \leq n^{\delta}$ for some $\delta \in (0,1)$, we see
\begin{align}\lbl{genju}
\frac{1}{n}\bme^T(\bd{I}_n-\H_a)\bme=\sigma^2 +O_p\big(n^{-\delta'}\big),
\end{align}
with $\delta'=\min\{1-\delta, 1/2\}$.

In lieu of the explanation before \eqref{jiutai}, $\{\bd{u}_j;\, q+1\leq j \leq p\}$ are independent of $\X_{a}$ and $\bme$, and hence are independent of the unit random vector $\bd{e}_2$. We then have from Lemma \ref{Aux_Gaussian} that  $(\bd{u}_{q+1}^T\bd{e}_2, \cdots, \bd{u}_{p}^T\bd{e}_2)^T$ has distribution $N(\bd{0}, \bd{\Sigma}_{bb\cdot a})$ and is also independent of $\|(\bd{I}_n-\H_a)\bme\|$.  By assumption, \eqref{assumption_A3} holds with ``$\bd{\Sigma}$'' replaced by ``$\bms_{b|a}$''. From Theorem \ref{theorem_1} and \eqref{youshayi} we have
\begin{align}\lbl{haochide}
\frac{\sum_{j=q+1}^{p} (\bd{u}_j^T\bd{e}_2)^2-(p-q)}{\sqrt{2\,\mbox{tr}(\bms_{b|a}^2)}} \to N(0, 1)
\end{align}
in distribution as $p\to \infty.$ From \eqref{huiguo}, we know $\frac{1}{n}\bme^T(\bd{I}_n-\H_a)\bme=\frac{n-q}{n}\hat{\sigma}^2$. By \eqref{zouzheqiao} and \eqref{haochide} we have
\begin{align*}
\frac{\frac{n\hat{\sigma}^{-2}}{n-q}\cdot \big(n^{-1}\hat{\bm\varepsilon}^{T} \X_b\X_b^{T} \hat{\bme}\big) -(p-q)}{\sqrt{2\,\mbox{tr}(\bms_{b|a}^2)}} \to N(0, 1).
\end{align*}
Use the fact $n/(n-q)\to 1$ to see
\begin{align}\lbl{shengliko}
%\frac{n}{n-q}\cdot
\frac{ n^{-1}\hat{\bm\varepsilon}^{T} \X_b\X_b^{T} \hat{\bme} -n^{-1}(n-q)(p-q)\hat{\sigma}^{2}}{\sqrt{2\hat{\sigma}^{4}\,\mbox{tr}(\bms_{b|a}^2)}} \to N(0, 1).
\end{align}
By Proposition \ref{sonfather},  $\widehat{\mbox{tr}\big(\bms_{b|a}^2\big)}/\mbox{tr}\big(\bms_{b|a}^2\big)\to 1$ in probability. We then arrive at
\begin{align*}
T^{(3)}_{sum}=\frac{n^{-1}\hat{\bm\varepsilon}^{T} \X_b\X_b^{T} \hat{\bme} -n^{-1}(n-q)(p-q)\hat{\sigma}^{2}}{\sqrt{2\hat{\sigma}^{4}\,\widehat{\mbox{tr}\big(\bms_{b|a}^2\big)}}} \to N(0, 1)
\end{align*}
through an application of the Slutsky lemma. The proof is completed. \hfill$\square$

\medskip

\noindent{\bf Proof of Theorem \ref{thlm}.}  Parts (i) and (ii) follow from Propositions S\ref{CLT_for_3} and S\ref{Extreme_for_3}, respectively.  The setting here is the same as those in the two propositions, so we will continue to use the same notation in the two propositions to prove  (iii) for the asymptotic independence. Recall the definition
\begin{align*}
T^{(3)}_{sum}=
\frac{n^{-1}\hat{\bm\varepsilon}^{T} \X_b\X_b^{T} \hat{\bme} -n^{-1}(n-q)(p-q)\hat{\sigma}^{2}}{\sqrt{2\hat{\sigma}^{4}\,\widehat{\mbox{tr}\big(\bms_{b|a}^2\big)}}}.
\end{align*}
By \eqref{huiguo} and \eqref{zouzheqiao}, we have
\begin{align*}
\frac{1}{n}\hat{\bm\varepsilon}^{T} \X_b\X_b^{T} \hat{\bme}=\frac{n-q}{n}\hat{\sigma}^2 \sum_{j=q+1}^{p} (\bd{u}_j^T\bd{e}_2)^2.
\end{align*}
Thus,
\begin{align*}
T^{(3)}_{sum}
=&\frac{n-q}{n}\cdot \frac{\sum_{j=q+1}^{p} (\bd{u}_j^T\bd{e}_2)^2-(p-q)}{\sqrt{2\widehat{\tr(\bms^2_{b|a})}}}\\
 = & \tilde{T}^{(3)}_{sum} +  (\omega_p-1)\cdot\tilde{T}^{(3)}_{sum}
-\frac{q}{n}\cdot \omega_p\tilde{T}^{(3)}_{sum}
\end{align*}
where
\begin{align*}
\tilde{T}^{(3)}_{sum}=: \frac{\sum_{j=q+1}^{p} (\bd{u}_j^T\bd{e}_2)^2-(p-q)}{\sqrt{2\tr(\bms^2_{b|a})}}\ \ \ \mbox{and}\ \ \ \omega_p:=\Big(\frac{\tr(\bms^2_{b|a})}{\widehat{\tr(\bms^2_{b|a})}}\Big)^{1/2}.
\end{align*}
By Proposition \ref{sonfather},  $\widehat{\mbox{tr}\big(\bms_{b|a}^2\big)}/\mbox{tr}\big(\bms_{b|a}^2\big)\to 1$ in probability. Hence $\omega_p \to 1$ in probability. Also, $\tilde{T}^{(3)}_{sum} \to N(0, 1)$ by \eqref{haochide}. This together with the assumption $q=O(n^{\delta})$ for some $\delta \in (0,1)$ implies that
\begin{align}\lbl{henzhongyao}
T^{(3)}_{sum}
=\frac{\sum_{j=q+1}^{p} (\bd{u}_j^T\bd{e}_2)^2-(p-q)}{\sqrt{2\,\tr(\bms^2_{b|a})}} +o_p(1).
\end{align}
On the other hand, based on  \eqref{alias2} and \eqref{woyaode},
\begin{align}\lbl{xinte}
& T^{(3)}_{max}-2\log(p-q)+\log\log(p-q)\nonumber\\
 = & H_{31}' +o_p(1) \nonumber\\
=& \max_{q+1\le j\le p}\big\{n^{-1}\sigma^{-2}\bme^{T} \wX_j\wX_j^{T} \bme\big\} -2\log(p-q)+\log\log(p-q) +o_p(1) \nonumber\\
=& \max_{q+1\leq j \leq p}\Big\{\frac{\|(\bd{I}_n-\H_a)\bme\|^2}{n\sigma^{2}}\cdot\big(\bd{u}_j^T\bm e_2\big)^2\Big\} -2\log(p-q)+\log\log(p-q) +o_p(1)
\end{align}
by\eqref{Haerxi}.
%that $\wX_{j}=(\bd{I}_n-\H_a)\bd{u}_j$.
Remember that $\bd{e}_2=(\bd{I}_n-\H_a)\bme/\|(\bd{I}_n-\H_a)\bme\|$ is independent of $\bd{u}_j$ because $\bd{u}_j, \bd{H}_a, \bme$ are independent. It follows that
\begin{align}\lbl{suibuzhao}
&\Big|\max_{q+1\leq j \leq p}\Big\{\frac{\|(\bd{I}_n-\H_a)\bme\|^2}{n\sigma^{2}}\cdot\big(\bd{u}_j^T\bm e_2\big)^2\Big\}- \max_{q+1\leq j \leq p}\big(\bd{u}_j^T\bm e_2\big)^2\Big| \nonumber\\
= & \Big|\frac{\|(\bd{I}_n-\H_a)\bme\|^2}{n\sigma^{2}}-1\Big|
\cdot\max_{q+1\leq j \leq p}\big(\bd{u}_j^T\bm e_2\big)^2.
\end{align}
According to \eqref{genju}, we know that
\begin{align*}
\frac{\|(\bd{I}_n-\H_a)\bme\|^2}{n\sigma^{2}}=\frac{1}{n\sigma^2}\bme^T(\bd{I}_n-\H_a)\bme=1  +O_p\big(n^{-\delta'}\big)
\end{align*}
with $\delta'=\min\{1-\delta, 1/2\}$.
As a consequence,
\begin{align}\lbl{taigaoxingle}
\Big|\frac{\|(\bd{I}_n-\H_a)\bme\|^2}{n\sigma^{2}}-1\Big|
%=O_p\Big(\frac{q}{n}+\frac{1}{\sqrt{n}}\Big)
=O_p\Big(\frac{1}{n^{\delta'}}\Big).
\end{align}
From the discussion between \eqref{genju} and \eqref{haochide}, it holds that
\begin{align}\lbl{compared}
(\bd{u}_{q+1}^T\bd{e}_2, \cdots, \bd{u}_{p}^T\bd{e}_2)^T \sim N(\bd{0}, \bd{\Sigma}_{bb\cdot a}).
\end{align}
By assumption, \eqref{assumption_A3}  holds with ``$\bd{\Sigma}$'' replaced by ``$\bms_{b|a}$'', we know Assumption \eqref{condition_2} also holds by the discussion below \eqref{assumption_A3}. Due to Theorem \ref{theorem_2}, we get
\begin{align*}
& \max_{q+1\leq j \leq p}(\bd{u}_j^T\bd{e}_2)^2-2\log (p-q) +\log\log (p-q) \ \mbox{converges weakly to a distribution}\\
& \mbox{with cdf}\ \exp\Big\{-\frac{1}{\sqrt{\pi}}e^{-x/2}\Big\}
\end{align*}
for any $x\in \mathbb{R}$ as $p\to \infty$. By assumption, $p=o(n^3)$, the above particularly implies  $\max_{q+1\leq j \leq p}(\bd{u}_j^T\bd{e}_2)^2=O(\log n)$. This together with \eqref{suibuzhao} and \eqref{taigaoxingle} concludes that
\begin{align*}
\max_{q+1\leq j \leq p}\Big\{\frac{\|(\bd{I}_n-\H_a)\bme\|^2}{n\sigma^{2}}\cdot\big(\bd{u}_j^T\bm e_2\big)^2\Big\}= \max_{q+1\leq j \leq p}\big(\bd{u}_j^T\bm e_2\big)^2 +o_p(1).
\end{align*}
According to \eqref{xinte}, we have
\begin{align*}
&T^{(3)}_{max}-2\log(p-q)+\log\log(p-q)\\
=& \max_{q+1\leq j \leq p}\big(\bd{u}_j^T\bm e_2\big)^2 -2\log(p-q)+\log\log(p-q)+o_p(1).
\end{align*}
Joining this with \eqref{henzhongyao} and \eqref{compared}, by Theorem \ref{theorem_3} and Lemma \ref{small_lemma}, we obtain $T^{(3)}_{sum}$ and $T^{(3)}_{max}-2\log(p-q)+\log\log(p-q)$ are asymptotically independent. \hfill$\square$

\end{document}